\documentclass[11pt]{article}

\usepackage[T1]{fontenc}
\usepackage[utf8]{inputenc}
\usepackage{lmodern}
\usepackage[margin=1in]{geometry}
\usepackage{microtype}
\usepackage{amsmath,amssymb,amsthm}
\usepackage{mathtools}
\usepackage{graphicx}
\usepackage{enumitem}
\usepackage{xcolor}
\usepackage{bbm}
\usepackage{etoc}

\definecolor{ReferencePink}{HTML}{C02678}
\usepackage[
  colorlinks=true,
  linkcolor=ReferencePink,
  citecolor=ReferencePink,
  urlcolor=ReferencePink,
  bookmarksnumbered=true
]{hyperref}
\usepackage[nameinlink,capitalize,noabbrev]{cleveref}

\theoremstyle{plain}
\newtheorem{theorem}{Theorem}[section]
\newtheorem{lemma}[theorem]{Lemma}
\newtheorem{proposition}[theorem]{Proposition}
\newtheorem{corollary}[theorem]{Corollary}
\theoremstyle{plain}
\newtheorem{definition}[theorem]{Definition}
\theoremstyle{plain}
\newtheorem{assumption}[theorem]{Assumption}
\theoremstyle{plain}

\theoremstyle{plain}
\newtheorem{remark}[theorem]{Remark}
\numberwithin{equation}{section}

\newcommand{\p}{\mathcal{P}}
\renewcommand{\subset}{\subseteq}
\renewcommand{\epsilon}{\varepsilon}

\renewcommand{\r}{\mathbb{R}}
\newcommand{\n}{\mathbb{N}}

\newcommand{\la}{\left\langle}
\newcommand{\ra}{\right\rangle}

\newcommand{\1}{\mathbbm{1}}

\title{Sharp Lower Bounds for Parallel Log-Concave Sampling}
\author{Evan Dogariu\vspace{0pt}\\Courant Institute, NYU\vspace{0pt}\\\texttt{ed2719@nyu.edu}\vspace{-7pt}}
\date{}

\begin{document}
\maketitle
\begin{abstract}
How many rounds of parallel queries are needed to sample from a smooth, strongly log-concave distribution on $\r^d$? For every fixed condition number $\kappa \in (1, \infty)$, we prove that any sampler making polynomially many queries per round requires $\Omega_\kappa(\log d)$ rounds to achieve constant total variation error in the worst case. The lower bound holds even if queries return derivatives of all orders. Together with the first-order parallel Picard algorithm of \cite{zhou2025parallel}, this establishes a depth complexity of $\Theta_\kappa(\log d)$ for parallel log-concave sampling with $\operatorname{poly}(d)$ width.

We then study centered Gaussian targets to understand how much width is needed at a given depth. From \cite{chewi2023query,zhou2026complexitygaussiansampling}, it is known that either one round of width $d$ or $\mathcal{O}(\sqrt{\kappa} \log d)$ rounds of constant width suffice for first-order Gaussian sampling. 
At fixed $\kappa$ we characterize the depth-width tradeoff interpolating between these regimes:
when using $ \lceil c \log d\rceil$ rounds, if $0 < c < c_\star(\kappa)$ then the optimal width at constant TV error is $\widetilde{\Theta}(d^{1-c/c_\star(\kappa)})$, while if $c \geq c_\star(\kappa)$ then it is $\Theta_\kappa(1)$. Here, $c_\star(\kappa)$ is an explicit constant with $c_\star(\kappa) \sim \sqrt{\kappa} / 8$ as $\kappa \to \infty$.
\end{abstract}

% Remove these three lines if a separate contents page is unnecessary.

\etocdepthtag.toc{main}

\etocsettagdepth{main}{subsection}
\etocsettagdepth{appendix}{section}
\tableofcontents

\clearpage
\section{Introduction}

How much sequential computation is necessary to sample from a high-dimensional probability distribution? Many sampling algorithms generate a trajectory whose successive steps depend on earlier oracle responses. Parallel sampling algorithms seek to shorten this dependence by evaluating the target potential and its derivative(s) at many points simultaneously. For smooth strongly log-concave distributions, recent algorithms achieve polylogarithmic depth \cite{anari2024fast}, and subsequently logarithmic depth \cite{zhou2025parallel}, using polynomially many queries per round. This raises a converse question: can larger batches reduce the number of rounds further, or is logarithmic depth an intrinsic requirement of sampling?

We resolve this question for smooth strongly log-concave distributions with fixed condition number $\kappa \in (1, \infty)$. For any fixed $\delta \in (0, 1)$, any parallel sampler achieving constant TV error using $\exp(d^{1-\delta})$ many queries per round requires $\Omega_{\kappa, \delta}(\log d)$ rounds in the worst case, matching the $\mathcal{O}_\kappa(\log d)$ upper bound of \cite{zhou2025parallel}. The lower bound holds even when the mode is known, the potential is $C^\infty$, and each query returns all derivatives of the potential. We also characterize the tradeoff between the number of rounds and the number of queries per round for centered Gaussian targets. In this case, we identify an explicit threshold at which the necessary batch size falls from a positive power of the dimension to a constant.

\paragraph{Problem setting.}
We consider sampling from a probability distribution $\pi$ on $\r^d$ with density proportional to $e^{-V}$, given oracle access to the potential $V$ and its gradient. Our regularity assumptions are as follows:

\begin{assumption}[Strong log-concavity and smoothness]
\label{assumption:stronglylogconcave}
Let $d \in \n$ and $\kappa \in [1,\infty)$. The target $\pi$ has density proportional to $e^{-V}$ for a differentiable potential $V:\r^d\to\r$ that is $\alpha$-strongly convex and has $\beta$-Lipschitz gradient, where
\begin{equation}
    0<\alpha\leq\beta<\infty,
    \qquad
    \frac{\beta}{\alpha}\leq\kappa.
\end{equation}
We refer to $\kappa$ as a bound on the condition number. When $V$ is twice differentiable, these conditions are equivalent to
\begin{equation}
        \alpha\1_d\preceq\nabla^2 V(z)\preceq\beta\1_d
    \qquad (\forall z\in\r^d).
\end{equation}
\end{assumption}
% We take the curvature bounds $\alpha$ and $\beta$ to be available to the sampler.
We also assume that the mode of $\pi$ is known, both to provide a starting point for sampling algorithms and to separate the difficulty of sampling from that of minimizing $V$. By translating the target and adding a constant to $V$, we may impose the following normalization:

\begin{assumption}[Known mode]
\label{assumption:knownmode}
In addition to Assumption~\ref{assumption:stronglylogconcave}, assume that
\begin{equation}
    \nabla V(0)=0,
    \qquad
    V(0)=0.
\end{equation}
\end{assumption}
Assumptions \ref{assumption:stronglylogconcave} and \ref{assumption:knownmode} yield the standard setting of smooth strongly log-concave sampling.
 For a general treatment of log-concave sampling, see the textbook \cite{chewi2023logconcave}.

\paragraph{Parallel oracle model.}
In each round, an algorithm submits a batch of query points and receives the corresponding values of $(V,\nabla V)$. The batch may depend on the algorithm's internal randomness and all responses from previous rounds, but its queries are chosen before any response from the current round is received. The algorithm then outputs a random point whose law must approximate $\pi$ in total variation (TV) distance, uniformly over the target class.

We call the number of rounds the \emph{depth} and the maximum number of queries in any round the \emph{width}.
Computation between oracle calls is unrestricted: our lower bounds concern the information available to the sampler and apply to arbitrary randomized algorithms without requiring a particular dynamics, discretization, or particle representation.

\paragraph{Optimal depth for parallel log-concave sampling.}
Parallel simulation methods have substantially reduced the adaptive complexity of sampling \cite{anari2024fast,zhou2025parallel}. The following consequence of the parallel Picard method of Zhou and Sugiyama gives the upper bound relevant to our setting:\footnote{For this upper bound we assume the algorithm has access to $\alpha, \beta$. This does not affect our lower bound.}

\begin{theorem}[Corollary 4.3 of \cite{zhou2025parallel}]
\label{theorem:upperboundalg}
Fix $\kappa\in(1,\infty)$. Under Assumptions~\ref{assumption:stronglylogconcave} and~\ref{assumption:knownmode}, there exists a parallel sampling algorithm that achieves TV error $ \leq 0.01$ using $\mathcal{O}_\kappa(d\log d)$ queries per round and $\mathcal{O}_\kappa(\log d)$ rounds.
\end{theorem}

Our main result shows that the dimension dependence of this round bound is optimal, even if the algorithm is allowed a much larger polynomial query budget:

\begin{theorem}[Corollary \ref{corollary:main}, informal]
\label{theorem:main}
Fix $\kappa\in(1,\infty)$. Under Assumptions~\ref{assumption:stronglylogconcave} and~\ref{assumption:knownmode}, any parallel sampling algorithm that achieves TV error $ \leq 0.99$ using $\operatorname{poly}(d)$ queries per round requires ${\Omega}_\kappa(\log d)$ rounds in the worst case.

More precisely, there is some constant $\zeta(\kappa)$ such that for any fixed $0 < c < \zeta(\kappa)$, any parallel sampling algorithm that achieves TV error $\leq 0.99$ with depth $\lceil c \log d\rceil$ requires worst-case width
\begin{equation}
\label{eqn:stretchedexponentialintro}
    \exp\left(\Omega_\kappa\left(d^{1-c/\zeta(\kappa)}\right)\right).
\end{equation}
\end{theorem}

\begin{remark}
The constants $0.01$ and $0.99$ are illustrative and can be taken arbitrarily close to $0$ and $1$, respectively, at the cost of changing the hidden constants. Similarly, the condition number $\kappa$ can be taken arbitrarily close to $1$. The hard instances have $C^\infty$ potentials, and the same lower bound holds when each query returns the potential and all of its derivatives at the queried point.
\end{remark}

Together, these results establish a worst-case round complexity of $\Theta_\kappa(\log d)$ for fixed-$\kappa$ log-concave sampling with polynomially many queries per round. This contrasts with Gaussian targets, which can be sampled exactly in one parallel round of $d$ gradient queries. Theorem \ref{theorem:main} therefore yields a separation between log-concave and Gaussian sampling in round complexity, which complements the separation in total query complexity established concurrently by Zhou and Suzuki \cite{zhou2026query}.

Theorems \ref{theorem:upperboundalg} and \ref{theorem:main} also identify two regimes at the logarithmic round scale. If the sampler uses $T = \lceil c \log d \rceil$ rounds for a fixed $c \in (0, \infty)$, then:
\begin{itemize}
    \item For any $c < \zeta(\kappa)$, Theorem \ref{theorem:main} rules out sampling with $\operatorname{poly}(d)$ queries per round. Informally, this means that a successful sampler cannot improve on a brute force solution (e.g. discretizing $\r^d$ and reconstructing $V$), see Remark \ref{remark:optimality}.
    \item If $c$ is sufficiently large, Theorem \ref{theorem:upperboundalg} provides a sampler using only $\mathcal{O}_\kappa(d \log d)$ queries per round. The sequence of interactions is long enough to allow a query-efficient algorithm.
\end{itemize}
Thus, allowing more rounds changes whether an efficient sampler exists. This raises a finer question: where does this transition occur, and how does the required number of queries decrease as the number of rounds approaches the transition? The width lower bound (\ref{eqn:stretchedexponentialintro}) suggests a stretched-exponential tradeoff between depth and width, but it is unknown if this is correct or how the true threshold for $c$ changes as $\kappa$ increases.

\paragraph{A sharp depth-width tradeoff for Gaussian sampling.}
We investigate this question in the Gaussian setting, where we can identify an explicit threshold and characterize the tradeoff leading up to it. As mentioned, centered Gaussians can be sampled using one round of width $d$; this is fully parallel but wasteful, and can be considered the brute force approach. Indeed, such targets can be sampled using $\mathcal{O}_\kappa(\log d)$ rounds of width 1 \cite{chewi2023query,zhou2026complexitygaussiansampling}. 

Our second result determines how additional depth reduces the required width. To state it, normalize the spatial scale and define
\begin{equation}
    \Pi_{d,\kappa}^{(\operatorname{Gaussian})}
    :=
    \left\{
        \mathcal{N}(0,\Sigma):
        \1_d\preceq\Sigma^{-1}\preceq\kappa\1_d
    \right\}.
\end{equation}
Throughout, $\log$ denotes the natural logarithm, tildes in asymptotic notation suppress factors polylogarithmic in $d$.

\begin{theorem}[Theorem~\ref{theorem:precisegaussian}, informal]
\label{theorem:gaussian}
Fix $\kappa\in(1,\infty)$ and let
\begin{equation}
    c_\star(\kappa)
    :=
    \frac{1}{
        4\log\!\left(
            \frac{\sqrt{\kappa}+1}{\sqrt{\kappa}-1}
        \right)
    }.
\end{equation}
Fix $c \in (0, \infty)$ and let $T := \lceil c \log d\rceil$ be the number of rounds. Then, as $d \to \infty$ the following hold:
    \begin{enumerate}
        \item[(i)] If $c > c_\star(\kappa)$, then there exists a parallel sampling algorithm for $\Pi_{d,\kappa}^{(\operatorname{Gaussian})}$ which achieves TV error 0.01 using 2 queries per round and $T$ rounds.
        \item[(ii)] If $c \leq c_\star(\kappa)$, then there exists a parallel sampling algorithm for $\Pi_{d,\kappa}^{(\operatorname{Gaussian})}$ which achieves TV error 0.01 using ${\mathcal{O}}_{\kappa}\left(d^{1-{c}/{c_\star(\kappa)}}\right)$ queries per round and $T$ rounds.
        \item[(iii)] If $c < c_\star(\kappa)$, then any parallel sampling algorithm for $\Pi_{d,\kappa}^{(\operatorname{Gaussian})}$ which achieves TV error 0.99 in $T$ rounds must have worst-case width $\widetilde{\Omega}\left(d^{1-{c}/{c_\star(\kappa)}}\right)$.
    \end{enumerate}
\end{theorem}
\begin{remark}
    The critical number of rounds is also $c_\star(\kappa) \log d$ in the regime where $(\kappa, d)$ grow jointly with $\kappa = d^{o(1)}$, see Section \ref{subsection:gaussiangrowingkappa}.
    Note that $c_\star(\kappa) \sim \sqrt{\kappa} / 8$ as $\kappa \to \infty$.
\end{remark}

For each fixed $c \in (0, \infty)$, these bounds characterize the smallest worst-case width needed to achieve constant TV error. Below the threshold, the required width is
\begin{equation}
    \widetilde{\Theta}\left(d^{1-c/c_\star(\kappa)}\right).
\end{equation}
At or above the threshold, constant width suffices. Thus, the Gaussian result identifies both the depth threshold $c_\star(\kappa) \log d$ for constant-width sampling and the tradeoff leading up to it. Since the depth is logarithmic, this also marks a transition in total query complexity: below the threshold a positive power of $d$ is necessary, while above it $\mathcal{O}(\log d)$ queries suffice.

\paragraph{Open questions.}
For general strongly log-concave targets, we have identified the
logarithmic round scale at which polynomial width becomes possible.
The Gaussian result suggests two finer questions about this transition.

The first concerns the location of the threshold and its dependence on $\kappa$. Taking $d \to \infty$ at each fixed $\kappa$ and then letting $\kappa \to \infty$, the coefficient $\zeta(\kappa)$ in our lower bound remains of constant order (see Corollary \ref{corollary:main}) while the coefficient obtained from the upper bound of \cite[Corollary 4.3]{zhou2025parallel} grows as $\kappa\log\kappa$. Hence, it remains open whether ill-conditioning forces the coefficient of $\log d$ needed for polynomial-width sampling to grow with $\kappa$ and, if so, at what rate. For comparison,
the Gaussian threshold for constant-width sampling has coefficient
$c_\star(\kappa)\sim\sqrt{\kappa}/8$. From a practical perspective, closing this gap would help us identify whether ill-conditioning is better handled by increasing the width or depth of parallel samplers.

The second question concerns how the required width varies with the
round budget $T=\lceil c\log d\rceil$. For fixed $0<c<\zeta(\kappa)$
our width lower bound (\ref{eqn:stretchedexponentialintro}) is stretched-exponential with exponent decreasing linearly with $c$, but
whether a sampler can attain this dependence remains open even at
$\kappa=2$ and constant TV error. At the other end of the tradeoff,
Theorem \ref{theorem:upperboundalg} achieves logarithmic depth with
near-linear width, while \cite{openai2026subpolynomial} achieves
$d^{o(1)}$ total queries. It remains open whether $d^{o(1)}$ width
suffices at logarithmic depth and, if not,
how many rounds are necessary for subpolynomial width.

\subsection{Related work}

\paragraph{Parallel algorithms for sampling.}
Parallel sampling algorithms differ in what they evaluate simultaneously. Along a single Langevin trajectory, parallel variants of the randomized midpoint method evaluate the drift at several intermediate times in one batch \cite{shen2019midpoint,yu2024midpoint}, while parallel Picard methods use the preceding trajectory estimate to update many time points in one round \cite{anari2024fast,zhou2025parallel}. ParaDiGMS applies the same idea to a diffusion model, evaluating a learned denoiser at several noise levels in parallel \cite{shih2023parallel}. Rejection sampling also admits parallel implementations: one can test independent candidate paths in parallel \cite{beskos2005exact,chen2026exact} or verify several proposed denoising steps in a batch \cite{anari2025autospeculation}. In interacting particle methods, each round updates many particles using information shared across the current ensemble -- examples include SVGD, LAWGD, the Radon-Wasserstein sampler, the ensemble Kalman sampler, and affine-invariant interacting Langevin dynamics \cite{LiuWang2016_SVGD,chewi2020svgd,HessChildsetal2026_RadonWasserstein,garbuno2019ensemble,garbuno2019affine}.

\paragraph{Parallel complexity of log-concave sampling.}
For strongly log-concave targets with known mode, Zhou and Sugiyama \cite[Corollary 4.3]{zhou2025parallel} give a sampler that achieves TV error $\epsilon$ using
\begin{equation}
    \mathcal{O}\left(\kappa(1+\log\kappa)\log\left(\frac{d\log(2\kappa)}{\epsilon^2}\right)\right)
    \text{ rounds}, \quad
    \mathcal{O}\left(\frac{\kappa^2 d}{\epsilon^2}\log\left(\frac{d\log(2\kappa)}{\epsilon^2}\right)\right)
    \text{ queries per round}.
\end{equation}
At fixed $\kappa$ and $\epsilon$, this gives $\mathcal{O}(\log d)$ rounds with $\mathcal{O}(d\log d)$ queries per round. On the lower-bound side, Zhou, Wang, and Sugiyama \cite{zhou2025adaptive} construct potentials that conceal a random partition of the coordinates and show that $\widetilde{\Omega}(d)$ rounds are necessary at exponentially small TV error. When $\epsilon = \exp(-\Theta(d))$, this gives a lower bound of $\widetilde{\Omega}(\log(1/\epsilon))$ rounds. Theorem \ref{theorem:main} shows that even at constant TV error $\Omega_\kappa(\log d)$ rounds are necessary with polynomially many queries per round, which settles the dependence on $d$ at fixed $\kappa$ and $\epsilon$. The optimal dependence on $\kappa$ and $\epsilon$, as well as the width needed at logarithmic depth, remain open.

\paragraph{Total query complexity of log-concave sampling.}
Counting all queries across rounds gives a different picture already for Gaussian targets. Chewi et al.\ \cite{chewi2023query} use a block Krylov reduction to prove dimension-dependent query lower bounds for Gaussian sampling, which also apply to the larger strongly log-concave class. For centered Gaussians, the known upper bound is $\mathcal{O}(\min\{d,\sqrt{\kappa}\log(d/\epsilon)\})$ matrix-vector queries, which the lower bound of Zhou and Suzuki \cite{zhou2026complexitygaussiansampling} matches for $4 \leq \kappa \leq d^{2-\delta}$ and any fixed $\delta \in (0,2)$. The two terms correspond to different strategies for using precision matrix queries: querying $d$ basis vectors recovers it in one parallel round of width $d$, while polynomial approximation uses a sequence of repeated matrix-vector products at constant width. Theorem \ref{theorem:gaussian} characterizes the depth-width tradeoff and identifies the precise threshold for constant-width Gaussian sampling.

For general strongly log-concave targets at $\kappa=2$ and constant TV error, OpenAI \cite{openai2026subpolynomial} gives a $d^{o(1)}$-query upper bound when the mode is known. Concurrently with the writing of this paper, Zhou and Suzuki \cite{zhou2026query} prove an $\exp(\Omega((\log d)^{1/3}))$ query lower bound even with a constant warm start. Their result separates Gaussian and strongly log-concave sampling in total query complexity, complementing the separation in round complexity given by Theorem \ref{theorem:main}. 

\paragraph{Parallel complexity of diffusion sampling.}
In diffusion sampling, the oracle estimates the score of the Gaussian-smoothed target at a noise level chosen by the algorithm. Randomized midpoint and Picard methods achieve polylogarithmic round complexity under sufficiently small mean-square score error and suitable Lipschitz assumptions \cite{gupta2024midpoints,zhou2025parallel}. However, mean-square accuracy alone permits large errors at query locations that are rare under the noisy target. In this oracle model, Xun and Price \cite{xun2026query} prove a $\widetilde{\Omega}(\sqrt{d})$ lower bound on total score queries by showing that a sampler with a polynomial query budget must probe $\widetilde{\Omega}(\sqrt{d})$ distinct noise levels. Kou and Wang \cite{kou2026lowerbounds} obtain lower bounds on parallel rounds by concealing successive coordinate blocks, proving a $\widetilde{\Omega}(d^{1/3})$ bound for near-isotropic Gaussian mixtures and a tight $\Theta(d)$ bound for a family of anisotropic boxes. In our setting, the oracle answers every query exactly and so the potential itself must conceal the information needed in later rounds.

\paragraph{Parallel convex optimization.}
In both optimization and sampling, a fundamental question is whether batches of parallel oracle queries can reduce the number of sequential rounds. For convex optimization on the Euclidean unit ball, the answer depends on the objective accuracy $\epsilon$ relative to the dimension. When $d$ is sufficiently large relative to $1/\epsilon$, polynomial-width algorithms still require $\Omega(\epsilon^{-2})$ rounds for $1$-Lipschitz nonsmooth objectives and $\Omega(\epsilon^{-1/2})$ rounds for $1$-smooth objectives, matching subgradient descent and accelerated gradient descent, respectively \cite{diakonikolas2020lower}. In the nonsmooth case, Bubeck et al.\ \cite{bubeck2019complexity} show that the subgradient rate remains optimal down to $\epsilon \approx d^{-1/4}$ up to logarithmic factors; at finer accuracy, their parallel algorithm uses $\widetilde{\mathcal{O}}(d^{1/3}\epsilon^{-2/3})$ rounds. By contrast, for strongly log-concave sampling parallel queries already give an asymptotic speedup at fixed condition number and constant TV error.

\paragraph{Zero-chain lower bounds.}
In the oracle lower-bound tradition of \cite{nemirovsky1983problem}, Nemirovski's parallel construction \cite{nemirovski1994parallel} hides directions in the slopes of affine functions. Their offsets determine which pieces can attain the maximum at a query point. Bubeck et al.\ \cite{bubeck2019complexity} shield these pieces with a convex wall at large query norms, while Diakonikolas and Guzm\'an \cite{diakonikolas2020lower} use local smoothing to preserve the same information restriction for smooth objectives. Carmon et al.\ \cite{carmon2020lower} build a smooth zero-chain directly: while a hidden coordinate remains near zero, a flat cutoff keeps the next link in the chain inactive.

For sampling, the construction depends on what an oracle reply can disclose about a hidden block. With a counting oracle, Anari, Gao, and Rubinstein \cite{anari2024parallel} use affine codes of increasing dimension constructed so that a query which detects constraints in one block can leave later blocks indistinguishable from uniform. With a potential oracle, Zhou, Wang, and Sugiyama \cite{zhou2025adaptive} use local smoothing: consecutive hidden blocks interact only when their coordinate sums differ by enough. In the diffusion construction of Kou and Wang \cite{kou2026lowerbounds}, the target blocks are independent and the ordering enters through the approximate score, which can suppress a later block's contribution when an earlier consistency test fails. Our construction encodes such a test in the potential itself using a smooth cutoff in the spirit of \cite{carmon2020lower}: alignment under one hidden rotation activates the interaction involving the next.

\subsection{Statement of AI usage}
Two mathematical ideas in this paper were suggested by GPT-6 Astra and subsequently developed and refined by the author. These are (1) the randomized correction step of Lemma \ref{lemma:gaussiancorrectionlemma} (which is used to attain the optimal width for Gaussian sampling) and (2) the suggestion to use geometrically-decaying block sizes to simplify the Hessian bounds in Lemma \ref{lemma:theconstruction} (the author's initial spherical-gate zero-chain construction used probabilistic Hessian control). GPT-6 Astra and GPT-6.1 Sol were also used to help revise the exposition and verify bibliographic details. The author takes full responsibility for all results, arguments, writing, and references in this paper.

\section{Proof ideas and intuition}
\label{sec:intuition}

For the lower bound of Theorem \ref{theorem:main}, we follow the standard \textit{zero-chain} principle from optimization theory \cite{nemirovsky1983problem,carmon2020lower}. The idea is to construct hard instances whose hidden information must be uncovered in a prescribed sequential order, unless the algorithm makes sufficiently many queries to bypass this order.
Related sequential information-hiding constructions underlie lower bounds for parallel optimization \cite{nemirovski1994parallel,bubeck2019complexity,diakonikolas2020lower} and parallel sampling \cite{zhou2025adaptive,kou2026lowerbounds}.
We encode this structure through gates: opening each gate requires hidden information, while closed gates conceal the information needed to open later gates.

In our parallel sampling setting, we will construct probability distributions for which any sampler which makes polynomially many queries per round can only open at most one gate per round with high probability. The main challenge is to preserve strong convexity and smoothness while maintaining this information-hiding property. In our construction, controlling the curvature leads us to choose geometrically decreasing block dimensions, which yields the $\Omega(\log d)$ round lower bound. In this section we provide more details while remaining informal.

\paragraph{Single gate.}
Let $n \geq 2$ and $\epsilon \in (0, 1)$.
A useful starting point is the centered Gaussian distribution $\r^{2n}$ with density $\pi_U \propto e^{-V_U}$, where we fix a hidden rotation $U \in O(n)$ and define the potential $V_U: \r^{2n} \to \r$ via
\begin{equation}
    V_U(x, y) :=  \frac{1}{2}(\|x\|^2 + \|y\|^2) - \varepsilon \la y, Ux \ra \quad (\forall x, y \in \r^n).
\end{equation}
The interaction encourages $y$ to align with $Ux$. Indeed, a typical target sample has normalized correlation $\epsilon$ with fluctuations of scale $n^{-1/2}$: for every $t \in (0, 1)$ we have
\begin{equation}
    \mathbb{P}_{(X_\star, Y_\star) \sim \pi_U}\left(|\rho_\star - \epsilon| > t\right) \leq 4\exp\left(-\frac{nt^2}{32}\right), \quad \rho_\star := \frac{2 \la Y_\star, UX_\star \ra}{\|X_\star\|^2 + \|Y_\star\|^2}.
\end{equation}
By contrast, if $U \sim \operatorname{Haar}(O(n))$ is uniformly chosen and $(X, Y)$ is independent of $U$, then the normalized correlation concentrates around 0: for every $t \in (0, 1)$ we have
\begin{equation}
    \mathbb{P}_{U,X,Y}(|\rho| > t) \leq 2\exp\left(-\frac{nt^2}{2}\right), \quad \rho := \frac{2 \la Y, UX \ra}{\|X\|^2 + \|Y\|^2}.
\end{equation}
Even if one were to make $\operatorname{poly}(n)$ parallel queries, a simple union bound shows that the probability of any query obtaining normalized correlation at least $\epsilon/2$ vanishes as $n \to \infty$. 
Thus, producing a typical sample requires learning something about the hidden rotation. 

\paragraph{Multiple gates.}
This Gaussian alone does not give a strong round lower bound: every gradient query provides information about $U$, and many parallel queries can learn it quickly. Our construction instead arranges several such sampling tasks in a chain using independent hidden rotations $U_1, \ldots, U_r$. Information about $U_1$ is available immediately, information about $U_2$ becomes available only at points exhibiting the alignment associated with $U_1$, information about $U_3$ requires alignment with $U_2$, and so on. These alignment requirements produce the effect of \textit{gates} and introduce couplings between consecutive blocks. 

An algorithm with information about $U_j$ can choose queries opening gate $j$ and therefore expose $U_{j+1}$. However, all queries in a parallel batch are chosen before any responses arrive. Using this new information to open the next gate therefore requires another round. As in the Gaussian case, concentration makes guessing the missing alignment exponentially unlikely in the block dimension, so a polynomially large batch cannot bypass this requirement. This sequential structure produces a zero-chain, for which we can prove query lower bounds.

A typical target sample satisfies all the alignments simultaneously, including the terminal alignment involving $U_r$. By contrast, an algorithm using fewer than $r$ rounds is unlikely to satisfy this terminal test. This yields an event whose probability is close to one under the target and close to zero under the algorithm’s output, yielding a TV separation.

Let us provide an example of such a chain. Write $2n_j$ for the dimension of block $j$ and express $z = (z_1, \ldots, z_r)$ with $z_j = (x_j, y_j) \in \r^{2n_j}$. As a heuristic, a natural gate mechanism can be found in the candidate potential
\begin{equation}
    V_U(z) := \frac{1}{2}\|z\|^2 - \epsilon \left(R_1(z_1) + \sum_{j=2}^r g(s_{j-1}(z_{j-1}))R_j(z_j)\right), \quad s_{j}(x_j, y_j) := \frac{2\la y_j, U_jx_j\ra}{\|x_j\|^2 + \|y_j\|^2},
\end{equation}
where $g$ is a $C^\infty$ threshold function and each $R_j$ carries information about $U_j$. For example, one might take $R_j(x_j, y_j) := \|x_j\|^2 + \|y_j\|^2 + 2\la y_j, U_jx_j\ra$. If $z$ is such that all gates $j-1, \ldots, r-1$ are closed (i.e. $g(s_{j-1}) = \ldots = g(s_{r-1}) = 0$ on a neighborhood of $z$) then the oracle responses agree with that of the truncated potential
\begin{equation}
    \widetilde{V}_{U, j-1} := \frac{1}{2}\|z\|^2 - \epsilon \left(R_1(z_1) + \sum_{\ell=2}^{j-1} g(s_{\ell-1}(z_{\ell - 1}))R_\ell(z_\ell)\right),
\end{equation}
which depends only on $U_1, \ldots, U_{j-1}$. Hence, as long as these gates remain closed no information about $U_j, \ldots, U_r$ is revealed, and so polynomially many queries cannot overcome the exponential difficulty of opening later gates. 

As mentioned, the main challenge is to implement this spherical alignment chain while preserving strong convexity and smoothness. For some loose calculations, at typical points $\|z_j\| \asymp \sqrt{n_j}$ and so the quadratic interaction satisfies
\begin{equation}
    |R_j(z_j)| \lesssim n_j, \quad \|\nabla R_j(z_j)\| \lesssim \sqrt{n_j}.
\end{equation}
Since $s_{j-1}$ is a normalized quadratic form, differentiation of its gate gives
\begin{equation}
    \|\nabla g(s_{j-1}(z_{j-1}))\| \lesssim n_{j-1}^{-1/2}, \quad \|\nabla^2 g(s_{j-1}(z_{j-1}))\|_{\operatorname{op}} \lesssim n_{j-1}^{-1}
\end{equation}
where derivatives are taken with respect to $z_{j-1}$ and constants depend on $g$. The product rule therefore introduces Hessian terms of the form
\begin{equation}
    \frac{n_j}{n_{j-1}} \quad \text{and}\quad \sqrt{\frac{n_j}{n_{j-1}}}.
\end{equation}
To suppress these contributions it is natural to take $n_{j-1} = kn_j$ for sufficiently large fixed $k$; with $\epsilon$ fixed and small enough the condition number is $\mathcal{O}(1)$ with respect to the dimension. The geometric decay of block dimensions yields a number of blocks $r$ which scales as $\log d$, where $d = 2\sum_{j=1}^r n_j$ is the total ambient dimension. This is the origin of the $\log d$ round lower bound.

Our formal construction modifies this potential to make the curvature bounds hold globally while retaining the zero-chain property. With geometric block sizes $n_j = k^{r-j}b$ and a base block size $b$, we get total dimension \begin{equation}d = 2b \frac{k^r - 1}{k-1}.\end{equation}
At depth $T = \lceil c \log d\rceil$, taking $r = T+1$ leaves room for $b \asymp_\kappa d^{1-c \log k}$, which yields the stretched-exponential width lower bound (\ref{eqn:stretchedexponentialintro}) via spherical concentration.

\paragraph{Gaussian bounds.}
For the Gaussian bounds of Theorem \ref{theorem:gaussian}, we instead use the fact that a first-order query is equivalent to a matrix-vector product. Indeed, for a target $\pi = \mathcal{N}(0,\Sigma)$ with precision matrix $A := \Sigma^{-1}$, we have
\begin{equation}
    V(x) = \frac{1}{2}\la x, Ax \ra, \quad \nabla V(x) = Ax.
\end{equation}
In $T$ rounds we can therefore evaluate $q(A)g$ for any vector $g$ and any polynomial $q$ of degree at most $T$. Parallel queries allow us to evaluate several such products simultaneously.

% For the upper bound, consider a sampler which outputs
% \begin{equation}
%     X = \sum_{j=1}^Q p_j(A)g_j,
% \end{equation}
% where $g_j \sim \mathcal{N}(0,\1_d)$ are independent and $p_j$ are fixed polynomials of degree at most $T$. Such a sampler can be implemented using $T$ rounds and $Q$ queries per round, and its output is Gaussian with covariance
% \begin{equation}
%     p(A) := \sum_{j=1}^Q p_j(A)^2, \quad \operatorname{deg}p \leq 2T.
% \end{equation}
% By Lemma \ref{lemma:gaussianlowerboundstep1} any sampler has this form, with the caveat that the $p_j$ may depend measurably on the $g_j$ and oracle responses (and so $X$ need not be Gaussian). Ignoring this caveat for clarity, we find that a $T$-round sampler aims to approximate the target covariance $A^{-1}$ by a polynomial $p(A)$ of degree at most $2T$.

For sampling, we therefore seek to approximate the target covariance $A^{-1}$ by a polynomial $p(A)$ whose degree is at most $2T$ (since polynomial-vector products of degree $T$ produce covariance which is polynomial in $A$ with degree $2T$).\footnote{This argument regarding degree $2T$ is informal: a general algorithm can choose to evaluate randomized adaptive polynomials and produce non-Gaussian outputs, which breaks the underlying covariance calculation. For clarity we ignore this here, while in our proofs we handle arbitrary algorithms using sphere covering and Gaussian concentration.} Gaussian TV bounds \cite[Theorem 1.1]{devroye2018total} imply
\begin{equation}
    \operatorname{TV}\left(\mathcal{N}(0,p(A)),\, \mathcal{N}(0,A^{-1})\right)
    \asymp \min\left\{1,\; \|Ap(A)-\1_d\|_{F}\right\}.
\end{equation}
The Frobenius norm is at most $\sqrt{d}$ times the operator norm, and our hard spectral construction makes this comparison sharp up to logarithmic factors when $\operatorname{deg}p \leq 2T$. Since the eigenvalues of $Ap(A)-\1_d$ are $\lambda p(\lambda)-1$, this leads to the uniform relative approximation problem
\begin{equation}
\label{eqn:intuitionpolyapproximation}
\mathcal{E}_T(\kappa) := \mathop{\mathrm{inf}\vphantom{\mathrm{sup}}}\limits_{\mathstrut \deg p \leq 2T}\;
    \sup_{\mathstrut \lambda \in [1, \kappa]}
    |\lambda p(\lambda) - 1|.
\end{equation}
Standard polynomial approximation results yield an optimal error scaling
\begin{equation}
    \mathcal{E}_T(\kappa) \asymp_\kappa \left(\frac{\sqrt{\kappa}-1}{\sqrt{\kappa}+1}\right)^{2T},
\end{equation}
see Lemma \ref{lemma:chebyshevapproximation}.
Comparing this with the $d^{-1/2}$ relative accuracy needed for the TV bound gives
\begin{equation}
    \mathcal{E}_T(\kappa) \asymp_\kappa d^{-1/2} \implies T = c_\star(\kappa) \log d + O_\kappa(1).
\end{equation}
In particular, the threshold $c_\star(\kappa)$ is a direct consequence of the error rate for the polynomial approximation problem (\ref{eqn:intuitionpolyapproximation}).
The three statements of Theorem \ref{theorem:gaussian} leverage this approximation problem in different ways:
\begin{itemize}
        \item Above the threshold $c > c_\star(\kappa)$, the optimal covariance polynomial is a sum of two squares and can be implemented with two queries per round. Its relative error is already $o(d^{-1/2})$, giving vanishing TV error.

    \item At or below the threshold $c \leq c_\star(\kappa)$, we use a randomized correction based on $n$ independent pairs of Gaussian probes. We reserve the last round to compute the rejection weights, so the covariance approximation has relative error $\mathcal{E}_{T-1}(\kappa)$. The corrected sampler uses $2n$ queries per round and has TV error at most $\sqrt{12d\mathcal{E}_{T-1}(\kappa)^2/n}$. For fixed $\kappa$, choosing $n$ of order $d\mathcal{E}_{T-1}(\kappa)^2 \asymp_\kappa d^{1-c/c_\star(\kappa)}$ therefore suffices for constant TV accuracy.

    \item For the matching lower bound, we combine the polynomial reduction of Lemma \ref{lemma:gaussianlowerboundstep1} with a fixed spectrum constructed from the hard points of the polynomial approximation lower bound of Lemma \ref{lemma:chebyshevapproximation}(ii).
\end{itemize}

\section{Setting}
As mentioned, we focus on the class of target probability measures which are absolutely continuous and strongly log-concave and log-smooth with known mode (located at 0 without loss of generality) and condition number $\kappa$, i.e. those satisfying Assumptions \ref{assumption:stronglylogconcave} and \ref{assumption:knownmode}. 
In this section we will briefly describe our notation and the class of parallel sampling algorithms we consider.

\subsection{Notation}
We denote the collection of natural numbers (not including 0) by $\n$, and we let $[n]:= \{1, \ldots, n\}$ for every $n \in \n$.
For $m \in \n$ we let $O(m)$ denote the space of orthogonal $m \times m$ matrices and we equip $O(m)$ with its unique rotation-invariant Haar probability measure, which we denote by $\operatorname{Haar}(O(m))$. We use standard asymptotic notation $\mathcal{O}(\cdot), \Omega(\cdot), o(\cdot)$ and use $\widetilde{\mathcal{O}}(\cdot), \widetilde{\Omega}(\cdot)$ to hide polylogarithmic dependence. We let $\n_0 := \n \cup \{0\}$, and for $f: \r^d \to \r$ we denote by $D^\alpha f$ the derivative with multi-index $\alpha \in \n_0^d$.

We let $\p(\r^d)$ denote the class of Borel probability measures on $\r^d$, and $\p_{\operatorname{a.c.}}(\r^d) \subset \p(\r^d)$ those which are absolutely continuous with respect to the Lebesgue measure. We equip these spaces with the total variation (TV) distance, which is given by
\begin{equation}
    \operatorname{TV}(\mu, \nu) := \sup_{S \subset \r^d \text{ Borel}} |\mu(S) - \nu(S)|, \quad (\forall \mu, \nu \in \p(\r^d)).
\end{equation}
For $V: \r^d \to \r$ measurable with $\int_{\r^d} e^{-V(y)}dy < \infty$, we write $\pi \propto e^{-V}$ to denote the absolutely continuous probability measure $\pi \in \p_{\operatorname{a.c.}}(\r^d)$ whose density w.r.t. Lebesgue is given by \begin{equation}\frac{d\pi}{dx}(x) = \frac{e^{-V(x)} }{\int_{\r^d} e^{-V(y)}dy}.\end{equation}
For each $d \in \n$ and $\kappa \in (1, \infty)$, we denote the class of smooth, strongly convex functions on $\r^d$ with condition number at most $\kappa$ and minimum value 0 at a unique mode 0 by
\begin{equation}
\mathcal{V}_{d,\kappa}
:=
\left\{
V\in C^{1,1}(\mathbb R^d):
\begin{array}{l}
\exists\; 0<\alpha\leq\beta<\infty
\text{ such that }
\alpha \1_d\preceq\nabla^2V\preceq\beta \1_d
\text{ a.e.},\\[2pt]
\beta/\alpha\leq\kappa,\quad
V(0)=0,\quad\nabla V(0)=0,
\end{array}
\right\},
\end{equation}
where $C^{1,1}(\r^d)$ denotes continuously differentiable functions on $\r^d$ with Lipschitz gradient, the Hessian is understood in the weak sense, and the constants $\alpha, \beta$ may depend on $V$. 
We denote the class of $d$-dimensional target probability measures satisfying Assumptions \ref{assumption:stronglylogconcave} and \ref{assumption:knownmode} with condition number at most $\kappa \in (1, \infty)$ by
\begin{equation}
    \Pi_{d,\kappa} := \left\{\pi \propto e^{-V}:\; V \in \mathcal{V}_{d,\kappa}\right\} \subset \p_{\operatorname{a.c.}}(\r^d).
\end{equation}
We also denote $\mathcal{V}_d := \bigcup_{\kappa \in (1, \infty)} \mathcal{V}_{d, \kappa}$ and $\Pi_d := \bigcup_{\kappa \in (1, \infty)} \Pi_{d, \kappa}$.

For the Gaussian case, for each $d \in \n$ and $\kappa \in (1, \infty)$ we denote the set of symmetric, positive definite matrices with eigenvalues in $[1, \kappa]$ by
\begin{equation}
    \mathcal{M}_{d,\kappa} := \left\{A \in \r^{d \times d}:\; A \text{ symmetric}, \; \1_d \preceq A \preceq \kappa \1_d \right\}.
\end{equation}
This generates a natural class of quadratic potentials
\begin{equation}
    \mathcal{V}_{d,\kappa}^{(\text{Gaussian})} := \left\{V_A: \r^d \to \r:\; V_A(x) = \frac{1}{2}x^\top A x \text{ for some } A \in \mathcal{M}_{d,\kappa}\right\} \subset \mathcal{V}_{d,\kappa}
\end{equation}
and the associated class of centered Gaussian measures
\begin{equation}
\Pi_{d,\kappa}^{(\text{Gaussian})} := \left\{\pi_A \propto e^{-V_A}:\; V_A \in \mathcal{V}_{d,\kappa}^{(\text{Gaussian})}\right\} \subset \Pi_{d,\kappa},
\end{equation}
where we may alternatively write $\pi_A = \mathcal{N}(0, A^{-1})$.

\subsection{Query model}
Let $V \in \mathcal{V}_{d}$ be the target potential. 
A parallel sampling algorithm's goal is to make simultaneous queries to $(V, \nabla V)$ (and maybe higher-order derivatives) in successive rounds and use the responses to produce a sample $Z \in \r^d$ whose law is close to $\pi \propto e^{-V}$. Bounds on number of rounds and the number of total queries may be imposed almost surely or in expectation.
% A parallel sampling algorithm's goal is to perform $T$ rounds of simultaneous queries to $(V, \nabla V)$ using at most $Q$ total queries, and to use this information to produce a sample $Z \in \r^d$ for which $\operatorname{Law}(Z)$ is close to $\pi \propto e^{-V}$. 
% Each round may depend on all prior rounds in addition to the sampler's internal randomness, and the final output can depend on the whole history in addition to the sampler's internal randomness. 
Computation between queries is unbounded and real number precision is allowed.
This captures randomized algorithms which simulate discretizations of ordinary/stochastic differential equations (including those with many interacting particles), importance or rejection sampling algorithms, etc.. 

To be precise, we define the class of randomized parallel sampling algorithms we consider in the following way. We will consider more general algorithms which are allowed access to the first $m$ derivatives of the potential, which we assume to be $C^m$. Taking $m = 1$ yields the usual first-order sampling oracle for $\mathcal{V}_{d,\kappa}$, while taking $m > 1$ yields a stronger class of samplers for these potentials under the additional assumption of $m$ times continuously-differentiability.

\begin{definition}[Parallel sampling algorithm]
\label{def:samplingalg}
    Let $d \in \n$, $m \in \n \cup \{\infty\}$, and $\kappa \in (1, \infty)$. 
A \textit{randomized $m$-order parallel sampling algorithm for $\mathcal{V}_{d,\kappa}$}, denoted $\mathcal{A}$, is specified by a probability space $(\Omega,  \mathcal{F}_{\mathcal{A}}, \mathbb{P}_{\mathcal{A}})$ and measurable rules satisfying the following for every instance $V \in \mathcal{V}_{d, \kappa} \cap C^m(\r^d)$:
    \begin{itemize}
    \item The algorithm samples a random seed $\omega \in \Omega$ with law $\mathbb{P}_{\mathcal{A}}$, which encodes all of the algorithm's internal randomness throughout its execution. 
    % \item The algorithm selects a random number of rounds $\mathcal{T} \in \n$ as a measurable function of its internal randomess $\omega$.
    \item In each round $t \in \n$, the algorithm selects a random number of queries $\mathcal{Q}_t \in \n$ and query locations $X_1^t, \ldots, X_{\mathcal{Q}_t}^t \in \r^d$ as measurable functions of $\omega$ and the history
    \begin{equation}
        \mathcal{H}_{t-1} := \left(\left(X_i^s, D^\alpha V(X_i^s)\right)_{\substack{i \in [\mathcal{Q}_s]\\ |\alpha| \leq m}}\right)_{s < t},
    \end{equation}
    where $\mathcal{H}_0$ is empty. It then receives the oracle responses $\left(X_i^t, D^\alpha V(X_i^t)\right)_{\substack{i \in [\mathcal{Q}_t] \\ |\alpha| \leq m}}$ simultaneously, which it appends to the history to form $\mathcal{H}_t$. 
    
    Here, the derivative indices range over $\alpha\in\mathbb N_0^d$ with $|\alpha|:=\sum_{j=1}^d\alpha_j\leq m$. When $m=\infty$, all multi-indices are included. Jets carry their product Borel $\sigma$-algebras. One query at a point returns the entire $m$-order jet at that point, including the value $D^0V=V$.
    \item After each round $t$, it chooses whether to stop as a measurable function of $\omega$ and the updated history $\mathcal{H}_t$. We denote the value of $t$ at which the algorithm chooses to stop by $\mathcal{T} \in \n$, and we assume $\mathcal{T} < \infty$ almost surely. The algorithm then chooses an output $Z \in \r^d$ as a measurable function of $\omega$ and the full history $\mathcal{H}_{\mathcal{T}}$. 
    \item For a given instance $V$, after randomized execution we denote by
    \begin{equation}
        Z(V) \in \r^d, \quad \mathcal{T}(V) \in \n, \quad \mathcal{Q}(V) := \sum_{t=1}^{\mathcal{T}(V)} \mathcal{Q}_t \in \n
    \end{equation}
   the random output, the random number of rounds, and the random number of total queries, respectively. These are all measurable functions of the algorithm's internal randomness $\omega$. We write $\mathcal{A}^V := \operatorname{Law}_{\mathbb{P}_{\mathcal{A}}}(Z(V)) \in \p(\r^d)$ to denote the probability distribution of outputs when the algorithm is executed on the given instance $V$.
    \end{itemize}
    The internal randomness $\mathbb{P}_{\mathcal{A}}$ and the measurable rules defining the algorithm must be fixed independently of $V \in \mathcal{V}_{d, \kappa} \cap C^m(\r^d)$ (though they may depend on $d, m,\kappa$); the only dependence on $V$ is through the oracle responses. 

    Fix $T, Q \in (0, \infty)$. We say that $\mathcal{A}$ \textit{uses at most $T$ rounds almost surely} if $\mathcal{T}(V) \leq T$ almost surely for all $V \in \mathcal{V}_{d, \kappa} \cap C^m(\r^d)$, and we say that $\mathcal{A}$ \textit{uses at most $T$ rounds in expectation} if
    \begin{equation}
        \sup_{V \in \mathcal{V}_{d, \kappa} \cap C^m(\r^d)}\mathbb{E}_{\mathcal{A}}[\mathcal{T}(V)] \leq T.
    \end{equation}
    Similarly, we say that $\mathcal{A}$ \textit{makes at most $Q$ total queries almost surely} if $\mathcal{Q}(V) \leq Q$ almost surely for all $V \in\mathcal{V}_{d, \kappa} \cap C^m(\r^d)$, and we say that $\mathcal{A}$ \textit{makes at most $Q$ total queries in expectation} if
    \begin{equation}
        \sup_{V \in \mathcal{V}_{d, \kappa} \cap C^m(\r^d)}\mathbb{E}_{\mathcal{A}}[\mathcal{Q}(V)] \leq Q.
    \end{equation}
\end{definition}

\begin{remark}
    Our log-concave lower bound will apply to $\infty$-order samplers, i.e. the hard instances are $C^\infty$ and sampling algorithms are allowed to view the entire jet $(D^\alpha V)_{\alpha \in \n_0^d}$ at each query point. Of course, this also provides a lower bound against $m$-order samplers for any $m \in \n$ (such as first-order or second-order algorithms) since the hard instances remain admissible and an $m$-order sampling algorithm can be simulated using an $\infty$-order sampling algorithm. 
\end{remark}

\begin{remark}
When we are interested in algorithms which only sample from centered Gaussian targets, we apply the exact same definition and replace $V \in \mathcal{V}_{d,\kappa} \cap C^m(\r^d)$ with $V \in \mathcal{V}_{d,\kappa}^{(\text{Gaussian})}$. In such cases we will only ever consider $m = 1$, since $m \geq 2$ allows reconstruction of the target measure in one query. For centered Gaussian sampling, if the covariance spectrum is supplied in advance then the algorithm’s rules may additionally depend measurably on that spectrum. 
% We denote the output law by $\mathcal{A}^A \in \p(\r^d)$ and the number of rounds and number of total queries by $\mathcal{T}(A)$ and $\mathcal{Q}(A)$, respectively, when executed on the centered Gaussian instance with inverse covariance $A \in \mathcal{M}_{d,\kappa}$. 
\end{remark}

% To summarize, a parallel sampling algorithm $\mathcal{A}$ produces a sample from the probability distribution $\mathcal{A}^V$ using at most $T$ rounds of parallel queries and at most $Q$ total queries to the $(V, \nabla V)$ oracle. We may enforce the $T, Q$ bounds to hold almost surely or in expectation. 

\section{Lower bound for log-concave measures}

In this section we will make precise the zero-chain intuition from Section \ref{sec:intuition}. Our hard instances will be random rotations of a construction which places spherical gates on coordinate blocks: closed gates provide no derivative signal, and a successful sampler must traverse these blocks and open their gates in sequential order.

We will first define the gates, a test set with which to show TV separation, and the hard potentials along the lines of the construction in Section \ref{sec:intuition}. We then use this construction to implement the zero-chain argument, yielding our parallel sampling lower bound.

\subsection{Hard instances via zero-chains}
\label{subsec:hardinstanceconstruction}
Fix $\kappa \in (1, \infty)$ and set
\begin{equation}
\label{eqn:defofrho2}
    \delta := \frac{\kappa-1}{\kappa+1}, \quad \epsilon := \frac{\delta}{50704}, \quad \rho := \frac{\delta}{25352 - \delta} = \frac{2\epsilon}{1-2\epsilon}.
\end{equation}
We also define
\begin{equation}
\label{eqn:defofk}
    k := \max\left\{2,\; \left\lceil k_0^2\right\rceil\right\}, \quad k_0 := \frac{134532}{25352} \left(1 + \frac{32}{\rho} + \frac{6400}{\rho^2}\right).
\end{equation}
The quantities $\epsilon, \rho, k$ depend only on $\delta$ (i.e. on $\kappa$), with $\epsilon, \rho \asymp \delta$ and $k \asymp \delta^{-4}$ as $\delta \downarrow 0$. Throughout the rest of this section we treat these parameters as fixed.

\paragraph{Defining the gates.} For fixed integers $b, r \in \n$ with $b,r\geq 2$ we define the sequence
\begin{equation}
    n_j := k^{r-j}b \quad (\forall j \in [r]), \quad d := 2\sum_{j=1}^r n_j = 2b\frac{k^r - 1}{k-1}.
\end{equation}
Our construction lives in dimension $d$, using $r$ many blocks each of dimension $2n_j$. The hard instance family will be indexed by separate rotations of each block, i.e. we take $U_j \in O(n_j)$ for each $j \in [r]$ and denote $U := (U_1, \ldots, U_r)$.

Write $z = (z_1, \ldots, z_r) \in \r^d$ to denote a full-dimensional vector with blocks $z_j \in \r^{2n_j}$, and within each block $j$ introduce the coordinates $z_j = (x_j, y_j)$ with $x_j, y_j \in \r^{n_j}$. We define the functions $s_1^U, \ldots, s_r^U: \r^d \to \r$ according to
\begin{equation}
    s_j^{U}(z) := \begin{cases}
        {2\la y_j, U_jx_j\ra}/{\|z_j\|^2} &  z_j \neq 0,\\
        0 & z_j = 0,
    \end{cases} \quad (\forall j \in [r]).
\end{equation}
We say that gate $j \in [r-1]$ is closed when $s_j^U(z) \leq \rho / 4$ and fully open when $s_j^U(z) \geq \rho / 2$, and we define the terminal set by
\begin{equation}
    S_U := \left\{z \in \r^d:\; s_r^U(z) \geq \rho / 2\right\} \subset \r^d.
\end{equation}
We will leave $b,r$ as free variables in order to show a parameterized lower bound in Proposition \ref{prop:parameterizedlowerbound}, after which we will choose $b, r$ carefully to yield Corollary \ref{corollary:main}.

\paragraph{Defining the potentials.} Before constructing the potentials, we first define some scalar threshold functions. Denote the smooth step function
\begin{equation}
\label{eqn:defofsigma}
    \sigma(t) := \frac{\eta(t)}{\eta(t) + \eta(1-t)}, \quad \eta(t) := \begin{cases}
        e^{-1/t}, & t > 0, \\ 0, & t \leq 0,
    \end{cases} \quad (\forall t \in \r).
\end{equation}
Thus, $\sigma(t)  = 0$ for $t \leq 0$, $\sigma(t) = 1$ for $t \geq 1$, and $\sigma\in C^\infty(\r).$ Using this, we define the cutoffs
\begin{equation}
\label{eqn:defofchiphi}
    \chi(u) := \sigma(16u - 1), \quad \phi(u) := u + (1-u) \sigma(4u - 1) \quad (\forall u \geq 0)
\end{equation}
and the gate
\begin{equation}
\label{eqn:defofg}
    g(t) := \sigma\left(\frac{t - \rho / 4}{\rho / 4}\right) \quad (\forall t \in \r).
\end{equation}
Lastly, for $A, B \geq 0$ we define
\begin{equation}
\label{eqn:defofH}
    H(A, B) := \begin{cases}
        B \phi(A/(kB)), & B > 0, \\ 0, & B = 0.
    \end{cases}
\end{equation}
These functions satisfy the following properties:
\begin{itemize}
    \item $\chi(u) = 0$ for $0 \leq u \leq 1/16$ and $\chi(u) = 1$ for $u \geq 1/8$.
    \item $\phi(u) = u$ for $0 \leq u \leq 1/4$ and $\phi(u)  = 1$ for $u \geq 1/2$. Also, $0 \leq \phi(u) \leq \min(1, 4u)$ for all $u \geq 0$.
    \item $g(t) = 0$ for $t \leq \rho / 4$ and $g(t) = 1$ for $t \geq \rho / 2$.
    \item $ 0 \leq H(A, B) \leq \min\left(4A/k,\; B\right)$ for all $A, B \geq 0$. Also, $H(A, B) = B$ if $A \geq kB/2$.
\end{itemize}
Using these, for each collection of rotations $U = (U_1, \ldots, U_r)$ with $U_j \in O(n_j)$ we define the potential $V_U: \r^d \to \r$ in the following way:
\begin{itemize}
    \item Define the function $F_1^U: \r^d \to \r$ on inputs $z = (z_1, \ldots, z_r) \in \r^d$ via \begin{equation}
    F_1^U(z) := \chi\left(\frac{\|z_1\|^2}{2n_1}\right) \|z_1\|^2(1+s_1^U(z)).
\end{equation}
Similarly, for $j \in \{2, \ldots, r\}$ define $F_j^U: \r^d \to \r$ via
\begin{equation}
    F_j^U(z) := \chi\left(\frac{\|z_{j-1}\|^2}{2n_{j-1}}\right) \chi\left(\frac{\|z_j\|^2}{2n_j}\right) \cdot H\left(\|z_{j-1}\|^2, \|z_j\|^2\right) \cdot (1 + s_j^U(z)) \cdot g(s_{j-1}^U(z)).
\end{equation}
Note that $F_1^U$ depends only on $z_1$, while for $j \in \{2, \ldots, r\}$ each $F_j^U$ depends only on $z_{j-1}, z_j$.
\item Define $V_U: \r^d \to \r$ according to 
\begin{equation}
    V_U(z) :=  \frac{1}{2}\|z\|^2 - \epsilon \sum_{j=1}^r F_j^U(z) \quad (\forall z \in \r^d).
\end{equation}
\end{itemize}

\paragraph{Properties of the hard instances.}
For each $U = (U_1, \ldots, U_r)$ with $U_j \in O(n_j)$, we have defined a hard instance according to a potential $V_U: \r^d \to \r$ and a test set $S_U \subset \r^d$. 
We state their properties in the following lemma, whose proof is deferred to Appendix \ref{subsec:proofoflemamtheconstruction}:

\begin{lemma}[Properties of $V_U,S_U$]
\label{lemma:theconstruction}
Fix $b, r \in \n$ with $b, r \geq 2$ and set $n_j := k^{r-j}b$ and $d := 2\sum_{j=1}^r n_j$. For each $j \in [r]$ fix an arbitrary $U_j \in O(n_j)$ and denote $U := (U_1, \ldots, U_r)$. Define the functions $V_U, F_1^U, \ldots, F_r^U: \r^d \to \r$ and the set $S_U \subset \r^d$ according to the above. Define also the truncations $\widetilde{V}_{U,0}, \ldots, \widetilde{V}_{U,r}: \r^d \to \r$ by
\begin{equation}
    \label{eqn:defofV_U}
    \widetilde{V}_{U,\tau}(z) := \frac{1}{2}\|z\|^2 - \epsilon \sum_{j=1}^\tau F_j^U(z) \quad (\forall \tau \in \{0, \ldots, r\})
\end{equation}
and note that $\widetilde{V}_{U,0}(\cdot) = \frac{1}{2}\|\cdot\|^2$ and $ \widetilde{V}_{U,r} = V_U$.
Then, the following properties hold:
\begin{enumerate}
\item[(i)] The functions $F_1^U, \ldots, F_r^U$ are all in $C^\infty(\r^d)$, and hence so is $\widetilde{V}_{U,\tau}$ for every $\tau \in \{0, \ldots, r\}$.
\item[(ii)] $(1-\delta) \1_d \preceq \nabla^2 \widetilde{V}_{U,\tau}(z) \preceq (1+\delta)\1_d$ for all $z \in \r^d$ and $\tau \in \{0, \ldots, r\}$.
\item[(iii)] $\widetilde{V}_{U,\tau}(0) = 0$ and $\nabla \widetilde{V}_{U,\tau}(0) = 0$ for every $\tau \in \{0, \ldots, r\}$.
 \item[(iv)] If $z \in \r^d$ and $j \in \{2, \ldots, r\}$ are s.t. $s_{j-1}^U(z) \leq \rho / 4$ then $D^\alpha F_j^U(z) = 0$ for all $\alpha \in \n_0^d$.
    \item[(v)] $S_U$ is Borel and $\pi_{V_U}(S_U) \geq 1 - 8r \exp(-{\rho^2 b}/{128})$, where $\pi_{V_U} \in \p(\r^d)$ is the probability measure with density $\pi_{V_U} \propto e^{-V_U}$ with respect to Lebesgue.
\end{enumerate}
\end{lemma}

Properties (i)-(iii) ensure that $V_U$ and the truncations $\widetilde{V}_{U,\tau}$ all lie in $C^\infty(\r^d) \cap \mathcal{V}_{d,\kappa}$.
    Property (iv) says that if gate $j-1$ is closed then the full jet $(D^\alpha F_j^U)_{\alpha \in \n_0^d}$ vanishes, which drives the adaptive-round hiding argument. Property (v) drives the TV separation, since our argument will show that a sampler without enough rounds typically fails to give enough mass to $S_U$.
\begin{remark}
Our explicit choices of $k, \rho, \epsilon$ satisfy the conditions
    \begin{equation}
    \label{eqn:conditionsonconstants}
      \epsilon \in \left(0, \frac{1}{4}\right), \quad \rho = \frac{2\epsilon}{1-2\epsilon}, \quad \epsilon \cdot \left(25352 + \frac{134532}{\sqrt{k}}\left(1 + \frac{32}{\rho} + \frac{6400}{\rho^2}\right)\right)  \leq {\delta},
    \end{equation}
    which are the inputs required to prove Lemma \ref{lemma:theconstruction}.
\end{remark}

\subsection{Derivation of lower bound}

Using this construction of $V_U: \r^d \to \r$ and $S_U \subset \r^d$, we may execute a standard zero-chain argument to show that parallel samplers encounter a query lower bound once we randomize $U$:

\begin{lemma}[Query lower bound]
\label{lemma:randomizedoraclebound}
    Let $U_j \sim \operatorname{Haar}(O(n_j))$ be independent random orthogonal matrices and $U = (U_1, \ldots, U_r)$, and define $V_U, S_U$ as in Lemma \ref{lemma:theconstruction}. 
    Fix $Q \in (0, \infty)$ and let $\mathcal{A}$ be a randomized $\infty$-order parallel sampling algorithm for $\mathcal{V}_{d,\kappa}$, independent of $U$, making at most $Q$ total queries in expectation (in the sense of Definition \ref{def:samplingalg}). Then,
    % Let $T, Q \in \n$ with $T < r$, and let $\mathcal{A}$ be any randomized adaptive $T$-round sampler (independent of $U$) which makes at most $Q$ simultaneous $(V_U, \nabla V_U)$ queries per round and outputs samples with law $\mathcal{A}_{T, Q}^U \in \p(\r^d)$. Then, it holds that
        \begin{equation}
            \mathbb{E}_U\left[\mathcal{A}^{V_U}(S_U)\right] \leq \mathbb{P}_{U,\mathcal{A}}(\mathcal{T}(V_U) \geq r) + Q r \exp\left(-\frac{\rho^2 b}{32}\right) +  \exp\left(-\frac{\rho^2 b}{8}\right),
        \end{equation}
        where the probability is taken over the product law of $U$ and $\mathcal{A}$'s internal randomness.
\end{lemma}
\begin{proof}[Proof sketch]
We present here an informal sketch under the simplifying assumption that the number of queries $Q_t$ in each round $t$ is deterministic and $\sum_t Q_t \leq Q$; the full proof treats adaptive random batch sizes and can be found in Appendix \ref{subsec:proofoflemmarandomizedoraclebound}.

There are $r-1$ gates (indexed by $j$), and we must show that it is unlikely for an algorithm to do better than opening one gate per round. For any $j \in \{2, \ldots, r\}$, by Lemma \ref{lemma:theconstruction}(iv) as long as all the algorithm's queries satisfy
\begin{equation}
\label{eqn:proofsketchtheclosedgates}
    s_{j-1}^U, \ldots, s_{r-1}^U \leq \rho /4,
\end{equation}
i.e. gates $j-1$ through $r-1$ are closed,
then the oracle responses for $V_U$ and the truncated potential $\widetilde{V}_{U, j-1}$ match. Hence, if the algorithm's execution on $V_U$ is coupled with its execution on $\widetilde{V}_{U, j-1}$ (by using the same internal randomness seed) then the resulting queries will be identical for as long as (\ref{eqn:proofsketchtheclosedgates}) holds. 
However, the execution on $\widetilde{V}_{U,j-1}$ is independent of $U_j$, and conditional on the seed and $U_1, \ldots, U_{j-1}$ its entire query list is fixed. Spherical concentration, a union bound over queries and later gates, and the expected query bound imply that
    \begin{equation}
        \mathbb{P}\left(\text{some query in round $j$ on $\widetilde{V}_{U,j-1}$ has $s_\ell^U > \rho / 4$ for some $\ell \geq j$}\right) \leq Q_jre^{-\rho^2 b/32},
    \end{equation}
    where $Q_j$ is the number of queries in round $j$.
    In words, running the algorithm with queries to $\widetilde{V}_{U,j-1}$ is exponentially unlikely to open any later gate $s_j^U, \ldots, s_{r-1}^U$. By the coupling argument and induction, for every round $t \in [r-1]$ we find that 
    \begin{equation}
    \label{eqn:proofquerylemmainformalsketch1}
        \mathbb{P}\left(\text{executions on $V_U$ and $\widetilde{V}_{U,t}$  differ by round $t$}\right) \leq \left(\sum_{j=1}^t Q_j\right) r e^{-\rho^2 b/32}.
    \end{equation}
    Lastly, since the output of the execution on $\widetilde{V}_{U,r-1}$ is independent of $U_r$, another application of spherical concentration implies
    \begin{equation}
        \label{eqn:proofquerylemmainformalsketch2}\mathbb{E}_U\left[\mathcal{A}^{\widetilde{V}_{U,r-1}}(S_U)\right] \leq e^{-\rho^2 b / 8}.
    \end{equation}
    Combining (\ref{eqn:proofquerylemmainformalsketch1}) and (\ref{eqn:proofquerylemmainformalsketch2}), if fewer than $r$ rounds are used then it is likely that the executions on $V_U$ and $\widetilde{V}_{U, r-1}$ to agree (with error probability $\leq Qre^{-\rho^2b/32}$). In such cases it is likely that the test gate defining $S_U$ is not opened (with error probability $\leq e^{-\rho^2b/8}$), yielding the claim. 
\end{proof}
% \begin{remark}
%     To be precise, at each round $t$ the algorithm selects queries $\{z_q^t\}_{q \in [Q_t]}$ with $Q_t \leq Q$ and receives exact responses $\left\{\left(V_U\left(z_q^t\right), \nabla V_U\left(z_q^t\right)\right)\right\}_{q \in [Q_t]}$. The queries may depend on all prior rounds adaptively, i.e. each $z_q^t$ is a measurable function of the full collection $\{\left(z_{h}^\tau, V_U\left(z_{h}^\tau\right), \nabla V_U\left(z_{h}^\tau\right)\right)\}_{\substack{h \in[Q_\tau],\; \tau < t}}$ as well as the algorithm's internal randomness. 
%     The algorithm's output distribution $\mathcal{A}_{T, Q}^U \in \p(\r^d)$ is conditional on the value of $U$ but is averaged over the algorithm's internal randomness.
% \end{remark}

These two lemmas combine in the natural way to yield 
our parameterized lower bound for parallel sampling of well-conditioned log-concave measures:
\begin{proposition}[Parameterized lower bound, log-concave]
\label{prop:parameterizedlowerbound}
Fix $\delta \in (0, 1)$ and set $\kappa = \frac{1+\delta}{1-\delta}$.
Then, there are constants $k = k(\delta) \in \n \setminus \{1\}$ and $\rho  = \rho(\delta) \in (0, 1)$ s.t. the following holds: for every pair of fixed $b, r \in \n$ with $b, r \geq 2$, if we denote $d = 2b \frac{k^r - 1}{k - 1} \in \n$ then there exists a Borel probability distribution $\mathcal{D}_{b, r}^{(\delta)}$ over $C^\infty(\r^d)$ (depending only on $b, r, \delta$) with the following properties:
    \begin{enumerate}
        \item[(i)] For almost every $V \sim \mathcal{D}_{b, r}^{(\delta)}$ we have $(1-\delta)\1_d \preceq \nabla^2 V(z) \preceq (1+\delta)\1_d$ for all $z \in \r^d$ and $V(0) = 0$, $\nabla V(0) = 0$. In particular, $\mathcal{D}_{b,r}^{(\delta)}$ is supported on $\mathcal{V}_{d, \kappa} \cap C^\infty(\r^d)$.
        \item[(ii)] Let $T, Q \in [1, \infty)$ with $T < r$, and let $\mathcal{A}$ be any randomized $\infty$-order parallel sampling algorithm for $\mathcal{V}_{d, \kappa}$ (independent of $V \sim \mathcal{D}_{b,r}^{(\delta)}$) which uses at most $T$ rounds almost surely and makes at most $Q$ total queries in expectation (in the sense of Definition \ref{def:samplingalg}).
        Then, it holds that
        \begin{equation}
            \mathbb{E}_{V \sim \mathcal{D}_{b,r}^{(\delta)}}\left[\operatorname{TV}\left(\mathcal{A}^V, \pi_V\right)\right] \geq 1 - 10Qr \exp\left(-\frac{\rho^2 b}{128}\right)
        \end{equation}
        where $\pi_V \in \p_{\operatorname{a.c.}}(\r^d)$ is the probability measure with density $\pi_V \propto e^{-V}$.
        \item[(iii)] If instead $\mathcal{A}$ is only assumed to use at most $T$ rounds in expectation, then we have
        \begin{equation}
            \mathbb{E}_{V \sim \mathcal{D}_{b,r}^{(\delta)}}\left[\operatorname{TV}\left(\mathcal{A}^V, \pi_V\right)\right] \geq 1 - \frac{T}{r} - 10Qr \exp\left(-\frac{\rho^2 b}{128}\right).
        \end{equation}
    \end{enumerate}
   
\end{proposition}
\begin{proof}[Proof sketch]
This follows by defining $k, \rho$ as in (\ref{eqn:defofk}) and (\ref{eqn:defofrho2}), taking independent $U_j \sim \operatorname{Haar}(O(n_j))$, and applying
\begin{equation}
\mathbb E_U\operatorname{TV}(\mathcal A^{V_U},\pi_{V_U})
\geq
\mathbb E_U\pi_{V_U}(S_U)
-
\mathbb E_U\mathcal A^{V_U}(S_U).
\end{equation}
The first term is bounded below by Lemma \ref{lemma:theconstruction}(v), while the second term is upper bounded by Lemma \ref{lemma:randomizedoraclebound}.
For the expected round bound, we use Markov's inequality to get
\begin{equation}
\mathbb P(\mathcal T(V_U)\geq r)\leq T/r.
\end{equation}
The full proof can be found in Appendix \ref{subsec:proofofparameterizedlowerboundprop}.
\end{proof}

Proposition \ref{prop:parameterizedlowerbound} directly implies the following asymptotic result, which Theorem \ref{theorem:main} summarizes. We include its proof here to show the choices of the relevant parameters. In particular, defining
\begin{equation}
    \zeta(\kappa) := \frac{1}{\log k(\kappa)} \in (0, \infty), \qquad a_\kappa := \frac{\rho(\kappa)^2}{2048 k(\kappa)} \in (0, \infty)
\end{equation}
with $k(\kappa)$ and $\rho(\kappa)$ as in (\ref{eqn:defofk}) and (\ref{eqn:defofrho2}), respectively, and
taking $r = \lceil c \log d\rceil + 1$ and $b \asymp_\kappa d^{1 - c/\zeta(\kappa)}$ yields the following asymptotic lower bound:
\begin{corollary}[Asymptotic lower bound, log-concave]
\label{corollary:main}
Fix $\kappa\in(1,\infty)$ and $c \in (0, \zeta(\kappa))$. For each $d\in\n$, there is a probability
distribution $\mathcal{D}_{d,\kappa,c}$ over
$C^\infty(\r^d)\cap\mathcal{V}_{d,\kappa}$ such that for every
$\beta\in(0,1)$ there is some $d_0=d_0(\kappa,c,\beta)$ for which the
following holds for all $d\geq d_0$:

Let $\mathcal{A}$ be any randomized $\infty$-order parallel sampling algorithm for $\mathcal{V}_{d,\kappa}$ which makes at most 
\begin{equation}
    \exp\left(a_\kappa d^{1-c/\zeta(\kappa)}\right)
\end{equation}
total queries in expectation and either uses at most $\lceil c \log d\rceil $ rounds almost surely or uses at most $\beta c\log d$ rounds in expectation. Then,
\begin{equation}
    \mathbb{E}_{V \sim \mathcal{D}_{d,\kappa,c}} \left[\operatorname{TV}(\mathcal{A}^V, \pi_V)\right] \geq 1 - \beta.
\end{equation}
\end{corollary}
\begin{proof}
Write $k = k(\kappa)$, $\rho = \rho(\kappa)$, and $\gamma := 1 - c/\zeta(\kappa) > 0$. Choose
\begin{equation}
    r := \lceil c \log d\rceil + 1, \quad b := \left\lfloor\frac{(k-1)d}{2(k^r-1)} \right\rfloor.
\end{equation}
Then, $2b(k^r-1)/(k-1) \leq d$. Since $r \leq c\log d + 2$ and $\log k = 1/\zeta(\kappa)$, we have $k^r \leq k^2 d^{1-\gamma}$. Using $k-1 \geq k/2$, we obtain
\begin{equation}
    b \geq \frac{d^\gamma}{4k} - 1 \geq \frac{d^\gamma}{8k}
\end{equation}
for all sufficiently large $d$. We may also take $d$ large enough that $b, r \geq 2$. Take the distribution over hard instances from Proposition \ref{prop:parameterizedlowerbound} and pad the remaining coordinates with standard Gaussians to obtain $\mathcal{D}_{d,\kappa,c}$.
Adding the same quadratic potential on the remaining coordinates to the full potential and each truncation preserves their curvature bounds and the oracle couplings. Projecting the output onto the hard coordinates therefore gives the same TV lower bounds in dimension $d$.

Set $Q := \exp(a_\kappa d^\gamma)$. By the lower bound on $b$ and the definition of $a_\kappa$, we have $\rho^2 b/128 \geq 2a_\kappa d^\gamma$. Hence, the error term in Proposition \ref{prop:parameterizedlowerbound} satisfies
\begin{equation}
    \eta_d := 10Qr \exp\left(-\frac{\rho^2 b}{128}\right) \leq 10r \exp\left(-a_\kappa d^\gamma\right).
\end{equation}
Since $\gamma > 0$ and $r = \mathcal{O}(\log d)$, we may choose $d_0(\kappa, c, \beta)$ sufficiently large that $\eta_d \leq \beta / r$ for all $d \geq d_0$. If $\mathcal{A}$ uses at most $\lceil c \log d\rceil = r-1$ rounds almost surely, then Proposition \ref{prop:parameterizedlowerbound}(ii) gives average TV error at least $1-\eta_d \geq 1-\beta$. If instead $\mathcal{A}$ uses at most $\beta c \log d$ rounds in expectation, then Proposition \ref{prop:parameterizedlowerbound}(iii) gives
\begin{equation}
     \mathbb{E}_{V \sim \mathcal{D}_{d,\kappa,c}} \left[\operatorname{TV}(\mathcal{A}^V, \pi_V)\right] \geq 1 - \frac{\beta c \log d}{r} - \eta_d \geq 1 - \frac{\beta(r-1)}{r} - \frac{\beta}{r} = 1-\beta.
\end{equation}
\end{proof}

To summarize, for fixed $\kappa \in (1, \infty)$ and $c \in (0, \zeta(\kappa))$, parallel sampling of $\mathcal{V}_{d,\kappa} \cap C^\infty(\r^d)$ with constant TV accuracy using almost surely at most $\lceil c \log d\rceil$ rounds requires worst-case width
\begin{equation}
\label{eqn:stretchedexponential}
    \exp\left(\Omega_\kappa\left(d^{1-c/\zeta(\kappa)}\right)\right).
\end{equation}
From (\ref{eqn:defofk}) it can be seen that $\zeta(\kappa) = \Theta\left((1 + \log \frac{\kappa+1}{\kappa-1})^{-1}\right)$ with universal constants; in particular, it vanishes as $\kappa \downarrow 1$ and approaches a constant as $\kappa \to \infty$.

\begin{remark}
\label{remark:optimality}
For a number of rounds $T \in \n$ which is fixed with respect to dimension, the same lower bound yields a tight width complexity of
\begin{equation}
    \exp\left(\Theta_{\kappa, T}(d)\right).
\end{equation}
In particular, as $c \downarrow 0$ the bound (\ref{eqn:stretchedexponential}) recovers the right width scaling. The remaining open questions are therefore (1) is the stretched-exponential shape of this scaling law correct and (2) what is the sharp threshold $\zeta_\star(\kappa)$ and its scaling with respect to $\kappa$?

To show the width complexity at a fixed depth $T \in \n$, taking $r=T+1$ and
    $b\asymp_{\kappa,T}d$ in Proposition \ref{prop:parameterizedlowerbound}
    gives a worst-case width lower bound of $\exp(\Omega_{\kappa,T}(d))$. Conversely, rejection sampling with
    proposal $\mathcal{N}(0,\alpha^{-1}\1_d)$ has acceptance probability
    at least $\kappa^{-d/2}$, and so a single batch of
    $\mathcal{O}(\kappa^{d/2}) = \exp(\mathcal{O}_\kappa(d))$ proposals suffices for constant TV accuracy.
\end{remark}

\section{Upper and lower bounds for Gaussian measures}

Recall that our centered Gaussian targets in $\r^d$ are indexed by $A \in \mathcal{M}_{d,\kappa}$, which are positive symmetric matrices $A \in \r^{d \times d}$ whose eigenvalues lie in $[1, \kappa]$. We write $V_A(x) = \frac{1}{2}x^\top Ax$ to denote the potential and $\pi_A = \mathcal{N}(0, A^{-1}) \propto e^{-V_A}$ the resulting Gaussian measure on $\r^d$, whose covariance is $A^{-1}$. A query to $(V_A, \nabla V_A)$ is equivalent to a matrix-vector product query for $A$, since the gradient oracle returns $\nabla V_A(x) = Ax$ which can be used to recover the value $V_A(x)$. 
% \evannote{adjust this?} A randomized first-order parallel sampling algorithm for $\mathcal{V}_{d,\kappa}^{(\operatorname{Gaussian})}$, denoted $\mathcal{A}$, produces outputs with law $\mathcal{A}^A \in \p(\r^d)$ using $\mathcal{T}(A)$ rounds and $\mathcal{Q}(A)$ total matrix-vector queries when executed on an instance $A \in \mathcal{M}_{d,\kappa}$.
% As before, we say $\mathcal{A}$ is a randomized first-order parallel sampling algorithm for $\mathcal{V}_{d,\kappa}^{(\operatorname{Gaussian})}$ using at most $T$ rounds almost surely and making at most $Q$ queries in expectation if
% \begin{equation}
%     \mathcal{T}(V_A) \leq T \text{ almost surely $\forall A \in \mathcal{M}_{d,\kappa}$}, \quad \sup_{A \in \mathcal{M}_{d,\kappa}} \mathbb{E}_{\mathcal{A}}[\mathcal{Q}(V_A)] \leq Q.
% \end{equation}
We denote the threshold
\begin{equation}
    c_\star(\kappa) := \frac{1}{4 \log \frac{\sqrt{\kappa} + 1}{\sqrt{\kappa} - 1}} \quad (\forall \kappa \in (1, \infty)).
\end{equation}

We prove three results regarding parallelized first-order Gaussian sampling using $\lceil c \log d\rceil$ rounds: an algorithm which succeeds when $c > c_\star(\kappa)$, an algorithm which succeeds when $c \leq c_\star(\kappa)$, and a matching lower bound (up to logarithmic factors) when $c < c_\star(\kappa)$. Taken together, these results characterize the depth-width tradeoff for all $c$.
Precise statements are given in the following result, which Theorem \ref{theorem:gaussian} summarizes:

\begin{theorem}
\label{theorem:precisegaussian}
    Fix $\kappa \in (1, \infty)$ and $c \in (0, \infty)$. For each dimension $d$, consider the class of randomized first-order parallel sampling algorithms for $\mathcal{V}_{d,\kappa}^{(\operatorname{Gaussian})}$ which use at most $\lceil c \log d\rceil$ rounds almost surely (in the sense of Definition \ref{def:samplingalg}). Then, the following hold as $d \to \infty$:
    \begin{enumerate}
        \item[(i)] If $c > c_\star(\kappa)$, then there is such an algorithm $\mathcal{A}$ which makes $2$ queries per round almost surely and satisfies
        \begin{equation}
            \sup_{V \in \mathcal{V}_{d,\kappa}^{(\operatorname{Gaussian})}} \operatorname{TV}\left(\mathcal{A}^V, \pi_V\right) \leq o_{\kappa, c}(1).
        \end{equation}
        \item[(ii)] If $c \leq c_\star(\kappa)$, then for any $\delta_d \in (0, 1)$ (which may depend on $d$) there is such an algorithm $\mathcal{A}$ which makes at most $\mathcal{O}\left(\left(\frac{\sqrt{\kappa} + 1}{\sqrt{\kappa} - 1}\right)^2d^{1-c/c_\star(\kappa)} / \delta_d^2\right)$ queries per round almost surely and satisfies
        \begin{equation}
            \sup_{V \in \mathcal{V}_{d,\kappa}^{(\operatorname{Gaussian})}} \operatorname{TV}\left(\mathcal{A}^V, \pi_V\right) \leq \delta_d.
        \end{equation}
        
        % \item[(ii)] If $c < c_\star(\kappa)$, then there is such an algorithm $\mathcal{A}$ which makes at most $d^{1-c/c_\star(\kappa)}(\log d)^{2}$ total queries almost surely and satisfies
        %  \begin{equation}
        %     \sup_{V \in \mathcal{V}_{d,\kappa}^{(\operatorname{Gaussian})}} \operatorname{TV}\left(\mathcal{A}^V, \pi_V\right) \leq o_{\kappa, c}(1) .
        % \end{equation}
        % More precisely, for any $c \leq c_\star(\kappa)$ and any $\delta_d \in (0, 1)$ (which may depend on $d$), the algorithm obtains uniform TV error $\leq \delta_d$ over $\mathcal{V}_{d,\kappa}^{(\operatorname{Gaussian})}$ when run with $\mathcal{O}\left(\left(\frac{\sqrt{\kappa} + 1}{\sqrt{\kappa} - 1}\right)^2d^{1-c/c_\star(\kappa)} / \delta_d^2\right)$ queries per round.

        \item[(iii)] If $c < c_\star(\kappa)$, then for every fixed $C, \delta \in (0, \infty)$, any such algorithm $\mathcal{A}$ which makes at most $Cd^{1-c/c_\star(\kappa)} / (\log d)^{1+\delta}$ total queries in expectation satisfies
        \begin{equation}
            \sup_{V \in \mathcal{V}_{d,\kappa}^{(\operatorname{Gaussian})}} \operatorname{TV}\left(\mathcal{A}^V, \pi_V\right) \geq 1-o_{\kappa, c,C,\delta}(1) .
        \end{equation}
        This lower bound holds even if $\mathcal{A}$ is supplied the covariance eigenvalues in advance.
    \end{enumerate}
       Thus, up to factors logarithmic in $d$, the optimal worst-case width for parallel sampling of
       $\mathcal{V}_{d,\kappa}^{(\operatorname{Gaussian})}$ with constant TV accuracy using almost surely at most $\lceil c \log d\rceil$ rounds is \begin{equation}
 \begin{cases}
               {\Theta}_{\kappa}(1) & c \geq c_\star(\kappa), \\
\widetilde{\Theta}\left(d^{1-c/c_\star(\kappa)}\right) & c < c_\star(\kappa).
            \end{cases}
        \end{equation}
\end{theorem}
\begin{proof}
The rest of this section is devoted to the proof of this result. 
For (i), we apply Proposition \ref{prop:gaussianupperboundsupercritical} with $T = \lceil c \log d\rceil$ to get that
\begin{equation}
    \sup_{V \in \mathcal{V}_{d,\kappa}^{(\operatorname{Gaussian})}} \operatorname{TV}(\mathcal{A}^V, \pi_V) \leq \sqrt{d^{1-{c}/{c_\star(\kappa)}}} \to 0
\end{equation}
since $c > c_\star(\kappa)$. For (ii) we apply Proposition \ref{prop:gaussianupperboundsubcritical} , and for (iii) we apply Corollary \ref{corollary:gaussianlowerbound}.
\end{proof}
\begin{remark}
For the general class of centered Gaussians with inverse covariance satisfying $\alpha \1_d \preceq A \preceq \beta \1_d$ for some $\alpha, \beta \in (0, \infty)$ with $\beta / \alpha \leq \kappa$, our upper bounds apply if the algorithm is allowed to know $\alpha, \beta$ (since it may rescale space accordingly). The lower bound with condition number $\kappa$ applies without change.
\end{remark}

\begin{remark}
    The hard instance for the lower bound of (iii) is a modification of the construction in \cite[Section 4.1]{zhou2026complexitygaussiansampling}: their construction uses a fixed spectrum of $d$ Chebyshev roots and a Haar-distributed eigenbasis, while we use a fixed spectrum of $\Theta(\log d)$ Chebyshev extrema each repeated with multiplicity $d / \Theta(\log d)$. As was the case in \cite{zhou2026complexitygaussiansampling}, our lower bound holds even if the algorithm is supplied the spectrum in advance.
\end{remark}

\subsection{Polynomial approximation of the inverse}
Throughout this section, for $r \in \n$ and $\kappa \in (1, \infty)$ we denote
\begin{equation}
    \epsilon_{r,\kappa} := \operatorname{sech}\left((2r-1) \log \frac{\sqrt{\kappa} + 1}{\sqrt{\kappa} - 1}\right).
\end{equation}
This quantity arises from standard bounds for uniform polynomial approximation of the function $\lambda \mapsto 1/\lambda$ \cite{kraus2012polynomial} and the usual minimax property of Chebyshev polynomials \cite[Proposition 9.5.3]{allaire2008numerical}. For a precise statement, we have the following:
\begin{lemma}
\label{lemma:chebyshevapproximation}
    Fix $r \in \n$ and $\kappa \in (1, \infty)$ and define
    \begin{equation}
    \label{eqn:chebyshevvalues}
        \lambda_j := \frac{\kappa + 1 - (\kappa - 1) \cos \left(\frac{j\pi}{2r-1}\right)}{2}, \quad j \in \{0, \ldots, 2r-1\}.
    \end{equation}
    Then, the following hold:
    \begin{enumerate}
        \item[(i)] There exist real polynomials $u, v$ of degree at most $r-1$ such that
        \begin{equation}
            \sup_{x \in [1, \kappa]} \left|1 - x(u(x)^2 + v(x)^2)\right| \leq \epsilon_{r,\kappa}.
        \end{equation}
        \item[(ii)] Every real polynomial $S$ of degree at most $2r-2$ satisfies
        \begin{equation}
            \max_{j \in \{0, \ldots, 2r-1\}} \left|1 - \lambda_j S(\lambda_j)\right| \geq \epsilon_{r,\kappa}.
        \end{equation}
        
    \end{enumerate}
\end{lemma}

In words, uniform approximation of the function $\lambda \mapsto 1/\lambda$ on the interval $[1, \kappa]$ using polynomials of degree $2r-2$ has optimal relative error $\epsilon_{r,\kappa}$. This error is achieved by a sum of the squares of two polynomials of degree $r-1$, while no polynomial of degree $2r-2$ can do better (even on the finite set $\{\lambda_0, \ldots, \lambda_{2r-1}\} \subset [1, \kappa]$). For completeness, we provide a proof in Appendix \ref{subsec:appendixchebyshevlemma}.

% Lemma \ref{lemma:chebyshevapproximation} forms the foundation of our query complexity upper and lower bounds for centered Gaussians. The intuition is that samplers with $T$ rounds and $K$ queries can only evaluate the action of polynomials of degree $T$ of the inverse covariance on $K$ many vectors. Related Gaussian sampling bounds leveraging such polynomial representations of samplers appear in \cite{chewi2023query} and \cite[Section 4]{zhou2026complexitygaussiansampling}.

% Related approaches of studying query complexity of centered Gaussian sampling using approximation of polynomials of the inverse covariance appear in \cite{chewi2023query} and \evannote{cite japanese fellas}.

\subsection{Upper bound for $c > c_\star(\kappa)$}
We start with an efficient algorithm for centered Gaussian sampling using $2$ queries per round in the regime where $c > c_\star(\kappa)$. The construction aims to approximate the map $A \mapsto A^{-1}$ using matrix-vector queries to $A$, which is implemented using the sum of two squares approximation from Lemma \ref{lemma:chebyshevapproximation}(i):

\begin{proposition}[Upper bound, $c > c_\star(\kappa)$]
\label{prop:gaussianupperboundsupercritical}
Fix $\kappa \in (1, \infty)$ and $d,T \in \n$.
There is a randomized parallel matrix-vector sampling algorithm for $\mathcal{M}_{d,\kappa}$, denoted $\mathcal{A}$, which uses exactly two matrix-vector queries per round such that when executed with exactly $T $ rounds it guarantees
    \begin{equation}
       \sup_{A \in \mathcal{M}_{d,\kappa}} \operatorname{TV}\left(\mathcal{A}^A, \mathcal{N}(0, A^{-1})\right) \leq \epsilon_{T+1,\kappa} \sqrt{\frac{d}{8(1-\epsilon_{T+1,\kappa})}} \quad .
    \end{equation}
    In particular, for every $d \geq 16$ we have that if $T > c_\star(\kappa) \log d$ then
    \begin{equation}
        \sup_{A \in \mathcal{M}_{d,\kappa}}  \operatorname{TV}\left(\mathcal{A}^A, \mathcal{N}(0, A^{-1})\right) \leq \sqrt{d^{1 - \frac{T }{ c_\star(\kappa) \log d}}}.
    \end{equation}
\end{proposition}
\begin{proof}
    Write $\epsilon  := \epsilon_{T+1, \kappa} \in (0, 1)$. By Lemma \ref{lemma:chebyshevapproximation}(i), there exist real polynomials $u, v$ of degree at most $T$ such that
    \begin{equation}
    \label{eqn:proofgaussiansupercriticalupperboundapproximation}
        \sup_{x \in [1, \kappa]} |1- x(u(x)^2 + v(x)^2)| \leq \epsilon.
    \end{equation}
    Draw $g, h \sim \mathcal{N}(0, \1_d)$ independently. In each round $t \in [T]$, query $A$ at $A^{t-1}g$ and $A^{t-1}h$ to receive $A^tg$ and $A^t h$. Using the values of $A^t g, A^th$ for $t \in [T]$, return the output
    \begin{equation}
        Z := u(A) g + v(A) h.
    \end{equation}
    This uses $T$ rounds and two queries per round, and we have
    \begin{equation}
        \mathcal{A}^A = \operatorname{Law}(Z) =\mathcal{N}(0, \Sigma), \quad \Sigma := u(A)^2 + v(A)^2.
    \end{equation}
    We apply the Gaussian relative entropy formula
    \begin{align}
        \operatorname{KL}(\mathcal{N}(0, \Sigma) \;\| \;\mathcal{N}(0, A^{-1})) &= \frac{1}{2}\left(\operatorname{tr}(A\Sigma) - d - \log \operatorname{det}(A\Sigma)\right) \\
        &= \frac{1}{2}\sum_{j=1}^d (\mu_j - 1 - \log \mu_j) \\
        &\leq \frac{1}{2}\sum_{j=1}^d \frac{(\mu_j-1)^2}{2\min\{\mu_j, 1\}},
    \end{align}
    where we denote the eigenvalues of $A\Sigma$ by $\mu_1, \ldots, \mu_d$ and use that $\mu - 1 - \log \mu \leq \frac{(\mu - 1)^2}{2\min\{\mu, 1\}}$ for $\mu > 0$. Denoting the eigenvalues of $A$ by $a_1, \ldots, a_d \in [1, \kappa]$, by (\ref{eqn:proofgaussiansupercriticalupperboundapproximation}) we know
    \begin{equation}
        \left|\mu_j - 1\right| = \left|1 - a_j (u(a_j)^2 + v(a_j)^2)\right| \leq \epsilon
    \end{equation}
    for all $j$. This also means $\min\{\mu_j, 1\} \geq 1 - \epsilon$, and so
    \begin{equation}
        \operatorname{KL}(\mathcal{N}(0, \Sigma) \;\| \;\mathcal{N}(0, A^{-1})) \leq \frac{d\epsilon^2}{4(1-\epsilon)}.
    \end{equation}
    The first claim then follows from Pinsker's inequality. 
    For the second claim, write 
    \begin{equation}
        \Lambda = \log \frac{\sqrt{\kappa}+1}{\sqrt{\kappa}-1} = \frac{1}{4c_\star(\kappa)}, \quad c := \frac{T}{\log d} > c_\star(\kappa).
    \end{equation}
   Since $T = c \log d$ and $d \geq 16$ we have
    \begin{equation}
        \epsilon_{T+1, \kappa} \leq 2e^{-(2T+1)\Lambda} \leq 2d^{-2c \Lambda} = 2d^{-c/(2c_\star(\kappa))} \leq 2d^{-1/2} \leq \frac{1}{2},
    \end{equation}
    where we used that $\operatorname{sech}(x) \leq 2e^{-x}$ for $x \geq 0$. Hence, the nonasymptotic bound gives
    \begin{equation}
         \sup_{A \in \mathcal{M}_{d,\kappa}}  \operatorname{TV}\left(\mathcal{A}^A, \mathcal{N}(0, A^{-1})\right) \leq \frac{\sqrt{d}}{2} \epsilon_{T+1, \kappa} \leq d^{\frac{1}{2} - \frac{c}{2c_\star(\kappa)}}. 
    \end{equation}
\end{proof}

\subsection{Upper bound for $c \leq c_\star(\kappa)$}

When there are fewer than $c_\star (\kappa) \log d$ rounds, direct application of the sum of two squares uniform approximation to $\lambda \mapsto 1/\lambda$ from Lemma \ref{lemma:chebyshevapproximation}(i) no longer guarantees vanishing TV error. However, we may apply a randomized correction using the following lemma:
\begin{lemma}
\label{lemma:gaussiancorrectionlemma}
    Fix $m, n \in \n$ with $n \geq 2$. Let $B \in \r^{m \times m}$ be symmetric with $\|B\|_{\operatorname{op}} \leq 1/3$, and let $G \in \r^{m \times n}$ have i.i.d. standard Gaussian entries. Conditional on $G$, draw $a \in \mathbb{S}^{n-1}$ with density proportional to
    \begin{equation}
        \exp\left(-\frac{1}{2}a^\top G^\top BGa\right)
    \end{equation}
    relative to the uniform probability distribution on $\mathbb{S}^{n-1}$. Then,
    \begin{equation}
        \operatorname{TV}\left(\operatorname{Law}(Ga), \;\mathcal{N}(0, (\1_m + B)^{-1})\right) \leq \sqrt{\frac{6}{n}}\|B\|_F. 
    \end{equation}
\end{lemma}
\begin{proof}[Proof sketch]
    Writing $\Gamma := \operatorname{Law}(G)$ and $\sigma$ for the uniform probability distribution on $\mathbb{S}^{n-1}$, we define
    \begin{equation}
        w(x) := \operatorname{det}(\1_m + B)^{1/2}e^{-x^\top Bx/2}, \quad \mathcal{Z}(G) := \int_{\mathbb{S}^{n-1}} w(Ga) d\sigma(a). 
    \end{equation}
    The conditional sampling procedure has joint density $w(Ga) / \mathcal{Z}(G)$ with respect to $\Gamma \otimes \sigma$. By contrast, if $(G', a')$ are drawn from the probability measure on $\r^{m \times n} \times \mathbb{S}^{n-1}$ with density $w(G'a')$ with respect to $\Gamma \otimes \sigma$ then $\operatorname{Law}(G'a') = \mathcal{N}(0, (\1_m + B)^{-1})$. Contractivity of total variation gives
    \begin{equation}
        \operatorname{TV}\left(\operatorname{Law}(Ga), \mathcal{N}(0, (\1_m + B)^{-1})\right) \leq \frac{1}{2}\mathbb{E}_{G \sim \Gamma}|\mathcal{Z}(G) - 1| \leq \frac{1}{2}\sqrt{\mathbb{E}_{G \sim \Gamma}[\mathcal{Z}(G)^2] - 1}.
    \end{equation}
    Some elementary Gaussian integration and matrix calculations give the second moment bound
    \begin{equation}
        \mathbb{E}_{G \sim \Gamma}[\mathcal{Z}(G)^2]  \leq \left(1 - \frac{3\|B\|_F^2}{n}\right)^{-1/2} \quad (\forall n > 3\|B\|_F^2).
    \end{equation}
    For $n \geq 6\|B\|_F^2$ this yields the claimed estimate, and if $n < 6\|B\|_F^2$ then the claim follows from $\operatorname{TV} \leq 1$.
    The full proof is given in Appendix \ref{subsec:appendixradomizedcorrectionlemma}.
\end{proof}

Using this, we may replicate the behavior of the sum of two squares approximation using $\lceil c \log d\rceil $ rounds of sufficient width in the regime $c \leq c_\star(\kappa)$:

\begin{proposition}[Upper bound, $c \leq c_\star(\kappa)$]
\label{prop:gaussianupperboundsubcritical}
    Fix $\kappa \in (1, \infty)$ and $c \in (0, c_\star(\kappa)]$. For all $d \in \n$ sufficiently large and any $\delta_d \in (0, 1)$ (which may depend on $d$), there is a randomized parallel matrix-vector sampling algorithm for $\mathcal{M}_{d,\kappa}$, denoted $\mathcal{A}$, which almost surely uses at most $T := \lceil c \log d\rceil $ rounds and at most
    \begin{equation}
        4 + \frac{96}{\delta_d^2} \left(\frac{\sqrt{\kappa}+1}{\sqrt{\kappa}-1}\right)^2 d^{1-c/c_\star(\kappa)}
    \end{equation}
    queries per round such that 
    \begin{equation}
        \sup_{A \in \mathcal{M}_{d,\kappa}} \operatorname{TV}\left(\mathcal{A}^A, \mathcal{N}(0, A^{-1})\right) \leq \delta_d.
    \end{equation}
\end{proposition}
\begin{remark}
    Note that at the critical round count $T = \lceil c_\star(\kappa) \log d\rceil$, this sampler achieves constant TV error with $\mathcal{O}\left(\left(\frac{\sqrt{\kappa}+1}{\sqrt{\kappa}-1}\right)^2\right)$ queries per round. Hence, $c_\star(\kappa)$ is the constant-width threshold.
\end{remark}
\begin{proof}
Write $\epsilon := \epsilon_{T,\kappa} \in (0, 1)$, $\delta := \delta_d$, and $\Lambda = \log \frac{\sqrt{\kappa} + 1}{\sqrt{\kappa}-1}$. By Lemma \ref{lemma:chebyshevapproximation}(i), there exist real polynomials $u, v$ of degree at most $T-1$ such that with $q := u^2 + v^2$ we have
\begin{equation}
\sup_{x \in [1, \kappa]} \left|1 - xq(x)\right|   \leq \epsilon. 
\end{equation}
For $d \geq d_0(\kappa, c)= \left\lceil \exp\left(\frac{1 + \operatorname{arccosh}(3)/\Lambda}{2c}\right)\right\rceil$ we have $\epsilon \leq 1/3$.
If we set
\begin{equation}
    n := \max\left\{2,\; \left \lceil\frac{12 d\epsilon^2}{\delta^2} \right\rceil\right\},
\end{equation}
then we may use that $\epsilon = \operatorname{sech}((2T-1)\Lambda) \leq 2e^{-(2T-1) \Lambda} \leq 2e^\Lambda d^{-2c\Lambda}$ to get
\begin{equation}
    n \leq 2 + \frac{12d\epsilon^2}{\delta^2} \leq 2 + \frac{48e^{2\Lambda} d^{1-4c\Lambda}}{\delta^2}  = 2 + \frac{48}{\delta^2}\left(\frac{\sqrt{\kappa}+1}{\sqrt{\kappa}-1}\right)^2 d^{1-c/c_\star(\kappa)}.
\end{equation}
Draw $g_j, h_j \sim \mathcal{N}(0, \1_d)$ i.i.d. for $j \in [n]$. Using $2n$ parallel queries in each of the $T$ rounds, we compute $A^t g_j, A^t h_j$ for $t \in [T]$ and $j \in [n]$ and use them to evaluate 
\begin{equation}
   Au(A)g_j, \quad Av(A)g_j, \quad Au(A) h_j, \quad Av(A)h_j.
\end{equation}
Using these vectors, we define the matrix $K \in \r^{n \times n}$ with entries
\begin{align}
    K_{ij} &:= \la u(A)g_i, Au(A) g_j\ra + \la v(A) g_i, Av(A) g_j\ra - \la g_i, g_j\ra \\
    &\quad +\la u(A)h_i, Au(A) h_j\ra + \la v(A) h_i, Av(A) h_j\ra - \la h_i, h_j\ra.
\end{align}
Since $q = u^2 + v^2$, we may write $ K = G^\top B G$ with
\begin{equation}
 G := \begin{bmatrix}
        g_1 & \cdots & g_n \\
        h_1 & \cdots & h_n
    \end{bmatrix} \in \r^{2d \times n}, \quad B := \begin{bmatrix}
        Aq(A) - \1_d & 0 \\ 0 & Aq(A) - \1_d
    \end{bmatrix} \in \r^{2d \times 2d}.
\end{equation}
Note that $\|B\|_{\operatorname{op}} \leq \epsilon \leq 1/3$ and so $\|B\|_F^2 \leq 2d\epsilon^2$. 
Conditional on $G$, draw $a \in \mathbb{S}^{n-1}$ with density proportional to
\begin{equation}
    \exp\left(-\frac{1}{2}a^\top Ka\right) =  \exp\left(-\frac{1}{2}a^\top G^\top B Ga\right)
\end{equation}
with respect to the uniform distribution on the sphere via rejection sampling. Use it to define the sample
\begin{equation}
    Z := \sum_{j=1}^n a_j  u(A)g_j -  \sum_{j=1}^n a_j v(A)h_j. 
\end{equation}
By Lemma \ref{lemma:gaussiancorrectionlemma}, we find that \begin{equation}
\label{eqn:proofofgaussiansubcriticalupperboundcontractivityofTV}
    \operatorname{TV}\left(\operatorname{Law}(Ga), \mathcal{N}(0, \operatorname{diag}(Aq(A), Aq(A))^{-1})\right) \leq \sqrt{\frac{12d\epsilon^2}{n}} \leq \delta
\end{equation}
by definition of $n$. If we define the map $f: \r^{2d} \to \r^d$ by $f(x, y) := u(A) x - v(A)y$ then $f(Ga) = Z$. Moreover, if $S \sim \mathcal{N}(0, \operatorname{diag}(Aq(A), Aq(A))^{-1})$ then
\begin{align}
  \operatorname{Law}(f(S)) &= \mathcal{N}\left(0, u(A)(Aq(A))^{-1}u(A) + v(A)(Aq(A))^{-1}v(A)\right) \\&= \mathcal{N}\left(0, (Aq(A))^{-1} (u(A)^2 + v(A)^2)\right) \\
  &= \mathcal{N}(0, A^{-1}).
\end{align}
Applying the contractivity of $\operatorname{TV}$ to  (\ref{eqn:proofofgaussiansubcriticalupperboundcontractivityofTV}) implies
\begin{equation}
    \operatorname{TV}(\operatorname{Law}(Z), \mathcal{N}(0, A^{-1})) \leq \delta.
\end{equation}
\end{proof}

\subsection{Lower bound for $c < c_\star(\kappa)$}
We now show a lower bound for parallel centered Gaussian sampling with $\lceil c \log d\rceil$ rounds in the $c < c_\star(\kappa)$ regime. In this subsection we will state two lemmas and show how they imply the claimed width lower bound. As before, we will prove a parameterized lower bound (Proposition \ref{prop:parameterizedlowerboundGaussian}) and its asymptotic consequence (Corollary \ref{corollary:gaussianlowerbound}).

 The first lemma reduces general centered Gaussian sampling algorithms to a canonical form in terms of polynomials of the inverse covariance. This is inspired by the revelation argument of \cite[Lemma 4.3]{zhou2026complexitygaussiansampling}, with the improvement that the polynomial degree is bounded by the number of rounds rather than the number of total queries.

\begin{lemma}[Canonical form for Gaussian samplers]
\label{lemma:gaussianlowerboundstep1}
    Fix $d, T, K \in \n$ and $\kappa \in (1, \infty)$. Let $\mathcal{A}$ be a randomized parallel matrix-vector algorithm for $\mathcal{M}_{d,\kappa}$ using at most $T$ rounds and at most $K$ total queries almost surely.\footnote{Recall that a randomized first-order parallel sampling algorithm for $\mathcal{V}_{d,\kappa}^{(\operatorname{Gaussian})}$ is a randomized parallel matrix-vector algorithm for $\mathcal{M}_{d,\kappa}$.} Define $\overline{\mathcal{A}}$ by drawing $U \sim \operatorname{Haar}(O(d))$ independently, running $\mathcal{A}$ with the simulated gradient oracle
    \begin{equation}
        x \mapsto U^\top AUx
    \end{equation}
    for each instance $A \in \mathcal{M}_{d,\kappa}$, and returning $U$ times the output. In particular, if $\overline{Z}(V_A) \in \r^d$ denotes the output of $\overline{\mathcal{A}}$ on the instance $A \in \mathcal{M}_{d,\kappa}$, then
    \begin{equation}
        \overline{Z}(V_A) := U Z(V_{U^\top A U}).
    \end{equation}
    Then $\overline{Z}(V_A) $ admits a representation
    \begin{equation}
        \overline{Z}(V_A) \overset{(d)}{=} \sum_{\ell = 1}^{K+1} p_\ell(A) g_\ell, \quad \operatorname{deg}p_\ell \leq T \quad (\forall A \in \mathcal{M}_{d,\kappa}),
    \end{equation}
    where $g_1, \ldots, g_{K+1} \sim \mathcal{N}(0, \1_d)$ are i.i.d. standard Gaussian and the coefficients of the $p_\ell$'s may depend measurably on the $g_\ell$'s, additional randomness, and oracle responses. Moreover, $\overline{\mathcal{A}}$ uses at most $T$ rounds and at most $K$ total queries almost surely, and for any rotation-invariant distribution $\mathcal{D}$ over $\mathcal{M}_{d,\kappa}$ we have
    \begin{equation}
        \mathbb{E}_{A \sim \mathcal{D}} \left[\operatorname{TV}\left(\operatorname{Law}\left(\overline{Z}(V_A)\right), \mathcal{N}(0, A^{-1})\right)\right] \leq    \mathbb{E}_{A \sim \mathcal{D}} \left[\operatorname{TV}\left(\operatorname{Law}\left({Z}(V_A)\right), \mathcal{N}(0, A^{-1})\right)\right].
    \end{equation}
\end{lemma}
\begin{proof}[Proof sketch]
    Fix $A \in \mathcal{M}_{d,\kappa}$. Suppose that $E \subset \r^d$ is the span of the vectors $Ux_i$, $AUx_i$ where $\{x_i\}_i$ enumerate the queries from all prior rounds. Process the current batch $x_1, \ldots, x_q$ in order, without revealing its responses yet. Conditional Haar symmetry allows us, for each $j \in [q]$, to construct a fresh standard Gaussian $g_j$ which is independent of the previous history (including the Gaussian vectors and rotated queries revealed earlier in this batch) such that
    \begin{equation}
        Ux_j \in E + \operatorname{span}\{Ux_1, \ldots, Ux_{j-1}, g_j\} \subset E + \operatorname{span}\{g_1, \ldots, g_j\}.
    \end{equation}
    Now reveal the responses $U^\top AUx_j$ for $j \in [q]$. Writing $F := E + \operatorname{span}\{g_1, \ldots, g_q\}$, the full revealed subspace satisfies
    \begin{equation}
        E' := E + \operatorname{span}\{Ux_j, AUx_j:\; j \in [q]\}\subset F + AF.
    \end{equation}
    Thus, each query adds one fresh Gaussian, and each round increases the possible degree in $A$ by at most one. Induction gives degree at most $T$. Applying the same revelation rule to $x=Z(V_{U^\top AU})$ reveals $\overline{Z}(V_A) = Ux$ and adds one final Gaussian, yielding the claimed representation with $K+1$ Gaussians. It can be shown by conditional independence that these Gaussians are mutually independent. 
    To finish, the total variation bound follows from convexity of TV under mixtures and rotation invariance of $\mathcal{D}$. See Appendix \ref{subsec:appendixrevelationlemma} for the full proof, which uses the Haar state formalism from \cite[Definition C.2]{zhou2026complexitygaussiansampling}.
\end{proof}

\begin{remark}
    In words, any centered Gaussian sampler can be replaced by a sampler admitting the above polynomial representation, without increasing its round or query bounds or its TV error averaged over a rotation-invariant distribution of centered Gaussian instances. We emphasize that the individual instances $A$ need not be isotropic; it is only required that the distribution $\operatorname{Law}(U^\top AU) = \mathcal{D}$ for $A \sim \mathcal{D}$ and every fixed $U \in O(d)$.
\end{remark}

The second lemma exhibits an approximation lower bound for such random polynomials, based on elementary Gaussian concentration and the deterministic polynomial approximation lower bound from Lemma \ref{lemma:chebyshevapproximation}(ii):

\begin{lemma}
\label{lemma:gaussianlowerboundstep2}
    Fix $\kappa \in (1, \infty)$ and integers $b, r \geq 2$, and set $d = 2br$. There exists a symmetric matrix $A_0 \in \r^{d \times d}$ and a Borel set $E \subset \r^d$, depending only on $\kappa, b, r$, with the following properties:
    \begin{enumerate}
        \item[(i)] We have $\1_d \preceq A_0 \preceq \kappa \1_d$, i.e. $A_0 \in \mathcal{M}_{d,\kappa}$, and
        \begin{equation}
            \pi_{A_0}(E) \geq 1 - 4r \exp\left(-\frac{b\epsilon_{r,\kappa}^2}{512}\right), \quad \pi_{A_0} := \mathcal{N}(0, A_0^{-1}).
        \end{equation}
        \item[(ii)] For every integer $n \geq 1$, i.i.d. $g_1, \ldots, g_n \sim \mathcal{N}(0, \1_d)$, and random real polynomials $p_1, \ldots, p_n$ of degree at most $r-1$, the vector
        \begin{equation}
            Z := \sum_{\ell = 1}^n p_\ell(A_0) g_\ell
        \end{equation}
        satisfies
        \begin{equation}
            \mathbb{P}(Z \in E) \leq 4r \exp\left(n \log 9 - \frac{b\epsilon_{r,\kappa}^2}{512}\right).
        \end{equation}
        The $p_\ell$'s may depend measurably on the $g_\ell$'s and additional randomness.
    \end{enumerate}
    In particular, every such polynomial output satisfies
    \begin{equation}
        \operatorname{TV}\left(\operatorname{Law}(Z), \mathcal{N}(0, A_0^{-1})\right) \geq 1 - 4r e^{-b\epsilon_{r,\kappa}^2 / 512} - 4re^{n \log 9 - b\epsilon_{r,\kappa}^2/512}.
    \end{equation}
\end{lemma}
\begin{proof}[Proof sketch]
Let $A_0$ be diagonal with the $2r$ distinct eigenvalues $\lambda_0, \ldots, \lambda_{2r-1}$ from (\ref{eqn:chebyshevvalues}), each repeated with multiplicity $b$. Writing $P_j$ to be the projection onto the eigenspace with eigenvalue $\lambda_j$ (so that $\operatorname{im}P_j$ has dimension $b$), define
\begin{equation}
    E :=  \left\{z \in \r^d:\; \max_{j}\left|\lambda_j \frac{\|P_jz\|^2}{b} - 1\right| \leq \frac{\epsilon_{r,\kappa}}{8}\right\} \subset \r^d.
\end{equation}
This is the event that every block's squared norm $\|P_jz\|^2$ is within relative error $\epsilon_{r,\kappa}/8$ of $b/\lambda_j$, which is its mean when $z \sim \mathcal{N}(0, A_0^{-1})$. Chi-square concentration gives (i). Denoting $Z := \sum_{\ell = 1}^n p_\ell(A_0)g_\ell$,  approximate orthogonality of Gaussian vectors within each eigenspace gives
\begin{equation}
    \frac{\|P_jZ\|^2}{b} \approx S(\lambda_j), \quad S(x) := \sum_{\ell = 1}^n p_\ell(x)^2.
\end{equation}
A covering argument on the sphere makes this approximation hold with relative error at most $\epsilon_{r,\kappa}/4$, simultaneously for any realization of the polynomials $p_1, \ldots, p_n$, except on an event of probability at most \begin{equation}
    4r\exp\left(n \log 9 - \frac{b\epsilon_{r,\kappa}^2}{512}\right).
\end{equation}
When the relative error is $\leq \epsilon_{r,\kappa}/4$, if $Z \in E$ then we would have
\begin{equation}
    \max_j |1 - \lambda_j S(\lambda_j)| \leq \epsilon_{r,\kappa}/2
\end{equation}
which would contradict the lower bound of Lemma \ref{lemma:chebyshevapproximation}(ii). This proves (ii), and the final assertion follows from TV testing on $E$. The full proof is given in Appendix \ref{subsec:appendixgaussianlowerboundstep2}.
\end{proof}

We may now apply these lemmas by (1) introducing random rotations of $A_0$ and (2) using Markov's inequality to convert to a lower bound for expected total queries. Doing so yields our parameterized lower bound for parallel first-order sampling of well-conditioned centered Gaussian measures:

\begin{proposition}[Parameterized lower bound, Gaussian]
\label{prop:parameterizedlowerboundGaussian}
    Fix $\kappa \in (1, \infty)$. For all fixed $b, r \in \n$ with $b, r \geq 2$, if we denote $d = 2br$ then there exists a Borel probability distribution $\mathcal{D}_{b,r}^{(\kappa)}$ over $C^\infty(\r^d)$ (depending only on $b, r, \kappa$) with the following properties:
    \begin{enumerate}
        \item[(i)] For almost every $V \sim \mathcal{D}_{b,r}^{(\kappa)}$ we have $V(z) = \frac{1}{2}z^\top Az$ for some symmetric $A \in \r^{d \times d}$ with $\1_d \preceq A \preceq \kappa \1_d$ and all $z \in \r^d$. In particular, $\mathcal{D}_{b,r}^{(\kappa)}$ is supported on $\mathcal{V}_{d,\kappa}^{(\operatorname{Gaussian})}$. 
        \item[(ii)] Let $T, Q \in (0, \infty)$ with $T < r$, and let $\mathcal{A}$ be any randomized first-order parallel sampling algorithm for $\mathcal{V}_{d,\kappa}^{(\operatorname{Gaussian})}$ (independent of $V \sim \mathcal{D}_{b,r}^{(\kappa)}$) which uses at most $T$ rounds almost surely and makes at most $Q$ queries in expectation (in the sense of Definition \ref{def:samplingalg}). Then, it holds that
        \begin{equation}
            \mathbb{E}_{V \sim \mathcal{D}_{b,r}^{(\kappa)}} \left[\operatorname{TV}(\mathcal{A}^V, \pi_V)\right] \geq 1 - \frac{1024(\log 9)}{b\epsilon_{r,\kappa}^2}Q - 40r \exp\left(-\frac{b\epsilon_{r,\kappa}^2}{1024}\right).
        \end{equation}
        This lower bound holds even if the algorithm is allowed to know the spectrum of $\nabla^2 V$ (the inverse covariance) in advance.
    \end{enumerate}
\end{proposition}
\begin{proof}[Proof sketch]
   The claim follows by Haar-conjugating the hard matrix from Lemma \ref{lemma:gaussianlowerboundstep2}, truncating the algorithm to a deterministic query budget via Markov's inequality, and applying Lemma \ref{lemma:gaussianlowerboundstep1}. The proof is given in Appendix \ref{subsec:appendixparameterizedgaussianprop}.
\end{proof}

As a direct consequence of Proposition \ref{prop:parameterizedlowerboundGaussian}, we get the following asymptotic statement which shows the optimality of the exponent in the upper bound from Proposition \ref{prop:gaussianupperboundsubcritical}. As before, we include the proof here to present the choices of the parameters: we take $r = \lceil c \log d\rceil + 1$ and $b \asymp d/r$ so that our hard spectrum contains $2\lceil c \log d\rceil +2$ distinct eigenvalues, each of multiplicity $b$. 
\begin{corollary}[Asymptotic lower bound, Gaussian]
\label{corollary:gaussianlowerbound}
Fix $\kappa \in (1, \infty)$ and $c \in (0, c_\star(\kappa))$. There is a constant $L = L(\kappa, c) > 0$ and, for each $d \in \n$, a probability distribution $\mathcal{D}_{d,\kappa,c}^{(\operatorname{Gaussian})}$ over $\mathcal{V}_{d,\kappa}^{(\operatorname{Gaussian})}$ such that the following holds:

    For every fixed $C,\delta \in (0, \infty)$, there exists $d_0 = d_0(\kappa, c, C, \delta)$ such that for all $d \geq d_0$, if $\mathcal{A}$ is any randomized first-order parallel sampling algorithm for $\mathcal{V}_{d,\kappa}^{(\operatorname{Gaussian})}$ which uses at most $ \lceil c \log d\rceil $ rounds almost surely and makes at most \begin{equation}C \frac{d^{1-c/c_\star(\kappa)}}{(\log d)^{1+\delta}}\end{equation} total queries in expectation, then
    \begin{equation}
        \mathbb{E}_{V \sim \mathcal{D}_{d,\kappa,c}^{(\operatorname{Gaussian})}} \left[\operatorname{TV}(\mathcal{A}^V, \pi_V)\right] \geq 1 - \frac{LC}{(\log d)^\delta}.
    \end{equation}
\end{corollary}
\begin{proof}
Write $\rho := \frac{\sqrt{\kappa} + 1}{\sqrt{\kappa} - 1}$ and $\gamma := 1 - \frac{c}{c_\star(\kappa)} > 0$. Choose \begin{equation}
    r := \lceil c \log d\rceil + 1,\quad b := \left\lfloor \frac{d}{2r}\right\rfloor.
\end{equation}
For sufficiently large $d$ we have $b, r \geq 2$, $2br \leq d$, and $\lceil c \log d\rceil < r$. Apply Proposition \ref{prop:parameterizedlowerboundGaussian} in dimension $2br$ and pad the remaining coordinates to dimension $d$ using standard Gaussians to obtain $\mathcal{D}_{d,\kappa,c}^{(\operatorname{Gaussian})}$. When $b < 2$ or $r < 2$, define $\mathcal{D}_{d,\kappa,c}^{(\operatorname{Gaussian})}$ arbitrarily. For sufficiently large $d$, we know
\begin{equation}
    b \geq \frac{d}{4r}, \quad r \leq (c+1) \log d, \quad \epsilon_{r,\kappa}^2 \geq e^{-2(2r-1)\log\rho} = \rho^{-4r + 2} , 
\end{equation}
where we used $\operatorname{sech}(u) \geq e^{-u}$. Since $r \leq c \log d + 2$ we have $-4r+2 \geq -4c\log d - 6$, which implies that
\begin{equation}
   b \epsilon_{r,\kappa}^2 \geq \frac{d}{4r}\rho^{-6}d^{-c/c_\star(\kappa)}  \geq \frac{d^{1-c/c_\star(\kappa)}}{4\rho^6(c+1) \log d}.
\end{equation}
Denote
\begin{equation}
    L(\kappa, c) := 8192\rho^6 (c+1) \log 9.
\end{equation}
If the algorithm makes at most
\begin{equation}
    Q := C \frac{d^\gamma}{(\log d)^{1+\delta}}
\end{equation}
total queries in expectation, then
\begin{equation}
    \frac{1024 \log 9}{b\epsilon_{r, \kappa}^2}Q \leq \frac{4096 \rho^6(c+1) \log 9 C}{(\log d)^\delta} = \frac{LC}{2(\log d)^\delta}. 
\end{equation}
Moreover,
\begin{equation}
\label{eqn:proofcorollarygaussianlowerbounddecaywrtd2}
    40r \exp\left(-\frac{b\epsilon_{r,\kappa}^2}{1024}\right) \leq 40(c+1) \log d \exp\left(-\frac{d^\gamma}{4096 \rho^6 (c+1)\log d}\right).
\end{equation}
Since $(\log d)^{1+\delta} \exp\left(-a\frac{d^\gamma}{\log d}\right) \to 0$ for any $a > 0$, we may choose $d$ sufficiently large that the right hand side of (\ref{eqn:proofcorollarygaussianlowerbounddecaywrtd2}) is $\leq \frac{LC}{2(\log d)^\delta}$. The claim follows from Proposition \ref{prop:parameterizedlowerboundGaussian}.
\end{proof}

\subsection{Growing condition number}
\label{subsection:gaussiangrowingkappa}

We now allow $\kappa$ to grow with $d$. Since the threshold coefficient $c_\star(\kappa)$ also grows, we write the number of rounds as $T=\lceil a c_\star(\kappa)\log d\rceil$ for fixed $a>0$. The two upper bounds of Theorem~\ref{theorem:precisegaussian} extend to arbitrary growth of $\kappa$. For polylogarithmic $\kappa$, the upper and lower bounds continue to match up to polylogarithmic factors. More generally, for $\kappa=d^{o(1)}$ they identify the same constant-width threshold and the same width exponent below it. Determining how the tradeoff changes for polynomially growing $\kappa$ remains open.

\begin{corollary}[Growing condition number]
\label{corollary:gaussiangrowingkappa}
Fix $a \in (0,\infty)$ and let $\kappa=\kappa(d)\to\infty$. For each dimension $d$, consider the class of randomized first-order parallel sampling algorithms for $\mathcal{V}_{d,\kappa}^{(\operatorname{Gaussian})}$ which use at most $T := \lceil a c_\star(\kappa)\log d\rceil$ rounds almost surely (in the sense of Definition~3.1). Then, the following hold as $d\to\infty$:
    \begin{enumerate}
        \item[(i)] If $a>1$, then there is such an algorithm $\mathcal{A}$ which makes $2$ queries per round almost surely and satisfies
        \begin{equation}
            \sup_{V \in \mathcal{V}_{d,\kappa}^{(\operatorname{Gaussian})}}
            \operatorname{TV}\left(\mathcal{A}^V,\pi_V\right)
            \leq d^{-(a-1)/2}.
        \end{equation}

        \item[(ii)] If $a\leq1$, then for any $\delta_d \in (0,1)$ (which may depend on $d$) there is such an algorithm $\mathcal{A}$ which makes at most $\mathcal{O}(d^{1-a}/\delta_d^2)$ queries per round almost surely and satisfies
        \begin{equation}
            \sup_{V \in \mathcal{V}_{d,\kappa}^{(\operatorname{Gaussian})}}
            \operatorname{TV}\left(\mathcal{A}^V,\pi_V\right)
            \leq \delta_d.
        \end{equation}

        \item[(iii)] If $a<1$ and $T\log T=o(d^{1-a})$, then for every fixed $C,\delta\in(0,\infty)$, any such algorithm $\mathcal{A}$ which makes at most $C (\log d)^{-\delta}d^{1-a}/T$ total queries in expectation satisfies
        \begin{equation}
            \sup_{V \in \mathcal{V}_{d,\kappa}^{(\operatorname{Gaussian})}}
            \operatorname{TV}\left(\mathcal{A}^V,\pi_V\right)
            \geq 1-o(1).
        \end{equation}
        This lower bound holds even if $\mathcal{A}$ is supplied the covariance eigenvalues in advance.
    \end{enumerate}
    In particular, if $\kappa\leq(\log d)^L$ for some fixed $L>0$, then the optimal worst-case width for parallel sampling of $\mathcal{V}_{d,\kappa}^{(\operatorname{Gaussian})}$ with constant TV accuracy using almost surely at most $\lceil a c_\star(\kappa)\log d\rceil$ rounds is
\begin{equation}
    \begin{cases}
        \Theta(1) & a \geq 1, \\
        \widetilde{\Theta}\left(d^{1-a}\right) & a < 1.
    \end{cases}
\end{equation}
\end{corollary}

\begin{proof}
Write $\Lambda := \log((\sqrt{\kappa}+1)/(\sqrt{\kappa}-1))$ so that $c_\star(\kappa)=\frac{1}{4\Lambda}$. (i) follows from Proposition \ref{prop:gaussianupperboundsupercritical}. For (ii), the choice of $T$ gives
\begin{equation}
    \varepsilon_{T,\kappa}
    = \operatorname{sech}((2T-1)\Lambda)
    \leq 2e^\Lambda d^{-a/2}.
\end{equation}
Since $\Lambda\to0$, for sufficiently large $d$, we have $\varepsilon_{T,\kappa}\leq1/3$. In the proof of Proposition~\ref{prop:gaussianupperboundsubcritical}, the sampler uses $n$ independent pairs of Gaussian probes, requiring $2n$ queries per round, and its output $Z$ satisfies
\begin{equation}
    \operatorname{TV}\left(\operatorname{Law}(Z),\mathcal{N}(0,A^{-1})\right)
    \leq \sqrt{\frac{12d\varepsilon_{T,\kappa}^2}{n}}
    \quad (\forall A \in \mathcal{M}_{d,\kappa}).
\end{equation}
Taking $n:=\max\{2,\lceil 12d\varepsilon_{T,\kappa}^2/\delta_d^2\rceil\}$ gives uniform TV error at most $\delta_d$. Since $\varepsilon_{T,\kappa}\leq 2e^\Lambda d^{-a/2}$, the width is at most $4+96e^{2\Lambda}d^{1-a}/\delta_d^2$.

For (iii), take $r=T+1$ and $b=\lfloor d/(2r)\rfloor$. Since $T\log T=o(d^{1-a})$, we have $T=o(d)$ and hence $b\geq d/(4r)$ for all sufficiently large $d$. Apply Proposition~\ref{prop:parameterizedlowerboundGaussian} in dimension $2br$ and pad the remaining coordinates with standard Gaussians as in Corollary~\ref{corollary:gaussianlowerbound}.

To bound the error terms in Proposition~\ref{prop:parameterizedlowerboundGaussian}, observe that $T\leq a\log d/(4\Lambda)+1$ and so $(2r-1)\Lambda=(2T+1)\Lambda\leq (a/2)\log d+3\Lambda$. Using $\operatorname{sech}(u)\geq e^{-u}$ for $u\geq0$, we obtain
\begin{equation}
    b\varepsilon_{r,\kappa}^2
    \geq \frac{d}{4r}e^{-2(2r-1)\Lambda}
    \geq \frac{e^{-6\Lambda}}{4r}d^{1-a}
    \gtrsim \frac{d^{1-a}}{T},
\end{equation}
where the last inequality uses $\Lambda\to0$ and $r=T+1$. Under the stated query budget, it follows that $Q/(b\varepsilon_{r,\kappa}^2)=\mathcal{O}(C/(\log d)^\delta)$. Also, $T\log T=o(d^{1-a})$ gives $b\varepsilon_{r,\kappa}^2/\log r\to\infty$, which means $r\exp(-b\varepsilon_{r,\kappa}^2/1024)=o(1)$. Both error terms in Proposition~\ref{prop:parameterizedlowerboundGaussian} therefore vanish, proving (iii).
\end{proof}

\bibliographystyle{alpha}
\bibliography{refs}

\clearpage
\appendix
\etocdepthtag.toc{appendix}
\begingroup
\renewcommand{\contentsname}{Appendix Contents}
\etocsettagdepth{main}{none}
\etocsettagdepth{appendix}{subsection}
\tableofcontents
\endgroup

\section{Proofs for log-concave lower bound}

\subsection{Proof of Lemma \ref{lemma:theconstruction}: Hessian bounds and gating conditions}
\label{subsec:proofoflemamtheconstruction}

We first show the following scalar derivative bounds on our cutoff functions:
\begin{lemma}
\label{lemma:scalarbounds}
    Let $\sigma, \chi, \phi, g, H$ be as defined in (\ref{eqn:defofsigma}), (\ref{eqn:defofchiphi}), (\ref{eqn:defofg}), and (\ref{eqn:defofH}). Then,
    \begin{align}
        % \|\sigma'\|_\infty \leq 8, \quad \|\sigma''\|_\infty \leq 400 \\
        \|\chi'\|_\infty \leq 128, \quad \|\chi''\|_\infty \leq 102400 \\
        \|\phi'\|_\infty \leq 25, \quad \|\phi''\|_\infty \leq 4864,
        \\
        \|g'\|_\infty \leq \frac{32}{\rho}, \quad \|g''\|_\infty \leq \frac{6400}{\rho^2}.
    \end{align}
    Also, $\operatorname{spt}(\phi'') \subset [1/4 ,1/2]$ and 
    \begin{align}
        \sup_{u \geq 0}|\phi(u) - u\phi'(u)| \leq 9.
        % , \quad \sup_{u \geq 0} \max \left\{u|\phi''(u)|, \; u^{3/2} |\phi''(u)|, \; 2u^2|\phi''(u)|\right\} \leq 2432.
    \end{align}
\end{lemma}
\begin{proof}
We claim that 
\begin{equation}
\label{eqn:proofofscalarlemmaboundonsigmas}
    \|\sigma'\|_\infty \leq 8, \quad \|\sigma''\|_\infty \leq 400.
\end{equation}
Before proving this, we first show how it implies the main result. By the chain rule and definitions of $\chi, g$, we have
\begin{equation}
    \|\chi'\|_\infty \leq 16 \cdot 8 = 128, \quad \|\chi''\|_\infty \leq 16^2 \cdot 400 = 102400,
\end{equation}
\begin{equation}
    \|g'\|_\infty \leq \frac{4}{\rho} \cdot 8 = \frac{32}{\rho}, \quad \|g''\|_\infty \leq \frac{16}{\rho^2} \cdot 400 = \frac{6400}{\rho^2}.
\end{equation}
Finally, differentiating $\phi(u) = u + (1-u) \sigma(4u-1)$ yields
\begin{align}
    \phi'(u) &= 1 - \sigma(4u-1) + 4(1-u)\sigma'(4u-1),\\ 
    \phi''(u) &= -8\sigma'(4u-1) + 16(1-u)\sigma''(4u-1).
\end{align}
Since $\phi(u) = u$ for $0 \leq u \leq 1/4$ and $\phi(u) = 1$ for $u \geq 1/2$, we have $\operatorname{spt}(\phi'') \subset [1/4, 1/2]$. On this transition interval $0 \leq 1-u \leq 3/4$ and so
\begin{equation}
    |\phi'(u)| \leq 1 + 4 \cdot \frac{3}{4} \cdot 8 = 25,
\end{equation}
\begin{equation}
    |\phi''(u)| \leq 8 \cdot 8 + 16 \cdot \frac{3}{4} \cdot 400 = 4864.
\end{equation}
Outside the transition interval $\phi'$ equals either 0 or 1, and so these bounds hold globally. Lastly,
\begin{equation}
\label{eqn:proofofscalarlemmaphi}
    \phi(u) - u\phi'(u) = \sigma(4u-1) - 4u(1-u)\sigma'(4u-1).
\end{equation}
For $u \in [1/4, 1/2]$, using $u(1-u) \leq 1/4$ gives
\begin{equation}
    | \phi(u) - u\phi'(u)| \leq 1 + 4 \cdot \frac{1}{4} \cdot 8 = 9.
\end{equation}
For $u \leq 1/4$ the expression (\ref{eqn:proofofscalarlemmaphi}) is 0, and for $u \geq 1/2$ it is 1, which proves all the claimed bounds. To show (\ref{eqn:proofofscalarlemmaboundonsigmas}), since $\sigma$ is constant outside $(0, 1)$ and $\sigma(1-t) = 1 - \sigma(t)$ it suffices to consider $0 < t < 1/2$. Set
\begin{equation}
    \ell(t) := \frac{1}{1-t} - \frac{1}{t}
\end{equation}
so that \begin{equation}\sigma' = \sigma(1-\sigma)\ell', \quad \sigma'' = \sigma(1-\sigma) [(1-2\sigma) (\ell')^2 + \ell''].\end{equation}
On the interval $(0, 1/2)$ we know
\begin{equation}
    \sigma(t)(1-\sigma(t)) \leq e^{\ell(t)} \leq e^{2-\frac{1}{t}}, \quad |\ell'(t)| \leq \frac{1}{t^2}+4, \quad |\ell''(t)| \leq \frac{2}{t^3} + 16.
\end{equation}
Writing $x = 1/t \geq 2$, we find
\begin{align}
    |\sigma'(t)| &\leq e^{2-x} (x^2 + 4) \leq 8, \\ |\sigma''(t)| &\leq e^{2-x} (x^4 + 2x^3 + 8x^2 + 32) \leq 256 + 54 + 64 \leq 400,
\end{align}
where we used that $\sup_{x \geq 0}x^p e^{-x} = (p/e)^p$.
\end{proof}

Next, we prove the following lemma regarding the functions $F_1^U, \ldots, F_r^U$, which ensures sufficient Hessian control and a Gaussian comparison when the radial cutoffs and gates are saturated:

\begin{lemma}
\label{lemma:intermediatelemma}
 Fix $U = (U_1, \ldots, U_r)$ where each $U_j \in O(n_j)$ is an orthogonal matrix, and let $F_1^U, \ldots, F_r^U: \r^d \to \r$ be defined as in Section \ref{subsec:hardinstanceconstruction}. Then, the following statements hold:
    \begin{enumerate}
    \item[(i)] The functions $F_1^U, \ldots, F_r^U$ are all infinitely differentiable.
        \item[(ii)] Denoting $K_g := 1 + \|g'\|_\infty + \|g''\|_{\infty}$, for all $z \in \r^d$ there are the Hessian operator norm bounds 
\begin{equation}
    \|\nabla^2_{z_1,z_1} F_1^U(z)\|_{\operatorname{op}} \leq 13124, \quad \max_{j \in \{2, \ldots, r\}}\|\nabla^2_{z_{j-1},z_{j-1}} F_j^U(z)\|_{\operatorname{op}} \leq \frac{77412 K_g}{k},  
\end{equation}
\begin{equation}
 \max_{j \in \{2, \ldots, r\}}\|\nabla^2_{z_{j-1},z_{j}} F_j^U(z)\|_{\operatorname{op}} \leq \frac{28560 K_g}{\sqrt{k}} , \quad \max_{j \in \{2, \ldots, r\}}\|\nabla^2_{z_{j},z_{j}} F_j^U(z)\|_{\operatorname{op}} \leq 25352.
\end{equation}
Here, $\nabla^2_{z_{i},z_{\ell}} F_j^U \in \r^{(2n_{i}) \times (2n_\ell)}$ (note that $F_j^U$ only depends on $(z_{j-1}, z_j)$ for $j \geq 2$).

\item[(iii)] For every $j \in [r]$ we have $0 \leq F_j^U(z) \leq \|z_j\|^2 (1 + s_j^U(z))$ for all $z \in \r^d$.
\item[(iv)] For every $z \in \r^d$, we have
\begin{equation}
    \chi\left(\frac{\|z_1\|^2}{2n_1}\right) = 1 \implies F_1^U(z) = \|z_1\|^2 (1 + s_1^U(z))
\end{equation}
and for each $j \in \{2, \ldots, r\}$ we have
\begin{align}
    \chi\left(\frac{\|z_{j-1}\|^2}{2n_{j-1}}\right) = \chi\left(\frac{\|z_{j}\|^2}{2n_{j}}\right) &= 1, \quad s_{j-1}^U(z) \geq \rho / 2, \quad \|z_{j-1}\|^2 \geq k \|z_j\|^2 / 2 \\&\quad \implies F_j^U(z) = \|z_j\|^2 (1+s_j^U(z)).
\end{align}
    \end{enumerate}
\end{lemma}
\begin{proof}
Throughout this proof, we suppress the fixed dependence on $U$ by writing $F_j := F_j^U$. We also write $s_j: \r^{2n_j} \to \r$ for the function
\begin{equation}
\label{eqn:proofofdaappendixdefinitionofs_j}
    s_j(x_j, y_j) := \begin{cases}
        \frac{2\la y_j, U_jx_j\ra}{\|x_j\|^2 + \|y_j\|^2} & (x_j, y_j) \neq 0,
        \\0 & (x_j, y_j) = 0.
    \end{cases}
\end{equation}
Thus, $s_j^U(z) = s_j(z_j)$. When a full vector $z = (z_1, \ldots, z_r)$ is understood, we abbreviate $s_j(z_j)$ by $s_j$. We will adopt these notational changes throughout this appendix.

    \textbf{(i)}. For the first interaction, note that 
    \begin{equation}
        \|z_1\|^2 (1 + s_1) = \|x_1\|^2 + \|y_1\|^2 + 2\la y_1, U_1 x_1\ra,
    \end{equation}
    which is a quadratic polynomial on $\r^{2n_1}$. Since $\chi \in C^\infty(\r)$, it follows that $F_1 \in C^\infty(\r^d)$. Fix $j \in \{2, \ldots, r\}$. Away from $z_{j-1} = 0$ and $z_j = 0$, every factor defining $F_j$ is infinitely differentiable. Moreover, $\chi\left(\frac{\|z_{j-1}\|^2}{2n_{j-1}}\right)$ is zero on a neighborhood of $z_{j-1} = 0$, and similarly $\chi\left(\frac{\|z_{j}\|^2}{2n_{j}}\right)$ is zero on a neighborhood of $z_j = 0$. Thus, $F_j \in C^\infty(\r^d)$.

    \textbf{(ii)}. 
    Fix $m \in \n$ and $U \in O(m)$ (we will take $m = n_j$ and $U = U_j$ later).
   Defining
    \begin{equation}
        M_U := \begin{bmatrix}
            0 & U^\top \\ U & 0,
        \end{bmatrix}
    \end{equation}
    we have that $\|M_U\|_{\operatorname{op}} \leq 1$ and $s(z) = \la z, M_Uz\ra / \|z\|^2$ for $z \in \r^{2m} \setminus \{0\}$. Differentiating gives
    \begin{equation}
        \nabla s(z) = \frac{2(M_Uz - s(z) z)}{\|z\|^2},
    \end{equation}
    \begin{equation}
        \nabla^2 s(z) = \frac{2M_U - 2s(z)\1_{2m}}{\|z\|^2} - \frac{4\left((M_Uz)z^\top + z(M_Uz)^\top\right)}{\|z\|^4} + \frac{8s(z)zz^\top}{\|z\|^4},
    \end{equation}
    and therefore we have the bounds
    \begin{equation}
        |s(z)| \leq 1, \quad \|\nabla s(z)\| \leq \frac{4}{\|z\|}, \quad \|\nabla^2 s(z)\|_{\operatorname{op}} \leq \frac{20}{\|z\|^2} \quad (\forall z \in \r^{2m}\setminus \{0\}).
    \end{equation}
    We also define the radial cutoff $c_m(z) := \chi\left(\frac{\|z\|^2}{2m}\right)$, whose derivatives are 
    \begin{equation}
        \nabla c_m(z) = \frac{1}{m}\chi'\left(\frac{\|z\|^2}{2m}\right)z, \quad \nabla^2 c_m(z) = \frac{1}{m}\chi'\left(\frac{\|z\|^2}{2m}\right)\1_{2m} + \frac{1}{m^2}\chi''\left(\frac{\|z\|^2}{2m}\right) zz^\top.
    \end{equation}
    We know that all derivatives of $\chi$ vanish outside $[1/16, 1/8]$, and on the annulus $\|z\|^2/m \in [1/8, 1/4]$ we apply Lemma \ref{lemma:scalarbounds} to see
    \begin{align}
        \|\nabla c_m(z)\| &\leq \frac{128}{\|z\|} \frac{\|z\|^2}{m} \leq \frac{128/4}{\|z\|}\\\|\nabla^2 c_m(z)\|_{\operatorname{op}} &\leq \frac{128}{\|z\|^2} \frac{\|z\|^2}{m} + \frac{102400}{\|z\|^2} \frac{\|z\|^4}{m^2} \leq \frac{128/4 + 102400 / 16}{\|z\|^2}.
    \end{align}
    Since any $z$ outside this annulus causes these derivatives to be 0, we find that
    \begin{equation}
    \label{eqn:proofofintermediatelemmaboundsonc_m}
        \|\nabla c_m(z)\| \leq \frac{32}{\|z\|}, \quad \|\nabla^2 c_m(z)\|_{\operatorname{op}} \leq \frac{6432}{\|z\|^2} \quad (\forall z \in \r^{2m} \setminus \{0\}).
    \end{equation}
    We first bound the Hessian of $F_1$. For all $z_1 \in \r^{2n_1}$, the quadratic polynomial
    \begin{equation}
        R(z_1) := \|z_1\|^2 (1+s_1(z_1)) = z_1^\top (\1_{2n_1} + M_{U_1})z_1
    \end{equation}
    satisfies the bounds
    \begin{equation}
        0 \leq |R(z_1)| \leq 2\|z_1\|^2, \quad \|\nabla R(z_1)\| \leq 4\|z_1\|, \quad \|\nabla^2 R(z_1)\|_{\operatorname{op}} \leq 4.
    \end{equation}
    Since $F_1(z) = R(z_1) c_{n_1}(z_1)$, the product rule and (\ref{eqn:proofofintermediatelemmaboundsonc_m}) imply
    \begin{equation}
        \|\nabla^2 F_1(z)\|_{\operatorname{op}} \leq 4 \cdot 1 + 2 \cdot 4\cdot 32  + 2 \cdot 6432 = 13124.
    \end{equation}
    Now, fix $j \in \{2, \ldots, r\}$ and for notation write
    \begin{equation}
        a = z_{j-1} \in \r^{2n_{j-1}}, \quad w = z_j \in \r^{2n_j}, \quad A = \|a\|^2, \quad B = \|w\|^2.
    \end{equation}
    If either $A$ or $B$ equals 0 then $F_j(\cdot) = 0$ on a neighborhood of $z$ and all derivatives vanish. Hence, we assume $A, B  >0$. Defining
    \begin{equation}
        C(a) := c_{n_{j-1}}(a) g(s_{j-1}(a)), \quad D(w) := c_{n_j}(w) (1+s_j(w)), \quad h(a, w) := H(A, B),
    \end{equation}
from Lemma \ref{lemma:scalarbounds} and (\ref{eqn:proofofintermediatelemmaboundsonc_m}) we have the bounds
\begin{align}
    0 \leq C \leq 1, \quad \|\nabla C\| \leq \frac{32 + 4\|g'\|_\infty}{\sqrt{A}} \leq \frac{32K_g}{\sqrt{A}}, \\ \|\nabla^2 C\|_{\operatorname{op}} \leq \frac{6432 + 20\|g'\|_\infty + 16 \|g''\|_\infty + 256 \|g'\|_\infty}{A}  \leq \frac{6432 K_g}{A}.
\end{align}
Similarly, 
\begin{align}
    0 \leq D \leq 2, \quad \|\nabla D\|\leq \frac{2\cdot 32 + 4}{\sqrt{B}} = \frac{68}{\sqrt{B}},\\
    \|\nabla^2 D\|_{\operatorname{op}} \leq \frac{2 \cdot 6432 + 20 + 2 \cdot 32 \cdot 4}{B} = \frac{13140}{B}.
\end{align}
Now, we know that $F_j = ChD$ and so by the product rule
    \begin{align}
    \label{eqn:proofofintermediatelemmaproductrule1}
        \|\nabla^2_a F_j\|_{\operatorname{op}} &\leq 2(h\|\nabla^2 C\|_{\operatorname{op}} + \|\nabla^2_ah\|_{\operatorname{op}} + 2\|\nabla C\| \|\nabla_a h\|) \\
        \|\nabla^2_w F_j\|_{\operatorname{op}} &\leq h\|\nabla^2 D\|_{\operatorname{op}} + 2\|\nabla^2_wh\|_{\operatorname{op}} + 2\|\nabla D\| \|\nabla_w h\| 
        \\ 
         \|\nabla_a\nabla_w F_j\|_{\operatorname{op}} &\leq 2\|\nabla C\| \|\nabla_wh\| + h\|\nabla C\| \|\nabla D\| + 2\|\nabla_a \nabla_w h\| + \|\nabla_a h\| \|\nabla D\|,
         \label{eqn:proofofintermediatelemmaproductrule3}
    \end{align}
    where we used that $D \leq 2$ and $C \leq 1$. 
It remains to bound the derivatives of $h$. Writing $u = \frac{A}{kB}$ and $H_A, H_B$ for partial derivatives of $H(A, B) = B\phi(u)$, we have that
\begin{align}
    H_A = \frac{\phi'(u)}{k}, \quad H_B = \phi(u) - u\phi'(u), \\ H_{AA} = \frac{\phi''(u)}{k^2B}, \quad H_{AB} = -\frac{u\phi''(u)}{kB}, \quad H_{BB} = \frac{u^2 \phi''(u)}{B}.
\end{align}
Along with the bounds on $\phi$ from Lemma \ref{lemma:scalarbounds}, we use the estimates
\begin{align}
\label{eqn:proofofinteremediatelemmaestimates1}
    0 \leq H(A, B) \leq \min\{4A/k, B\}, 
    \\
   \phi'(u) = 0 \text{ for $u \geq 1/2$}, \quad \phi(u) - u\phi'(u)= 0 \text{ for $u \leq 1/4$}
   \label{eqn:proofofinteremediatelemmaestimates2}
\end{align}
which follow from the definitions.
    The chain rule yields
    \begin{align}
        \|\nabla_a h\| &= 2\sqrt{A}|H_A| \leq \frac{50\sqrt{A}}{k}, \\
        \|\nabla_w h\| &= 2\sqrt{B}|H_B| \leq 18 \sqrt{B},\\
        \|\nabla^2_a h\|_{\operatorname{op}} &\leq 2|H_A| + 4A|H_{AA}| \leq \frac{50 + 2 \cdot 4864}{k} = \frac{9778}{k},
        \\\|\nabla^2_w h\|_{\operatorname{op}} &\leq 2|H_B| + 4B|H_{BB}| \leq 18 + 4864 = 4882,
        \\ \|\nabla_a \nabla_w h\|_{\operatorname{op}} &= 4\sqrt{AB}|H_{AB}| = \frac{4u^{3/2}|\phi''(u)|}{\sqrt{k}} \leq \frac{9728}{\sqrt{k}},
    \end{align}
    where the last bound uses $u \leq 1/2$ whenever $\phi''(u) \neq 0$. The estimates (\ref{eqn:proofofinteremediatelemmaestimates1}) and (\ref{eqn:proofofinteremediatelemmaestimates2}) also imply
    \begin{align}
        \frac{|h|}{\sqrt{AB}} &\leq \frac{\sqrt{(4A/k)B}}{\sqrt{AB}} = \frac{2}{\sqrt{k}},
        \\ \frac{\|\nabla_a h\|}{\sqrt{B}} &= \frac{2\sqrt{u}|\phi'(u)|}{\sqrt{k}} \leq \frac{25\sqrt{2}}{\sqrt{k}} \leq \frac{36}{\sqrt{k}},\\
        \frac{\|\nabla_w h\|}{\sqrt{A}} &= \frac{2|\phi(u) - u\phi'(u)|}{\sqrt{ku}} \leq \frac{36}{\sqrt{k}}. 
    \end{align}
    Applying the product rule bounds (\ref{eqn:proofofintermediatelemmaproductrule1})-(\ref{eqn:proofofintermediatelemmaproductrule3}), we find
    \begin{align}
        \|\nabla^2_a F_j\|_{\operatorname{op}} &\leq 2 \left(\frac{4A}{k} \frac{6432K_g}{A} + \frac{9778}{k} + 2 \frac{32K_g}{\sqrt{A}} \frac{50\sqrt{A}}{k}\right) \\&\leq \frac{2(4\cdot 6432 + 9778 + 2 \cdot 32 \cdot 50)K_g}{k} \\&= \frac{77412K_g}{k}
    \end{align}
    since $h \leq 4A/k$. Similarly, with $h \leq B$ we get 
    \begin{equation}
        \|\nabla^2_w F_j\|_{\operatorname{op}} \leq B \frac{13140}{B} + 2 \cdot 4882 + 2\frac{68}{\sqrt{B}} 18\sqrt{B} = 25352.
    \end{equation}
    For the mixed block, the weighted estimates give
    \begin{align}
        \|\nabla_a \nabla_w F_j\|_{\operatorname{op}} &\leq 2\frac{32K_g}{\sqrt{A}} \frac{36\sqrt{A}}{\sqrt{k}} + \frac{2\sqrt{AB}}{\sqrt{k}} \frac{32K_g}{\sqrt{A}}\frac{68}{\sqrt{B}} + 2\frac{9728}{\sqrt{k}} + \frac{36\sqrt{B}}{\sqrt{k}} \frac{68}{\sqrt{B}} \\
        &\leq \frac{(2\cdot 32 \cdot 36 + 2 \cdot 32 \cdot 68 + 2 \cdot 9728 + 36 \cdot 68)K_g}{\sqrt{k}} \\
        &= \frac{28560K_g}{\sqrt{k}}.
    \end{align}
    \textbf{(iii)}. We know that $s_j \in [-1, 1]$ and so $1 + s_j \geq 0$. Since $0 \leq \chi \leq 1$ we know
    \begin{equation}
        0 \leq F_1(z) \leq \|z_1\|^2 (1 + s_1).
    \end{equation}
    For every $j \in \{2, \ldots, r\}$ we know that all the factors defining $F_j$ are nonnegative. Using that $0 \leq g \leq 1$ and $H(\|z_{j-1}\|^2, \|z_j\|^2) \leq \|z_j\|^2$, we find
\begin{equation}
        0 \leq F_j(z) \leq \|z_j\|^2 (1 + s_j) \quad (\forall j \in [r]).
    \end{equation}
    \textbf{(iv)}. If $\chi\left(\frac{\|z_1\|^2}{2n_1}\right) = 1$ then $F_1(z) = \|z_1\|^2 (1+s_1)$ by definition. Fix $j \in \{2, \ldots, r\}$ and suppose that $\chi\left(\frac{\|z_{j-1}\|^2}{2n_{j-1}}\right)  = \chi\left(\frac{\|z_j\|^2}{2n_j}\right) = 1$, that $s_{j-1} \geq \rho / 2$, and that $\|z_{j-1}\|^2 \geq k\|z_j\|^2 / 2$. Since $g(t) = 1$ for $t \geq \rho / 2$ and $H(A, B) = B$ if $A \geq kB/2$, the claim follows from the definition of $F_j$.
\end{proof}

We will also need the following standard chi-square estimate:
\begin{lemma}[Equation (2.19) of \cite{Wainwright2019}]
    \label{lemma:chisquare}
    Let $m \in \n$ and $\xi \sim \mathcal{N}(0, \1_m)$ so that $\|\xi\|^2 \sim \chi_m^2$. Then, for all $u \in (0, 1)$ we have 
    \begin{equation}
        \mathbb{P}\left(\left|\frac{\|\xi\|^2}{m} - 1\right| > u\right) \leq 2\exp\left(-\frac{mu^2}{8}\right) .
    \end{equation}
\end{lemma}

We may now turn to the proof of Lemma \ref{lemma:theconstruction} using the above intermediate results. In this proof, we again denote $F_j := F_j^U$, define $s_j: \r^{2n_j} \to \r$ as in (\ref{eqn:proofofdaappendixdefinitionofs_j}), and abbreviate $s_j(z_j)$ by $s_j$ when its argument can be inferred from context.

\begin{proof}[Proof of Lemma \ref{lemma:theconstruction}]
Properties (i)–(iii) for $\tau=0$ follow immediately from $\widetilde{V}_{U,0}(z)=\frac{1}{2}\|z\|^2$. We therefore assume $\tau \in [r]$.

Property (i) follows from Lemma \ref{lemma:intermediatelemma}(i). For (iii), since the radial cutoffs $\chi$ ensure that each $F_j$ is identically zero in a neighborhood of $z=0$ we know $\widetilde{V}_{U, \tau}(\cdot) = \frac{1}{2}\|\cdot\|^2$ on a neighborhood of the origin for every $\tau \in [r]$.
% By the strong convexity of $\widetilde{V}_{U, \tau}$ from (ii) we know this critical point is unique. 
For (iv), fix $j \in \{2, \ldots, r\}$ and suppose that $z \in \r^d$ is s.t. $s_{j-1} \leq \rho / 4$. If $z_{j-1}, z_j \neq 0$ then $g(s_{j-1}) = 0$ by definition and $g^{(m)}(s_{j-1}) = 0$ for all $m \in \n$ since $g$ is constant on $(-\infty, \rho/4]$. On the other hand, if $z_{j-1} = 0$ or $z_j = 0$ then the radial $\chi$ cutoff is 0 on a neighborhood of $z$. In either case, $F_j$ and all its derivatives vanish at $z$. It remains to prove (ii) and (v).

\textbf{(ii)}. Fix $\tau \in [r]$ and write
\begin{equation}
    A_\tau(z) := \nabla^2 \sum_{j=1}^\tau F_j(z).
\end{equation}
This is a symmetric block-tridiagonal matrix, since each $F_j(z)$ only depends on $z_{j-1}, z_j$ for $j \geq 2$ and only on $z_1$ for $j=1$. By Lemma \ref{lemma:intermediatelemma}(ii), its diagonal blocks have operator norm at most $25352 + \frac{77412K_g}{k}$ and its off-diagonal blocks have operator norm at most $\frac{28560K_g}{\sqrt{k}}$. In particular,
\begin{equation}
    \left|\la v, A_\tau(z) v\ra\right| \leq \left(25352 + \frac{77412K_g}{k}\right) \sum_{j=1}^r \|v_j\|^2 + 2 \frac{28560K_g}{\sqrt{k}} \sum_{j=2}^{r} \|v_j\| \|v_{j-1}\| \quad (\forall v \in \r^d),
\end{equation}
where the factor of two occurs since each off-diagonal block occurs together with its transpose. Using $2\|v_{j-1}\|\|v_j\| \leq \|v_{j-1}\|^2 + \|v_j\|^2$ implies 
\begin{equation}
       \left|\la v, A_\tau(z) v\ra\right| \leq \left(25352 + \frac{77412K_g}{k}\right) \|v\|^2 + 2\frac{28560K_g}{\sqrt{k}} \|v\|^2,
\end{equation}
and since $A_\tau(z)$ is symmetric this implies
\begin{equation}
    \|A_\tau(z)\|_{\operatorname{op}} \leq 25352 + \frac{77412K_g}{k} + \frac{57120K_g}{\sqrt{k}} \leq 25352 + \frac{134532K_g}{\sqrt{k}}.
\end{equation}
Since $K_g \leq 1 + \frac{32}{\rho} + \frac{6400}{\rho^2}$ by Lemma \ref{lemma:scalarbounds} and we choose $k, \rho, \epsilon$ to satisfy (\ref{eqn:conditionsonconstants}), we find
\begin{equation}
\|\nabla^2 \widetilde{V}_{U, \tau}(z) - \1_d\|_{\operatorname{op}} = \epsilon\|A_\tau(z)\|_{\operatorname{op}} \leq \delta.
\end{equation}
\textbf{(v)}. Writing $z_r = (x_r, y_r)$ we find that
\begin{equation}
    S_U = \left\{z \in \r^d:\; z_r \neq 0,\; 2 \la y_r, U_rx_r\ra \geq \frac{\rho}{2}\|z_r\|^2\right\},
\end{equation}
and so $S_U$ is the intersection of an open set and a closed set (and is therefore Borel measurable). Define 
\begin{equation}
    R_j(z_j) := \|z_j\|^2 (1+s_j(z_j))
\end{equation}
and introduce the reference potential
\begin{equation}
    G_U(z) := \sum_{j=1}^r \left[\frac{1}{2}\|z_j\|^2 - \epsilon R_j(z_j)\right].
\end{equation}
Since $0 \leq R_j(z_j) \leq 2\|z_j\|^2$ and $\epsilon < 1/4$, we know $e^{-G_U}$ is integrable.
By Lemma \ref{lemma:intermediatelemma}(iii) we know that $V_U(z) \geq G_U(z)$ for every $z \in \r^d$. Let us now define the set
\begin{equation}
    E_U := \left\{z \in \r^d:\; \chi\left(\frac{\|z_j\|^2}{2n_j}\right) = 1,\; s_j \geq \frac{\rho}{2} \; \forall j \in[r] \text{ and  } \|z_{j-1}\|^2 \geq \frac{k}{2}\|z_j\|^2 \;\forall j \in \{2, \ldots, r\} \right\}.
\end{equation}
By Lemma \ref{lemma:intermediatelemma}(iv) we know that $V_U(z) = G_U(z)$ for every $z \in E_U$. Also, by definition we have $E_U \subset S_U$. We claim that if $\gamma_U \in \p(\r^d)$ is the probability measure with density $\gamma_U \propto e^{-G_U}$ then
\begin{equation}
\label{eqn:proofgammameasurebound}
    \gamma_U(E_U) \geq 1 - 8r \exp\left(-\frac{\rho^2 b}{128}\right).
\end{equation}
This would imply the claim (v), since if $\mathcal{Z}_V := \int_{\r^d} e^{-V_U(z)} dz$ and $\mathcal{Z}_G := \int_{\r^d} e^{-G_U(z)} dz$ then $\mathcal{Z}_V \leq \mathcal{Z}_G$ since $V_U \geq G_U$ and so 
\begin{align}
    \int_{S_U} e^{-V_U(z)}dz &\geq \int_{E_U} e^{-V_U(z)} dz = \int_{E_U}e^{-G_U(z)}dz = \mathcal{Z}_G \cdot \gamma_U(E_U) \\&\geq  \mathcal{Z}_V \cdot \gamma_U(E_U) \geq  \mathcal{Z}_V \cdot \left( 1 - 8r \exp\left(-\frac{\rho^2 b}{128}\right)\right).
\end{align}
To show (\ref{eqn:proofgammameasurebound}), for every $j \in [r]$ let us define the coordinates
\begin{equation}
    p_j := \frac{x_j + U_j^\top y_j}{\sqrt{2}}, \quad q_j := \frac{x_j - U_j^\top y_j}{\sqrt{2}}.
\end{equation}
This is an orthogonal change of coordinates, which means
\begin{equation}
    \|z_j\|^2 = \|p_j\|^2 + \|q_j\|^2, \quad 2 \la y_j, U_j x_j\ra = \|p_j\|^2 - \|q_j\|^2
\end{equation}
and so
\begin{equation}
    G_U(z) = \sum_{j=1}^r \left[\frac{1-4\epsilon}{2}\|p_j\|^2 + \frac{1}{2}\|q_j\|^2\right].
\end{equation}
Under $\gamma_U$ the vectors $p_j, q_j$ are mutually independent centered Gaussians with covariances $\frac{1}{1-4\epsilon}\1_{n_j}$ and $\1_{n_j}$, respectively. In particular, $X_j := (1-4\epsilon) \|p_j\|^2$ and $Y_j := \|q_j\|^2$ are independent $\chi^2_{n_j}$ random variables. Moreover, on the event
\begin{equation}
    \left|\frac{X_j}{n_j} - 1\right| \leq \frac{\rho}{4}, \quad \left|\frac{Y_j}{n_j} - 1\right| \leq \frac{\rho}{4}
\end{equation}
we have $\|p_j\|^2 \geq \frac{n_j(1-\rho/4)}{1-4\epsilon}$ and $\|q_j\|^2 \leq n_j(1+\rho/4)$; since $(a-b)/(a+b)$ is increasing in $a$ and decreasing in $b$, this implies
\begin{equation}
    s_j = \frac{\|p_j\|^2 - \|q_j\|^2}{\|p_j\|^2 + \|q_j\|^2} \geq \frac{(1-4\epsilon)^{-1} (1-\rho/4) - (1+\rho/4)}{(1-4\epsilon)^{-1} (1-\rho/4) + (1+\rho/4)} = \frac{\rho - \rho/4}{1 - \rho^2/4} > \frac{\rho}{2},
\end{equation}
where the equality follows from $\rho = \frac{2\epsilon}{1-2\epsilon} = \frac{(1-4\epsilon)^{-1} - 1}{(1-4\epsilon)^{-1} + 1}$.
Applying Lemma \ref{lemma:chisquare} with $m = n_j$ and $u = \rho/4$ therefore implies
\begin{equation}
\label{eqn:proofdaunionbound1}
    \gamma_U\left(s_j < \frac{\rho}{2}\right) \leq \mathbb{P}\left(\left|\frac{X_j}{n_j} - 1\right| > \frac{\rho}{4}\right) + \mathbb{P}\left(\left|\frac{Y_j}{n_j} - 1\right| > \frac{\rho}{4}\right) \leq 4\exp\left(-\frac{\rho^2 n_j}{128}\right).
\end{equation}
Next, by applying Lemma \ref{lemma:chisquare} to $X_j,Y_j$ with $u = 1/4$ yields
\begin{equation}
    \label{eqn:proofdaunionbound2}\gamma_U\left(\mathcal{E}_j\right) \geq 1 - 4 \exp\left(-\frac{n_j}{128}\right), \quad \mathcal{E}_j := \left\{z \in \r^d:\; \frac{3}{4} \mu n_j \leq \|z_j\|^2 \leq \frac{5}{4} \mu n_j\right\}
\end{equation}
with $\mu = (1-4\epsilon)^{-1} + 1$. Since $\mu \geq 2$, on $\mathcal{E}_j$ we know $\frac{\|z_j\|^2}{2n_j} \geq \frac{3}{4}$ and so $\chi\left(\frac{\|z_j\|^2}{2n_j}\right) = 1$. Furthermore, if $j \in \{2, \ldots, r\}$ and $\mathcal{E}_{j-1} \cap \mathcal{E}_j$ holds then since $n_{j-1} = kn_j$ we find
\begin{equation}
    \frac{\|z_{j-1}\|^2}{k\|z_j\|^2} = \frac{\|z_{j-1}\|^2 / n_{j-1}}{\|z_{j}\|^2 / n_j} \geq \frac{3/4}{5/4} = \frac{3}{5} > \frac{1}{2}.
\end{equation}
In particular, we have shown that
\begin{equation}
\bigcap_{j=1}^r \left(\mathcal{E}_j \cap \left\{z \in \r^d:\; s_j \geq \frac{\rho}{2}\right\}\right) \subset E_U.
\end{equation}
Therefore, combining $n_j \geq b$ and (\ref{eqn:proofdaunionbound1}) and (\ref{eqn:proofdaunionbound2}) with a union bound gives 
\begin{align}
    \gamma_U(E_U^C) \leq \sum_{j=1}^r\left( 4 \exp\left(-\frac{b}{128}\right) + 4 \exp\left(-\frac{\rho^2 b}{128}\right)\right) \leq 8r \exp\left(-\frac{\rho^2 b}{128}\right).
\end{align}
\end{proof}

\subsection{Proof of Lemma \ref{lemma:randomizedoraclebound}: zero-chain lower bound via coupling}
\label{subsec:proofoflemmarandomizedoraclebound}

We will make use of the following standard spherical cap estimate due originally to Keith Ball, see \cite{Tkocz2012} for a clear self-contained proof:
\begin{lemma}
\label{lemma:concentration}
    Fix an integer $m \geq 2$ and $x, y \in \r^m$, and let $W \sim \operatorname{Haar}(O(m))$. Define
    \begin{equation}
        s_W(x, y) := \begin{cases}
            {2 \la y, W x\ra}/(\|x\|^2 + \|y\|^2), & (x, y) \neq 0, \\ 0, & (x, y)= 0,
        \end{cases}
    \end{equation}
    Then, for all $u \in (0, 1)$ we have
    \begin{equation}
        \mathbb{P}\left(s_W(x, y) \geq u\right) \leq  \exp\left(-\frac{mu^2}{2}\right).
    \end{equation}
\end{lemma}
\begin{proof}
    The claim is trivial if either $x = 0$ or $y = 0$, so we assume they are both nonzero. If we denote $\hat{x} := x / \|x\|$, $\hat{y} := y / \|y\|$, and $\alpha := 2 \|x\|\|y\| / (\|x\|^2 + \|y\|^2)$, then
    \begin{equation}
        s_W(x, y) = \alpha \la \hat{y}, W\hat{x}\ra
    \end{equation}
    and since $\alpha \in (0, 1]$ we know
    \begin{equation}
        \mathbb{P}(s_W(x, y) \geq u) \leq \mathbb{P}(\la \hat{y}, W\hat{x}\ra \geq u).
    \end{equation}
    Since $W$ is Haar distributed, we know $W \hat{x}$ is distributed uniformly on the sphere, and since $\hat{y}$ is a fixed unit vector we know by rotational invariance that
    \begin{equation}
         \mathbb{P}(\la \hat{y}, W\hat{x}\ra \geq u) = \mathbb{P}_{X \sim \operatorname{Unif}(\mathbb{S}^{m-1})} (X_1 \geq u).
    \end{equation}
    The claim follows from \cite{Tkocz2012}.
\end{proof}

We can now proceed with the proof of the query lower bound in Lemma \ref{lemma:randomizedoraclebound}. In this proof, we again denote $F_j := F_j^U$ and define $s_j: \r^{2n_j} \to \r$ as in (\ref{eqn:proofofdaappendixdefinitionofs_j}).

\begin{proof}[Proof of Lemma \ref{lemma:randomizedoraclebound}]
Draw the instance $U \sim \otimes_{j=1}^r \operatorname{Haar}(O(n_j))$ and algorithm seed $\omega \sim \mathbb{P}_{\mathcal{A}}$ independently.
Define the potential $V_U\in C^\infty(\r^d)$ as detailed in Lemma \ref{lemma:theconstruction} along with the truncated potentials
\begin{equation}
     \widetilde{V}_{U, \tau}(z) := \frac{1}{2}\sum_{j=1}^r \|z_j\|^2 - \epsilon \sum_{j=1}^\tau F_j(z) \quad (\forall \tau \in [r-1])
 \end{equation}
 as in (\ref{eqn:defofV_U}). For convenience, we set $\widetilde{V}_{U,0}(z) := \frac{1}{2}\|z\|^2$. 
We will compare the sampler's rollout on the true potential $V_U$ with its rollouts on the alternate potentials $\widetilde{V}_{U,\tau}$. Importantly, we couple these executions to use the same internal randomness $\omega$. Probabilities, expectations, and almost sure statements are taken over $U$ and the common seed $\omega$ throughout this proof.

Using $t$ to index rounds and $\tau \in \{0, \ldots, r-1\}$ to index truncations, let us denote the queries, batch sizes, histories, and output of the execution on the true potential $V_U$ by
\begin{equation}
    X_i^t, \quad \mathcal{Q}_t, \quad \mathcal{H}_t, \quad Z,
\end{equation}
and the corresponding quantities in the execution on the truncated potential $\widetilde{V}_{U,\tau}$ by
\begin{equation}
\label{eqn:proofoflemma2taurandomvariables}
    \widetilde{X}_i^{t,\tau}, \quad \widetilde{\mathcal{Q}}_t^{(\tau)}, \quad \widetilde{\mathcal{H}}_t^\tau, \quad \widetilde{Z}^\tau.
\end{equation}
By assumption, these executions all stop almost surely. After an execution stops, we take its subsequent batches to be empty (with batch size 0 and $[0]$ interpreted as the empty set). Hence, we may define the total number of queries by
\begin{equation}
    \mathcal{Q} := \mathcal{Q}(V_U) = \sum_{t \in \n} \mathcal{Q}_t, \quad \widetilde{\mathcal{Q}}^{(\tau)} := \mathcal{Q}(\widetilde{V}_{U,\tau}) = \sum_{t \in \n} \widetilde{\mathcal{Q}}_t^{(\tau)}.
\end{equation}
Note that the full and truncated potentials lie in $\mathcal{V}_{d,\kappa} \cap C^\infty(\r^d)$ with $\kappa = \frac{1+\delta}{1-\delta}$ by Lemma \ref{lemma:theconstruction}(i)-(iii). 
The expected query assumption then yields
 \begin{equation}
 \label{eqn:proofoflemma2expectedqueryupperbound}
   \mathbb{E}[\mathcal{Q}] \leq Q, \quad  \mathbb{E}[\widetilde{\mathcal{Q}}^{(\tau)}] \leq Q \quad (\forall \tau \in \{0, \ldots, r-1\}).
 \end{equation}
 Let us define the $\sigma$-algebras
 \begin{equation}
     \mathcal{F}_0 := \sigma(\omega), \quad \mathcal{F}_\tau := \sigma(\omega, U_1, \ldots, U_\tau).
 \end{equation}
 Since $\widetilde{V}_{U, \tau}$ only depends on the first $\tau$ random rotations and the sampler's execution depends measurably on $\omega$ and $\widetilde{V}_{U,\tau}$, the random variables (\ref{eqn:proofoflemma2taurandomvariables}) are all $\mathcal{F}_\tau$-measurable. For each block $j \in [r]$, we denote the coordinates $X^t_{i,j}, \widetilde{X}^{t,\tau}_{i,j} \in \r^{2n_j}$ and define the event
 \begin{equation}
     \mathcal{C}_j := \left\{\exists t \in \n, \; i \in [\widetilde{\mathcal{Q}}_t^{(j-1)}]:\; s_j(\widetilde{X}_{i,j}^{t,j-1}) > \rho / 4\right\}.
 \end{equation}
 Conditional on $\mathcal{F}_{j-1}$, the rollout's query list $\left((\widetilde{X}_{i}^{t,j-1})_{i \in [\widetilde{\mathcal{Q}}_t^{(j-1)}]}\right)_{t \in \n}$ is fixed while $U_j$ remains Haar distributed. By Lemma \ref{lemma:concentration} and a union bound over all the queries, we find that
 \begin{equation}
     \mathbb{P}(\mathcal{C}_j \mid \mathcal{F}_{j-1}) \leq \sum_{t \in \n} \widetilde{\mathcal{Q}}_t^{(j-1)} \exp\left(-\frac{n_j \rho^2}{32}\right) = \widetilde{\mathcal{Q}}^{(j-1)} \exp\left(-\frac{n_j \rho^2}{32}\right).
 \end{equation}
 Taking expectations and using (\ref{eqn:proofoflemma2expectedqueryupperbound}) implies that
 \begin{equation}
     \mathbb{P}(\mathcal{C}_j) \leq Q \exp\left(-\frac{n_j \rho^2}{32}\right) \quad (\forall j \in [r]).
 \end{equation}
 Define for each $t \in [r-1]$ the bad event
 \begin{equation}
    \mathcal{B}_t := \left\{\exists i \in [\mathcal{Q}_t], \; j \in \{t+1, \ldots, r\}:\; s_{j-1}(X_{i,j-1}^t) > \rho / 4\right\}.
 \end{equation}
 We claim that \begin{equation}
 \label{eqn:proofoflemma2inclusiontoprove}
      \bigcup_{t=1}^{r-1}\mathcal{B}_t\subset \bigcup_{j=1}^{r-1} \mathcal{C}_j.
 \end{equation}
Indeed, suppose that the event $ \bigcup_{t=1}^{r-1}\mathcal{B}_t$ occurs, let $t_\star$ be the first occurrence, and let $j_\star \in \{t_\star+1, \ldots, r\}$ be a witnessing index. At every earlier round $t < t_\star$ we know that 
 \begin{equation}
     D^\alpha F_{j}(X_i^t) = 0 \quad (\forall  i \in [\mathcal{Q}_t], \; j \in \{t+1, \ldots, r\},\; \alpha \in \n_0^d),
 \end{equation}
 where we apply Lemma \ref{lemma:theconstruction}(iv) since $s_{j-1}(X_{i,j-1}^t) \leq \rho / 4$ for all such $t,i,j$. 
 Since we have
 \begin{equation}
     V_U - \widetilde{V}_{U, j_\star - 2} = -\epsilon \sum_{j = j_\star - 1}^r F_j
 \end{equation}
 and we know $j_\star - 1 \geq t_\star > t$,
 we get the jet equality \begin{equation}\left(D^\alpha V_U(X_i^t)\right)_{\alpha \in \n_0^d} = \left( D^\alpha \widetilde{V}_{U, j_\star - 2}(X_i^t)\right)_{\alpha \in \n_0^d} \end{equation} for every query $i \in [\mathcal{Q}_t]$ of every round $t < t_\star$. By induction, this means that the rollout on the true potential $V_U$ and the rollout on the truncated potential $\widetilde{V}_{U, j_\star - 2}$ have identical queries, responses, and stopping decisions through round $t_\star - 1$. They consequently choose the same batch in round $t_\star$, i.e. $\mathcal{Q}_{t_\star} = \widetilde{\mathcal{Q}}_{t_\star}^{(j_\star - 2)}$ and \begin{equation}X_{i}^{t_\star} = \widetilde{X}_i^{t_\star, j_\star - 2} \quad (\forall i \in [\mathcal{Q}_{t_\star}]).\end{equation}
 If we let $i_\star \in [\mathcal{Q}_{t_\star}]$ be a witnessing query index for the event $\mathcal{B}_{t_\star}$, then
 \begin{equation}
     s_{j_\star - 1}\left(\widetilde{X}_{i_\star, j_\star - 1}^{t_\star, j_\star - 2}\right) = s_{j_\star - 1}\left({X}_{i_\star, j_\star - 1}^{t_\star}\right) > \rho / 4.
 \end{equation}
This means that the event $\mathcal{C}_{j_\star - 1}$ occurs, and so the inclusion (\ref{eqn:proofoflemma2inclusiontoprove}) is proved. It follows that
\begin{equation}
    \mathbb{P}\left(\bigcup_{t=1}^{r-1} \mathcal{B}_t\right) \leq \sum_{j=1}^{r-1} \mathbb{P}(\mathcal{C}_j) \leq Q \sum_{j=1}^{r-1} \exp\left(-\frac{n_j\rho^2}{32}\right) \leq Qr \exp\left(-\frac{\rho^2 b}{32}\right),
\end{equation}
where we used that $n_j \geq b$. 

 Now, suppose that  $\bigcap_{t=1}^{r-1} \mathcal{B}_t^c$ occurs. This means that
 \begin{equation}
     s_{r-1}(X_{i,r-1}^t) \leq \rho / 4 \quad (\forall t \in [r-1],\; i \in [\mathcal{Q}_t]),
 \end{equation}
 which by Lemma \ref{lemma:theconstruction}(iv) implies that 
\begin{equation}\left(D^\alpha V_U(X_i^t)\right)_{\alpha \in \n_0^d} = \left( D^\alpha \widetilde{V}_{U, r-1}(X_i^t)\right)_{\alpha \in \n_0^d} \quad (\forall t \in [r-1], i \in [\mathcal{Q}_t]). \end{equation} 
If we additionally know that the true rollout stops before round $r$ (i.e. $\mathcal{T}(V_U) < r$), then induction shows that the true rollout and the truncated $\widetilde{V}_{U,r-1}$ rollout have identical responses, histories, and stopping decisions. Since they use the same seed their outputs also agree, and so
\begin{equation}
\label{eqn:proofoflemma2couplingbound}
    \mathbb{P}(Z \neq \widetilde{Z}^{r-1}) \leq \mathbb{P}(\mathcal{T}(V_U) \geq r) + Qr\exp\left(-\frac{\rho^2b}{32}\right).
\end{equation}
Lastly, since $\widetilde{Z}^{r-1}$ is $\mathcal{F}_{r-1}$-measurable it is independent of $U_r$. By Lemma \ref{lemma:concentration}, we find
\begin{equation}
    \mathbb{P}\left(s_r\left(\widetilde{Z}^{r-1}_r\right) \geq \rho / 2 \mid \mathcal{F}_{r-1}\right) \leq \exp \left(-\frac{\rho^2 b}{8}\right).
\end{equation}
Taking expectations and using the definition of the terminal set $S_U \subset \r^d$, we find
\begin{equation}
    \label{eqn:proofoflemma2terminalbound}\mathbb{P}\left(\widetilde{Z}^{r-1} \in S_U\right) \leq \exp\left(-\frac{\rho^2b}{8}\right).
\end{equation}
Combining (\ref{eqn:proofoflemma2couplingbound}) and (\ref{eqn:proofoflemma2terminalbound}) yields the claim since
\begin{align}
    \mathbb{E}_U\left[\mathcal{A}^{V_U} (S_U)\right]& = \mathbb{P}(Z \in S_U) \\&\leq \mathbb{P}(Z \neq \widetilde{Z}^{r-1}) + \mathbb{P}(\widetilde{Z}^{r-1} \in S_U)\\&
    \leq \mathbb{P}(\mathcal{T}(V_U) \geq r) + Qr \exp\left(-\frac{\rho^2 b}{32}\right) + \exp\left(-\frac{\rho^2b}{8}\right).
\end{align}
\end{proof}

\subsection{Proof of Proposition \ref{prop:parameterizedlowerbound}: parameterized log-concave lower bound}
\label{subsec:proofofparameterizedlowerboundprop}
\begin{proof}[Proof of Proposition \ref{prop:parameterizedlowerbound}]
Equip $C^\infty(\mathbb R^d)$ with its usual topology of uniform convergence of all derivatives on compact sets. The map $\Phi: \prod_{j=1}^r O(n_j) \to C^\infty(\r^d)$ sending $U \mapsto V_U$ is continuous: on every compact set, $V_U$ and each of its derivatives depend continuously on the matrix entries of $U$, and the radial $\chi$ cutoffs remove all singularities uniformly in $U$. In particular, $\Phi$ is Borel measurable and so
\begin{equation}
    \mathcal{D}_{b,r}^{(\delta)} := \Phi_\# \left(\otimes_{j=1}^r \operatorname{Haar}(O(n_j))\right)
\end{equation}
is a Borel probability distribution on $C^\infty(\r^d)$. By Lemma \ref{lemma:theconstruction}, for each $U \in \prod_{j=1}^r O(n_j)$ we know that for all $z \in \r^d$
\begin{equation}
    (1-\delta) \1_d \preceq \nabla^2 V_U(z) \preceq (1+\delta) \1_d,  \quad V_U(0) = 0, \quad \nabla V_U(0) = 0,
\end{equation}
which is part (i) of the claim.
Now, by definition of the total variation, for all $U$ we have
\begin{equation}
    \operatorname{TV}(\mathcal{A}^{V_U}, \pi_{V_U}) \geq \pi_{V_U}(S_U) - \mathcal{A}^{V_U}(S_U).
\end{equation}
Taking an expectation over $U$, Lemma \ref{lemma:theconstruction}(v) and Lemma \ref{lemma:randomizedoraclebound} yield 
\begin{align}
    \mathbb{E}_U \left[\operatorname{TV}(\mathcal{A}^{V_U}, \pi_{V_U})\right] &\geq 1 - \mathbb{P}_{U,\mathcal{A}}(\mathcal{T}(V_U) \geq r) - 8r \exp\left(-\frac{\rho^2 b}{128}\right) \\&\quad - Qr\exp\left(-\frac{\rho^2 b}{32}\right) - \exp\left(-\frac{\rho^2 b}{8}\right) \\
    &\geq 1 - \mathbb{P}_{U,\mathcal{A}}(\mathcal{T}(V_U) \geq r) -10 Qr \exp\left(-\frac{\rho^2 b}{128}\right),
\end{align}
where the second inequality follows from $Q \geq 1$, $r \geq 2$. Under the assumptions of (ii) we know $\mathcal{T}(V_U) \leq T < r$ almost surely, while under the assumptions of (iii) we know \begin{equation}
    \mathbb{P}_{U,\mathcal{A}}(\mathcal{T}(V_U) \geq r) \leq \frac{\mathbb{E}_{U,\mathcal{A}}[\mathcal{T}(V_U)]}{r} \leq \frac{T}{r}
\end{equation}
by Markov's inequality. The claims (ii) and (iii) follow.
\end{proof}

\section{Proofs for Gaussian upper and lower bounds}

\subsection{Proof of Lemma \ref{lemma:chebyshevapproximation}: polynomial approximation of $\lambda \mapsto 1/\lambda$}
\label{subsec:appendixchebyshevlemma}
\begin{proof}[Proof of Lemma \ref{lemma:chebyshevapproximation}]
For notation, write $\epsilon = \epsilon_{r,\kappa}$ and
\begin{equation}
    N := 2r-1, \quad \Lambda := \log\frac{\sqrt{\kappa }+1}{\sqrt{\kappa}-1}, \quad s(x) := \frac{\kappa + 1 - 2x}{\kappa - 1}, \quad \delta := s(0) = \cosh(\Lambda).
\end{equation}
    Let $C_N$ denote the degree-$N$ Chebyshev polynomial, which is characterized by $C_N(\cos \theta) = \cos(N \theta)$ and satisfies $C_N(\cosh t) = \cosh(Nt)$. 
    By definition of $\Lambda$, we know
    \begin{equation}
        \cosh \Lambda = \frac{1}{2}\left(\frac{\sqrt{\kappa} + 1}{\sqrt{\kappa} - 1} + \frac{\sqrt{\kappa} - 1}{\sqrt{\kappa} + 1}\right)  = \frac{\kappa + 1}{\kappa - 1} = \delta
    \end{equation}
    and hence
    \begin{equation}
        C_N(\delta) = \cosh(N \Lambda) = \epsilon^{-1}.
    \end{equation}
    For (i), define the function
    \begin{equation}
        q(x) := \frac{1-C_N(s(x)) / C_N(\delta)}{x}. 
    \end{equation}
    Since the numerator is a polynomial of degree $N$ with a root at 0, $q$ is a polynomial of degree $N-1=2r-2$. Since $s([1, \kappa]) = [-1, 1]$, for any $x \in [1, \kappa]$ we may write $s(x) = \cos(\theta)$ for some $\theta \in [0, \pi]$ to get
    \begin{equation}
        |C_N(s(x))| = |\cos(N\theta)| \leq 1. 
    \end{equation}
    This implies that
    \begin{equation}
       \sup_{x \in [1, \kappa]} |1 - xq(x)| = \sup_{x \in [1, \kappa]} \frac{|C_N(s(x))|}{C_N(\delta)} \leq \epsilon_{r,\kappa} . 
    \end{equation}
    We claim that $q$ is nonnegative on $\r$.  Indeed, since $N$ is odd we know 
    \begin{equation}
        \sup_{|t| \leq 1}C_N(t) \leq 1, \quad C_N'(t) > 0 \quad (\forall |t| \geq 1). 
    \end{equation}
    Since $\delta > 1$ and $C_N(\delta) > 1$, it follows that
    \begin{equation}
        C_N(t) < C_N(\delta) \iff t < \delta.
    \end{equation}
    Because $s$ is decreasing and $s(0) = \delta$, the numerator $1 - C_N(s(x)) / C_N(\delta)$ has the same sign as $x$. Thus, $q(x) > 0$ for all $x \neq 0$ and so $q$ is nonnegative.
    By \cite[Proposition 1.2.1]{marshall2008positive}, we may write $q(x) = u(x)^2 + v(x)^2$ for real polynomials $u, v$. Since the leading terms of the squares of two real polynomials cannot cancel, we must have $\operatorname{deg}u, \operatorname{deg}v \leq r-1$. 

    For (ii), the definition of the nodes $\lambda_0, \ldots, \lambda_N$ implies that
    \begin{equation}
        s(\lambda_j) = \cos(j \pi / N), \quad C_N(s(\lambda_j)) = (-1)^j.
    \end{equation}
    Suppose by way of contradiction that some real polynomial $S$ of degree at most $N-1$ satisfies
    \begin{equation}
        |1 - \lambda_j S(\lambda_j)| < \epsilon \quad (\forall j \in \{0, \ldots, N\}). 
    \end{equation}
    Use it to define the polynomial
    \begin{equation}
        D(x) := C_N(s(x)) - C_N(\delta) (1-xS(x)),
    \end{equation}
    which has degree at most $N$. At each node, we know
    \begin{equation}
        (-1)^j D(\lambda_j) = 1 - \frac{(-1)^j (1-\lambda_j S(\lambda_j))}{\epsilon} > 0,
    \end{equation}
    and so $D(\lambda_j)$ is nonzero with sign $(-1)^j$. In particular, along the nodes $\lambda_0 < \lambda_1 < \ldots < \lambda_N$ the polynomial $D$ must alternate signs, which means it must have $N$ roots in $(\lambda_0, \lambda_N)$. However, we also have $D(0) = 0$, and since $1 = \lambda_0 > 0$ this is an additional distinct root. Therefore, $D$ is a nonzero polynomial of degree at most $N$ with $N+1$ distinct roots. This is a contradiction, which proves (ii).
\end{proof}

\subsection{Proof of Lemma \ref{lemma:gaussiancorrectionlemma}: randomized corrections via rejection sampling}
\label{subsec:appendixradomizedcorrectionlemma}
\begin{proof}[Proof of Lemma \ref{lemma:gaussiancorrectionlemma}]
    Write $\Gamma$ to denote the standard Gaussian on $\r^{m \times n}$ so that $G \sim \Gamma$, let $\sigma$ denote the uniform probability distribution on $\mathbb{S}^{n-1}$, and write
    \begin{equation}
\nu := \mathcal{N}(0, (\1_m + B)^{-1}).
    \end{equation}
    The density of $\nu$ relative to $\mathcal{N}(0, \1_m)$ is given by
    \begin{equation}
        w(x) := \operatorname{det}(\1_m + B)^{1/2} e^{-x^\top B x/2} \quad (x \in \r^m).
    \end{equation}
    Let us define the normalizer
    \begin{equation}
        \mathcal{Z}(G) := \int_{\mathbb{S}^{n-1}} w(Gu) d\sigma(u).
    \end{equation}
    We draw $(G, a) \sim P$ from a probability measure $P$ on $\r^{m \times n} \times \mathbb{S}^{n-1}$ with joint law
\begin{equation}
    dP(G, a) := \frac{w(Ga)}{\mathcal{Z}(G)} d\Gamma(G) d\sigma(a).
\end{equation}
We now define a second measure $Q$ on $\r^{m \times n} \times \mathbb{S}^{n-1}$ with joint law
\begin{equation}
    dQ(G, a) := {w(Ga)}d\Gamma(G) d\sigma(a).
\end{equation}
We claim that $Q$ is a probability measure and that $Ga$ has law $\nu$ under $Q$. To see this, for any Borel $E \subset \r^m$ we have
\begin{align}
    Q(\{(G, a):\; Ga \in E\}) &= \int_{\mathbb{S}^{n-1}} \int_{\r^{m \times n}} \1_E(Ga)w(Ga) d\Gamma(G) d\sigma(a) \\
    &= \int_{\mathbb{S}^{n-1}} \mathbb{E}_{x \sim \mathcal{N}(0, \1_m)}[\1_E(x) w(x)] d\sigma(a) \\
    &= \mathbb{E}_{x \sim \mathcal{N}(0, \1_m)}[\1_E(x) w(x)]
    \\ &= \nu(E),
\end{align}
where we used that $Ga$ is a standard Gaussian for each fixed unit vector $a$. In particular,
\begin{equation}
\label{eqn:proofsubcriticalupperboundlemmaexpectationofZ}
    Q(\mathbb{R}^{m \times n} \times \mathbb{S}^{n-1}) = \mathbb{E}_{G \sim \Gamma}[\mathcal{Z}(G)] = 1.
\end{equation}
Now, by applying contractivity of $\operatorname{TV}$ to the map $(G, a) \mapsto Ga$ we get
\begin{align}
    \operatorname{TV}\left(\operatorname{Law}_{(G, a) \sim P}(Ga), \nu\right) &\leq \operatorname{TV}(P, Q) \\
    &= \frac{1}{2} \int_{\r^{m \times n}} \int_{\mathbb{S}^{n-1}} \left|\frac{w(Ga)}{\mathcal{Z}(G)} - w(Ga)\right|  d\sigma(a) d\Gamma(G)\\
    &= \frac{1}{2} \int_{\r^{m \times n}} \int_{\mathbb{S}^{n-1}} w(Ga)   d\sigma(a) \left|\frac{1}{\mathcal{Z}(G)} - 1\right| d\Gamma(G) \\
    &= \frac{1}{2}\int_{\r^{m \times n}} \mathcal{Z}(G) \left|\frac{1}{\mathcal{Z}(G)} - 1\right| d\Gamma(G) \\
    &= \frac{1}{2} \mathbb{E}_{G \sim \Gamma}|\mathcal{Z}(G) - 1|
\\ &\leq \frac{1}{2} \sqrt{\mathbb{E}_{G \sim \Gamma} [(\mathcal{Z}(G) - 1)^2]} \\
&= \frac{1}{2} \sqrt{\mathbb{E}_{G \sim \Gamma} [\mathcal{Z}(G)^2] - 1},
\label{eqn:proofsubcriticalupperboundlemmaexpectationofZboundinterms}
\end{align}
where we used (\ref{eqn:proofsubcriticalupperboundlemmaexpectationofZ}) to get last line. 
To compute the second moment, note that if $G \sim \Gamma$ and $a, a' \in \mathbb{S}^{n-1}$ are fixed with $\theta := \la a, a'\ra$ then \begin{equation}
    \operatorname{Law}((Ga, Ga')) = \mathcal{N}\left(0, \Sigma_\theta\right), \quad \Sigma_\theta := \begin{bmatrix}
        \1_m & \theta \1_m \\
\theta \1_m & \1_m
    \end{bmatrix}.
\end{equation}
% If $a, a' \in \mathbb{S}^{n-1}$ are fixed then $(Ga + Ga')/\sqrt{2}$ and $(Ga-Ga')/\sqrt{2}$ are independent with laws $\mathcal{N}(0, (1 + \la a, a'\ra)\1_m)$ and $\mathcal{N}(0, (1 - \la a, a'\ra)\1_m)$, respectively. 
Hence, if $\Theta$ is the law of $\la a, a'\ra$ with independent $a, a' \sim \sigma$ then we have
\begin{align}
    \mathbb{E}_{G \sim \Gamma}[\mathcal{Z}(G)^2] &=  \operatorname{det}(\1_m + B)\int_{[-1, 1]}\mathbb{E}_{(x, y) \sim \mathcal{N}(0, \Sigma_\theta)} [e^{-x^\top Bx/2} e^{-y^\top By/2}] d\Theta(\theta) \\
    &=\operatorname{det}(\1_m + B) \int_{[-1, 1]}\mathbb{E}_{z \sim \mathcal{N}(0, \Sigma_\theta)} \left[\exp\left(-z^\top \widetilde{B}z/2\right)\right] d\Theta(\theta) \\
    &= \operatorname{det}(\1_m + B)\int_{[-1, 1]} \mathbb{E}_{q \sim \mathcal{N}(0, \1_{2m})}\left[\exp\left(-q^\top \Sigma_\theta^{1/2} \widetilde{B}\Sigma_\theta^{1/2}q/2\right)\right]d \Theta(\theta) \\
    &= \operatorname{det}(\1_m + B) \int_{[-1, 1]} \operatorname{det}(\1_{2m} + \Sigma_\theta^{1/2}\widetilde{B}\Sigma_\theta^{1/2})^{-1/2}d\Theta(\theta),
\end{align}
where we denote $\widetilde{B} := \operatorname{diag}(B, B)$ and computed the Gaussian integral. Using that $\operatorname{det}(\1_k + XY) = \operatorname{det}(\1_k + YX)$, we know
\begin{equation}
     \operatorname{det}(\1_{2m} + \Sigma_\theta^{1/2}\widetilde{B}\Sigma_\theta^{1/2}) = \operatorname{det}\left(\begin{bmatrix}
         \1_m + B & \theta B \\ \theta B & \1_m + B
     \end{bmatrix}\right).
\end{equation}
Since changing coordinates from $(x, y)$ to $(\frac{x-y}{\sqrt{2}}, \frac{x+y}{\sqrt{2}})$ makes this matrix block diagonal, we find
\begin{align}
     \operatorname{det}(\1_{2m} + \Sigma_\theta^{1/2}\widetilde{B}\Sigma_\theta^{1/2}) &= \operatorname{det}(\1_m + (1+\theta)B)\operatorname{det}(\1_m + (1-\theta)B) \\&= \operatorname{det}\left((\1_m + B)^2 - \theta^2 B^2\right) \\
     &= \operatorname{det}((\1_m + B)^2) \operatorname{det}(\1_m - \theta^2 (B(\1_m + B)^{-1})^2).
\end{align}
Hence, if we denote $M := B(\1_m + B)^{-1}$ then
\begin{align}
     \mathbb{E}_{G \sim \Gamma}[\mathcal{Z}(G)^2] &= \mathbb{E}_{\theta \sim \Theta} \operatorname{det}(\1_m - \theta^2 M^2)^{-1/2}.
\end{align}
Writing $b_1, \ldots, b_m$ for the eigenvalues of $B$, since $\|B\|_{\operatorname{op}} \leq 1/3$ we know
\begin{align}
    \|M\|_{\operatorname{op}} &= \max_{j \in [m]} \left|\frac{b_j}{1+b_j}\right| \leq \max_{j \in [m]}\frac{|b_j|}{1-|b_j|} \leq \frac{1}{2} \\\|M\|_F^2 &= \sum_{j \in [m]} \frac{b_j^2}{(1+b_j)^2} \leq \frac{9}{4}\sum_{j \in [m]}b_j^2 = \frac{9}{4}\|B\|_F^2.
\end{align}
In particular, the eigenvalues of $\theta^2M^2$ belong to $[0, 1/4]$ and so
\begin{equation}
    -\frac{1}{2}\log \operatorname{det}(\1_m - \theta^2 M^2) \leq \frac{2}{3}\theta^2 \operatorname{tr}(M^2) = \frac{2}{3}\theta^2 \|M\|_F^2 \leq \frac{3}{2}\theta^2\|B\|_F^2
\end{equation}
since $-\log(1-x) \leq 4x/3$ for $x \in [0, 1/4]$. Thus,
\begin{equation}
    \mathbb{E}_{G \sim \Gamma}[\mathcal{Z}(G)^2] \leq \mathbb{E}_{\theta \sim \Theta}\exp\left(\frac{3}{2}\theta^2\|B\|_F^2\right).
\end{equation}
Using that $\theta \sim \Theta$ is distributed as the first coordinate of a uniform vector on $\mathbb{S}^{n-1}$ by rotational invariance, we know $R\theta \sim \mathcal{N}(0, 1)$ with $R^2 \sim \chi_n^2$ drawn independently. Since $\mathbb{E}[R^2] = n$ we find
\begin{align}
    \mathbb{E}_{\theta \sim \Theta} \exp(t\theta^2) &= \mathbb{E}_{\theta \sim \Theta} \exp\left(\mathbb{E}_{R^2 \sim \chi^2_n} \frac{t\theta^2 R^2}{n}\right) \\&\leq \mathbb{E}_{\xi \sim \mathcal{N}(0, 1)} \exp\left(\frac{t\xi^2}{n}\right) \\
    &= \frac{1}{\sqrt{1-2t/n}}
\end{align}
for all $t \in [0, n/2)$, where the second line is Jensen's inequality. Taking $t = \frac{3}{2}\|B\|_F^2$, we conclude that for any $n > 3\|B\|_F^2$ we have
\begin{equation}
    \mathbb{E}_{G \sim \Gamma}[\mathcal{Z}(G)^2] \leq \left(1 - \frac{3\|B\|_F^2}{n}\right)^{-1/2}. 
\end{equation}
Combining with (\ref{eqn:proofsubcriticalupperboundlemmaexpectationofZboundinterms}) yields that whenever $ n > 3\|B\|_F^2$,
\begin{equation}
  \operatorname{TV}\left(\operatorname{Law}_{(G, a) \sim P}(Ga), \nu\right)    \leq  \frac{1}{2} \sqrt{\left(1 - \frac{3\|B\|_F^2}{n}\right)^{-1/2} - 1}.
\end{equation}
If $n \geq 6\|B\|_F^2$ then $x := 3\|B\|_F^2/n \in [0, 1/2]$, in which case we use $(1-x)^{-1/2} - 1 \leq 2x$ to find that
\begin{equation}
    \operatorname{TV}\left(\operatorname{Law}_{(G, a) \sim P}(Ga), \nu\right)    \leq \frac{1}{2} \sqrt{2x} = \sqrt{\frac{3}{2n}} \|B\|_F.
\end{equation}
If $n < 6\|B\|_F^2$ then we simply use
\begin{equation}
    \operatorname{TV}\left(\operatorname{Law}_{(G, a) \sim P}(Ga), \nu\right)  \leq 1 < \sqrt{\frac{6}{n}}\|B\|_F,
\end{equation}
and so the claimed result always holds.
\end{proof}

\subsection{Proof of Lemma \ref{lemma:gaussianlowerboundstep1}: polynomial representation of Gaussian samplers}
\label{subsec:appendixrevelationlemma}

\begin{proof}[Proof of Lemma \ref{lemma:gaussianlowerboundstep1}]
    Fix $A \in \mathcal{M}_{d,\kappa}$. By padding with zero queries, assume the algorithm makes exactly $K$ total queries in $T$ rounds. Suppose $\mathcal{H}$ is a $\sigma$-algebra and $E \subset \r^d$ is a random subspace for which $E$ and the operator $U^\top|_E$ are $\mathcal{H}$-measurable and
    \begin{equation}
        \label{eqn:proofrevelationconditionalhaarproperty}\operatorname{Law}\left(U^\top|_{E^\perp} \mid \mathcal{H}\right) = \operatorname{Haar}(\operatorname{Iso}(E^\perp, (U^\top E)^\perp)),
    \end{equation}
    where $\operatorname{Iso}(V, W)$ denotes the set of linear isometries from $V$ onto $W$. Thus, $(\mathcal{H}, E)$ are a Haar state according to \cite[Definition C.2]{zhou2026complexitygaussiansampling} with $O = U^\top$.
    We apply Facts C.3-C.4 of \cite[Appendix C.1]{zhou2026complexitygaussiansampling} with $O = U^\top$ to get the following revelation rules:
    \begin{enumerate}
        \item[(1)] For any $\mathcal{H}$-measurable query $x \in \r^d$, we may reveal $Ux$ and construct a random vector $g \in \r^d$ such that
    \begin{equation}
        \operatorname{Law}(g \mid \mathcal{H}) = \mathcal{N}(0, \1_d), \quad Ux \in E + \operatorname{span}\{g\}.
    \end{equation}
    This vector $g$ has the property that if $E' := E + \operatorname{span}\{Ux\}$ and $\mathcal{H}' := \mathcal{H} \vee \sigma(g, Ux)$, then $(\mathcal{H}', E')$ is a Haar state.
    \item[(2)] For any $\mathcal{H}'$-measurable vector $b \in \r^d$, we may reveal $U^\top b$ and set $E'' := E' + \operatorname{span}\{b\}$ and $\mathcal{H}'' := \mathcal{H}' \vee \sigma(U^\top b)$, in which case $(\mathcal{H}'', E'')$ is a Haar state.
    \end{enumerate}
    We apply these rules to show the main result. Let $k_t \in \n \cup \{0\}$ denote the cumulative number of queries through round $t \in [T]$, inclusive, and write $x_1, \ldots, x_K$ for the queries in chronological order (with an arbitrary order chosen for queries in the same parallel batch). 
    Define the subspaces
    \begin{align}
        E_t &:= \operatorname{span}\left\{Ux_i, AUx_i:\; 1 \leq i \leq k_t\right\}, \\ S_t &:= \operatorname{span}\left\{A^j g_i:\; 1 \leq i \leq k_t,\; 0 \leq j \leq t\right\}.
    \end{align}
    We construct independent standard Gaussian vectors $g_1, \ldots, g_K \sim \mathcal{N}(0, \1_d)$ and histories $\mathcal{H}_t$ such that $E_t \subset S_t$ and the conditional Haar property (\ref{eqn:proofrevelationconditionalhaarproperty}) holds with $(\mathcal{H}, E) = (\mathcal{H}_t, E_t)$ for all $t \in [T]$, i.e. each $(\mathcal{H}_t, E_t)$ is a Haar state.

    To begin, we set $k_0 := 0$ and $E_0 = S_0 = \{0\}$ and let $\mathcal{H}_0$ be the $\sigma$-algebra generated by the algorithm's internal randomness. Since this is independent of $U$, we know $(\mathcal{H}_0, E_0)$ is a Haar state. Suppose that the claim holds through round $t-1$ for some $t \in [T]$, i.e. $E_{t-1} \subset S_{t-1}$ and $(\mathcal{H}_{t-1}, E_{t-1})$ is a Haar state. The batch size $k_t - k_{t-1}$ and the queries $x_{k_{t-1} + 1}, \ldots, x_{k_t}$ are all $\mathcal{H}_{t-1}$-measurable. Apply revelation rule (1) successively to these queries, before revealing any of their responses. Before processing $x_i$, the revealed subspace is
    \begin{equation}
        E_{t-1} + \operatorname{span}\{Ux_j:\; k_{t-1} < j < i\}.
    \end{equation}
    Consequently, induction over the queries in this batch gives
    \begin{equation}
    \label{eqn:proofhaarstateUxinclusion}
        Ux_i \in E_{t-1} + \operatorname{span}\{g_j:\; k_{t-1} < j \leq i\}.
    \end{equation}
    Each $g_i$ has conditional law $\mathcal{N}(0, \1_d)$ given the preceding history, which contains all previously constructed Gaussians.  After revealing $Ux_i$ for each $i \in \{k_{t-1}+1, \ldots, k_t\}$, each vector $AUx_i$ is measurable with respect to the current history. Apply revelation rule (2) successively with $b := AUx_i$, at each step revealing the oracle responses $U^\top AUx_i$. Let $\mathcal{H}_t$ denote the resulting history, i.e.
    \begin{equation}
        \mathcal{H}_t := \mathcal{H}_{t-1} \vee \sigma(g_i, Ux_i, U^\top AUx_i:\; k_{t-1} < i \leq k_t).
    \end{equation}
    The revealed subspace is now
    \begin{equation}
        E_{t-1} + \operatorname{span}\{Ux_i, AUx_i:\; k_{t-1} < i \leq k_t\} = E_t,
    \end{equation}
    and by the revelation rules $(\mathcal{H}_t, E_t)$ is a Haar state. If we denote
    \begin{equation}
        F_t := E_{t-1} + \operatorname{span}\{g_i:\; k_{t-1 }< i \leq k_t\}
    \end{equation}
    then by (\ref{eqn:proofhaarstateUxinclusion}) we know 
    \begin{equation}
        E_t \subset F_t + AF_t \subset S_t
    \end{equation}
    since $E_{t-1} \subset S_{t-1}$ by the inductive hypothesis. This completes the induction. Lastly, since each $g_i$ has
    \begin{equation}
        \operatorname{Law}(g_i \mid \sigma(g_j:\; j < i)) = \mathcal{N}(0, \1_d)
    \end{equation}
    and $g_1 \sim \mathcal{N}(0, \1_d)$,
    we know that the sequence $g_1, \ldots, g_K \sim \mathcal{N}(0, \1_d)$ is i.i.d..\footnote{Fixing a deterministic $i \in [K]$ and letting $\mathcal{F}_i$ denote the entire history immediately before constructing $g_i$, we know $\sigma(g_1, \ldots, g_{i-1}) \subset \mathcal{F}_i$. By revelation rule (1), $\mathbb{P}(g_i \in B \mid \mathcal{F}_i) = \gamma_d(B)$ for each Borel $B \subset \r^d$, where $\gamma_d := \mathcal{N}(0, \1_d)$. Hence, for $B_1, \ldots, B_i \subset \r^d$ Borel we have
    \begin{equation}
        \mathbb{P}(g_1 \in B_1, \ldots, g_i \in B_i) = \mathbb{E}\left[\prod_{j < i}\1_{g_j \in B_j} \mathbb{P}(g_i \in B_i \mid \mathcal{F}_i) \right] = \gamma_d(B_i) \mathbb{P}(g_1 \in B_1, \ldots, g_{i-1} \in B_{i-1}).
    \end{equation}
    Induction proves that $g_1, \ldots, g_i \sim \gamma_d$ i.i.d..}

    Since $Z(V_{U^\top AU})$ is $\mathcal{H}_T$-measurable and $(\mathcal{H}_T, E_T)$ is a Haar state, we apply revelation rule (1) once more with $x = Z(V_{U^\top AU})$. This gives a standard Gaussian $g_{K+1}$, independent of $g_1, \ldots, g_K$, such that
    \begin{equation}
        \overline{Z}(V_A) = UZ(V_{U^\top AU}) \in E_T + \operatorname{span}\{g_{K+1}\} \subset \operatorname{span}\{A^jg_i:\; 1 \leq i \leq K+1,\; 0 \leq j \leq T\}.
    \end{equation} 
    Consequently, we must have
    \begin{equation}
        \overline{Z}(V_A) \overset{(d)}{=} \sum_{i=1}^{K+1}p_i(A) g_i, \quad \operatorname{deg}p_i \leq T.
    \end{equation}
    The polynomial coefficients may be chosen measurably using the pseudoinverse of the matrix whose columns are the vectors $A^jg_i$. Each simulated query to $U^\top AU$ at $x$ uses one query to $A$ at $Ux$, and queries within the same round may be simulated in parallel. Finally, writing $U_\#$ for the pushforward by $z \mapsto Uz$, since $U$ is independent of the algorithm's internal randomness we know
    \begin{equation}
        \operatorname{Law}(\overline{Z}(V_A) \mid U) = U_\# \mathcal{A}^{V_{U^\top AU}}.
    \end{equation}
    Averaging over $U \sim H$ with $H := \operatorname{Haar}(O(d))$, we have
    \begin{equation}
        \operatorname{Law}(\overline{Z}(V_A)) = \int_{O(d)} U_\# \mathcal{A}^{V_{U^\top AU}} dH(U).
    \end{equation}
    Furthermore, since $U_\# \pi_{U^\top AU} = \pi_A$ for every $U \in O(d)$ we also have
    \begin{equation}
        \pi_A =  \int_{O(d)} U_\# \pi_{U^\top AU}dH(U).
    \end{equation}
    By convexity of TV under mixtures and rotation invariance,
    \begin{align}
        \operatorname{TV}(\operatorname{Law}(\overline{Z}(V_A)), \pi_A) &\leq \mathbb{E}_U \left[\operatorname{TV}(U_\# \mathcal{A}^{V_{U^\top AU}}, U_\# \pi_{U^\top AU})\right] \\
        &=  \mathbb{E}_U \left[\operatorname{TV}( \mathcal{A}^{V_{U^\top AU}}, \pi_{U^\top AU})\right].
    \end{align}
    Averaging over $\mathcal{D}$ yields the claim.
\end{proof}

\subsection{Proof of Lemma \ref{lemma:gaussianlowerboundstep2}: lower bound for polynomial samplers}
\label{subsec:appendixgaussianlowerboundstep2}

\begin{proof}[Proof of Lemma \ref{lemma:gaussianlowerboundstep2}]
    Writing $N := 2r-1$ and $\epsilon := \epsilon_{r,\kappa}$, let us define
    \begin{equation}
        A_0 := \operatorname{diag}(\lambda_0 \1_b, \ldots, \lambda_N \1_b), \quad \lambda_j := \frac{\kappa + 1 - (\kappa - 1) \cos(j \pi / N)}{2} \quad (\forall j \in \{0, \ldots, N\}).
    \end{equation}
    There are $N + 1 = 2r$ blocks, and so $A_0$ has dimension $d=2br$. Since
    \begin{equation}
        1 = \lambda_0 < \lambda_1 < \ldots < \lambda_N = \kappa,
    \end{equation}
    we have $\1_d \preceq A_0 \preceq \kappa \1_d$. Denoting the projections onto the blocks by $P_0, \ldots, P_N$, we define
    \begin{equation}
        E := \left\{z \in \r^d:\; \left|\lambda_j \frac{\|P_jz\|^2}{b}-1\right| \leq \frac{\epsilon}{8} \quad \forall j \in \{0, \ldots, N\}\right\}.
    \end{equation}
    This is a closed set, and is therefore Borel. 

    For (i), if $X \sim \pi_{A_0} = \mathcal{N}(0, A_0^{-1})$ then
    \begin{equation}
        \lambda_j \|P_jX\|^2 \sim \chi_b^2 \quad (\forall j \in \{0, \ldots, N\}).
    \end{equation}
    By applying the $\chi^2$ concentration estimate of Lemma \ref{lemma:chisquare} with $u = \epsilon / 8$, $m=b$, and taking a union bound over the $2r$ blocks, we get
    \begin{equation}
        \pi_{A_0}(E^c) \leq 4r \exp\left(-\frac{b\epsilon^2}{512}\right).
    \end{equation}
    To show (ii), we first apply Lemma \ref{lemma:chebyshevapproximation}(ii) to see
    \begin{equation}
\label{eqn:proofgaussianlowerboundlemma2polynomialobstruction}
        \max_{0 \leq j \leq N} |1 - \lambda_j S(\lambda_j)| \geq \epsilon \quad (\forall \text{ polynomials $S$ with $\operatorname{deg}S \leq N-1$}).
    \end{equation}
    Fix $n \in \n$ and a polynomial output
    \begin{equation}
        Z = \sum_{\ell = 1}^n p_\ell (A_0) g_\ell, \quad \operatorname{deg}p_\ell \leq r-1.
    \end{equation}
    Set 
    \begin{equation}
        G := [g_1, \ldots, g_n] \in \r^{d \times n}, \quad S(x) := \sum_{\ell = 1}^n p_\ell(x)^2
    \end{equation}
    so that $\operatorname{deg}S \leq 2r-2 = N-1$, and consider the event
    \begin{equation}
        \mathcal{G} := \left\{\left\|\frac{1}{b}G^\top P_j G - \1_n\right\|_{\operatorname{op}} \leq \frac{\epsilon}{4} \quad (\forall j \in \{0, \ldots, N\})\right\}.
    \end{equation}
    Defining $\mathbf{p}_j = (p_1(\lambda_j), \ldots, p_n(\lambda_j))^\top$, observe that $P_j Z =  P_j G\mathbf{p}_j$ while $S(\lambda_j) = \|\mathbf{p}_j\|^2$. Hence, on the event $\mathcal{G}$ we apply the operator norm bound to get that
    \begin{equation}
        \left(1 - \frac{\epsilon}{4}\right)S(\lambda_j) \leq \frac{\|P_jZ\|^2}{b} \leq \left(1 + \frac{\epsilon}{4}\right)S(\lambda_j).
    \end{equation}
    If $Z \in E$ as well, then
    \begin{equation}
        \frac{1 - \epsilon / 8}{1 + \epsilon / 4} \leq \lambda_j S(\lambda_j) \leq \frac{1 + \epsilon / 8}{1 - \epsilon / 4}
    \end{equation}
    and so we find
    \begin{equation}
        \max_{j \in \{0, \ldots, N\}} |1 - \lambda_j S(\lambda_j)| \leq \frac{\epsilon / 8 + \epsilon / 4}{1 - \epsilon / 4} \leq \frac{\epsilon}{2},
    \end{equation}
    contradicting (\ref{eqn:proofgaussianlowerboundlemma2polynomialobstruction}). Thus, $\mathcal{G} \subset \{Z \notin E\}$ and so
    \begin{equation}
        \mathbb{P}(Z \in E) \leq \mathbb{P}(\mathcal{G}^c).
    \end{equation}
    To bound this probability, take a fixed $1/4$-net $\mathcal{N}$ of $\mathbb{S}^{n-1}$ with at most $9^n$ points. For each symmetric matrix $B$, we know
    \begin{equation}
        \|B\|_{\operatorname{op}} \leq 2 \max_{v \in \mathcal{N}} |v^\top Bv|.
    \end{equation}
    Since we know $\|P_j Gv\|^2 \sim \chi_b^2$ for each $v \in \mathcal{N}$, we apply Lemma \ref{lemma:chisquare} and take a union bound over the net and the $2r$ blocks to find
    \begin{align}
        \mathbb{P}(\mathcal{G}^c) &\leq \sum_{j=0}^N \sum_{v \in \mathcal{N}} \mathbb{P}\left(\left|\frac{\|P_jGv\|^2}{b} - 1\right| > \frac{\epsilon}{8}\right)  \\
        &\leq 4r 9^n \exp\left(-\frac{b\epsilon^2}{512}\right).
    \end{align}
    This proves (ii), after which the main claim follows from 
    \begin{equation}
        \operatorname{TV}(\operatorname{Law}(Z), \pi_{A_0}) \geq \pi_{A_0}(E) - \mathbb{P}(Z \in E).
    \end{equation}
\end{proof}

\subsection{Proof of Proposition \ref{prop:parameterizedlowerboundGaussian}: parameterized Gaussian lower bound}
\label{subsec:appendixparameterizedgaussianprop}
\begin{proof}[Proof of Proposition \ref{prop:parameterizedlowerboundGaussian}]
    Fix $b, r, \kappa$, set $d = 2br$, and let $A_0$, $E$ be given by Lemma \ref{lemma:gaussianlowerboundstep2}. Draw $U \sim \operatorname{Haar}(O(d))$, define
    \begin{equation}
        V_U(z) := \frac{1}{2}z^\top U^\top A_0 Uz ,
    \end{equation}
    and let $\mathcal{D}_{b,r}^{(\kappa)} := \operatorname{Law}(V_U)$. The map $U \mapsto V_U$ is continuous into $C^\infty(\r^d)$, so this law is Borel. Moreover,
    \begin{equation}
        \nabla^2 V_U = U^\top A_0 U, \quad \1_d \preceq U^\top A_0 U \preceq \kappa \1_d,
    \end{equation}
    which proves (i). For (ii), let us abbreviate
    \begin{equation}
        x := b\epsilon_{r,\kappa}^2, \quad K := \left\lfloor \frac{x}{1024 \log 9}\right\rfloor \geq 1,
    \end{equation}
    where we may assume $K \geq 1$ since if $\frac{x}{1024 \log 9} < 1$ then
    \begin{equation}
        40re^{-x/1024} > \frac{40r}{9} > 1
    \end{equation}
    and so the claimed lower bound is trivial. Let $\mathcal{A}_{[K]}$ be the truncated algorithm which follows $\mathcal{A}$, but returns 0 before submitting any batch that would make its cumulative query count exceed $K$. This algorithm uses at most $K$ total queries and at most $r-1$ rounds almost surely. Couple it with $\mathcal{A}$ by using the same internal randomness, which means their outputs agree whenever $\mathcal{A}$ uses at most $K$ queries. By Markov's inequality,
    \begin{align}
        \mathbb{E}_{V \sim \mathcal{D}_{b,r}^{(\kappa)}}\left[\operatorname{TV}(\mathcal{A}^{V}, \mathcal{A}_{[K]}^{V})\right] &= \mathbb{E}_{U}\left[\operatorname{TV}(\mathcal{A}^{V_U}, \mathcal{A}_{[K]}^{V_U})\right] \\ &\leq \mathbb{P}_{U, \mathcal{A}}\left(\mathcal{Q}(V_U) > K\right) \\&\leq \frac{\mathbb{E}_{U, \mathcal{A}}[\mathcal{Q}(V_U)]}{K+1} \leq \frac{Q}{K+1} \\&\leq \frac{1024 \log 9}{b \epsilon_{r,\kappa}^2}Q.
        \label{eqn:proofgaussianlowerboundpropcouplingbound}
    \end{align}
    Now, apply Lemma \ref{lemma:gaussianlowerboundstep1} to $\mathcal{A}_{[K]}$ to get the symmetrized algorithm $\overline{\mathcal{A}_{[K]}}$, and let $Y := \overline{Z}(V_{A_0})$ be its output on the instance $A_0$.
    Since the round count is at most $r-1$, we have the distributional equality
    \begin{equation}
        Y \overset{(d)}{=} \sum_{\ell = 1}^{K+1} p_\ell(A_0) g_\ell, \quad \operatorname{deg}p_\ell \leq r-1.
    \end{equation}
    By definition of the symmetrized algorithm, its output at $A_0$ satisfies
    \begin{equation}
        Y \overset{(d)}{=} U Z_{[K]}(V_U),
    \end{equation}
    where we denote the output of $\mathcal{A}_{[K]}$ on instance $V_U$ by $Z_{[K]}(V_U)$ and $U \sim \operatorname{Haar}(O(d))$ is independent of the algorithm's internal randomness. Thus,
    \begin{equation}
        \operatorname{Law}(Y) = \int_{O(d)} U_\# \mathcal{A}_{[K]}^{V_U} dH(U),
    \end{equation}
    where $H$ denotes $\operatorname{Haar}(O(d))$ and $U_\#$ denotes the pushforward by $z \mapsto Uz$. On the other hand, $
        \pi_{V_U} = \mathcal{N}(0, U^\top A_0^{-1}U)$ and so 
        \begin{equation}
            U_\# \pi_{V_U} = \mathcal{N}(0, A_0^{-1}) \quad (\forall U \in O(d)).
        \end{equation}
        Hence, by convexity of TV under mixtures and invariance under orthogonal transformations,
        \begin{align}
            \operatorname{TV}(\operatorname{Law}(Y), \mathcal{N}(0, A_0^{-1})) &= \operatorname{TV}\left(\int_{O(d)} U_\# \mathcal{A}_{[K]}^{V_U}dH(U),  \int_{O(d)} U_\# \pi_{V_U} dH(U)\right) \\&\leq 
            \int_{O(d)} \operatorname{TV}\left(U_\# \mathcal{A}_{[K]}^{V_U}, U_\# \pi_{V_U}\right)dH(U) \\
            &= \mathbb{E}_U \left[\operatorname{TV}\left( \mathcal{A}_{[K]}^{V_U}, \pi_{V_U}\right)\right] \\
            &= \mathbb{E}_{V \sim \mathcal{D}_{b,r}^{(\kappa)}} \left[\operatorname{TV}(\mathcal{A}_{[K]}^V, \pi_V)\right].
        \end{align}
        Combining this with (\ref{eqn:proofgaussianlowerboundpropcouplingbound}) and using the reverse triangle inequality,
        \begin{align}
            \mathbb{E}_{V \sim \mathcal{D}_{b,r}^{(\kappa)}} \left[\operatorname{TV}(\mathcal{A}^V, \pi_V)\right] &\geq \mathbb{E}_{V \sim \mathcal{D}_{b,r}^{(\kappa)}} \left[\operatorname{TV}(\mathcal{A}_{[K]}^V, \pi_V)\right] - \mathbb{E}_{V \sim \mathcal{D}_{b,r}^{(\kappa)}} \left[\operatorname{TV}(\mathcal{A}^V, \mathcal{A}_{[K]}^V)\right] \\&\geq \operatorname{TV}(\operatorname{Law}(Y), \mathcal{N}(0, A_0^{-1})) - \frac{1024 \log 9}{b\epsilon_{r,\kappa}^2}Q.
        \end{align}
        We now apply Lemma \ref{lemma:gaussianlowerboundstep2} to $Y$ with $n = K+1$ to get
        \begin{equation}
            \operatorname{TV}(\operatorname{Law}(Y), \mathcal{N}(0, A_0^{-1})) \geq 1 - 4re^{-x/512} - 4r \exp\left((K+1) \log 9 - \frac{x}{512}\right).
        \end{equation}
        By definition of $K$,
        \begin{equation}
            (K+1 ) \log 9 - \frac{x}{512}\leq \log 9 - \frac{x}{1024}.
        \end{equation}
        Therefore, (ii) follows from
        \begin{align}
            \operatorname{TV}(\operatorname{Law}(Y), \mathcal{N}(0, A_0^{-1})) &\geq 1 - 4re^{-x/512} - 36r e^{-x/1024} \\&\geq  1 - 40re^{-b\epsilon_{r,\kappa}^2 / 1024}.
        \end{align}
\end{proof}
\end{document}